\documentclass{article}
\usepackage{authblk}
\usepackage[margin=1in]{geometry}
\usepackage{xcolor}
\usepackage{amsmath, amsfonts, amssymb, amsthm}
\usepackage{listings}
\usepackage{hyperref}
\usepackage{todonotes}
\usepackage{booktabs}
\usepackage{tabularx}
\usepackage{lineno}
\usepackage{orcidlink}
\usepackage{makecell}
\usepackage{float}
\usepackage{rotating}
\usepackage{booktabs}
\usepackage{multirow}
\usepackage[flushleft]{threeparttable}
\usepackage[american]{babel}
\usepackage{csquotes}
\usepackage{courier}
\usepackage{enumitem}
\usepackage{lineno}

\DeclareMathOperator{\E}{E}
\DeclareMathOperator{\V}{V}
\DeclareMathOperator{\Prob}{P}

\usepackage[
    backend=biber,
    style=apa,
    sorting=nyt,
    uniquelist=true
]{biblatex}
\DeclareBibliographyCategory{dataset}

\addtocategory{dataset}{rioux_paquette_data_2014}
\addtocategory{dataset}{mulla_data_2026}
\AtEveryBibitem{%
  \ifcategory{dataset}
    {\printtext{[dataset]\space}}
    {}%
}
\title{A Geometric Perspective for Marginal $R^2$: Estimation and Inference in Generalized Linear Mixed Models}

\author[1]{Hao Qin \orcidlink{0009-0000-8261-6148}\thanks{Corresponding author}}
\author[1]{Ayumi Mizuno \orcidlink{0000-0003-0822-5637}}
\author[1]{Kyle Morrison \orcidlink{0000-0002-3700-2398}}
\author[1]{Allison K. White \orcidlink{0009-0006-5398-026X}}
\author[1]{Shinichi Nakagawa \orcidlink{0000-0002-7765-5182}\thanks{Corresponding author}}

\affil[1]{\footnotesize
Department of Biological Sciences, Faculty of Science,
University of Alberta, Edmonton, AB, Canada
}

\affil[]{\footnotesize
Email addresses:
Hao Qin, \texttt{hq7@ualberta.ca};
Ayumi Mizuno, \texttt{amizuno@ualberta.ca};
Kyle Morrison, \texttt{khmorri1@ualberta.ca};
Allison K. White, \texttt{akwhite1@ualberta.ca};
Shinichi Nakagawa, \texttt{snakagaw@ualberta.ca}
}

\affil[]{\footnotesize
Correspondence:
\texttt{hq7@ualberta.ca};
\texttt{snakagaw@ualberta.ca}
}

\begin{document}
\maketitle

\section*{Abstract}
Generalized linear mixed models (GLMMs) are widely used across scientific disciplines, yet quantifying response variation explained by auxiliary information remains challenging for non-Gaussian outcomes and hierarchical dependence. Existing $R^2$ measures provide useful summaries, but their variance-based constructions are not generally linked to a common geometric interpretation, a fixed-design marginal estimand, and corresponding statistical inference. We propose marginal quasi-$R^2$ through two distinct operations. First, marginalization integrates over latent random effects, retaining their uncertainty in the marginal response moments and specifying the model for explanatory assessment. Second, a Quasi-GLM formulation supplies a local working information metric under which orthogonal projection yields an Pythagorean decomposition of explained and unexplained variation. We define a fixed-design population estimand, establish consistency and asymptotic normality of its estimator, and derive cluster-robust sandwich standard errors and confidence intervals under stated regularity conditions. Simulations assess estimation accuracy and interval coverage across response families and random-effects structures. Two ecological applications illustrate the importance of reporting estimation uncertainty alongside marginal quasi-$R^2$. The method is implemented in the R package \texttt{quasiR2}.

\hfill\break
\noindent \textbf{Keywords:} coefficient of determination; variance decomposition; marginal effects; generalized linear mixed models; explanatory power

\newpage

\section{Introduction}
Generalized linear mixed models (GLMMs) provide a flexible statistical framework for non-Gaussian responses and hierarchical dependence structures, and are widely used in ecology, sociology, and medicine \parencite{jiang_linear_2007, bolker_generalized_2009, mizuno_promoting_2025}. A natural question arising from a fitted GLMM is how much of the response variation can be explained by the available auxiliary information, commonly summarized through the coefficient of determination, $R^2$. Extending this interpretation beyond Gaussian linear models, however, requires careful consideration of the statistical structure in which explained variation is defined. GLMMs specify response distributions conditional on latent random effects, whereas marginal response distributions are obtained by integrating over their distribution. This marginalization incorporates latent heterogeneity into the marginal moment structure, which need not preserve the mean and variance relationships of the original conditional model. For Gaussian identity-link models with zero-mean random effects, the marginal mean retains the fixed-effect linear predictor, making the conventional sum-of-squares interpretation more readily applicable \parencite{rao_small_2015}. Under nonlinear links the marginal mean generally differs from the inverse-link transformation of the conditional fixed-effect predictor, and the associated variance structure may also change \parencite{nakagawa_coefficient_2017}. These considerations motivate a quasi-model perspective, in which the marginalized response is characterized through its moment structure without requiring a fully specified marginal distribution, as in quasi-likelihood methods \parencite{wedderburn_quasi-likelihood_1974} and generalized estimating equations (GEE; \textcite{liang_longitudinal_1986}). Consequently, defining explained variation requires specifying both the marginal statistical model and the geometric structure under which the decomposition is constructed.

Several measures of marginal $R^2$ for GLMMs have been proposed, including the distribution-specific variance decomposition of \textcite{nakagawa_general_2013}, the extensions of \textcite{jaeger_r2_2019}, and the matrix-based formulations of \textcite{piepho_coefficient_2019, piepho_adjusted_2023}. These approaches provide useful summaries of explained variation through fixed-effect contributions, variance components, or covariance-based comparisons, but leave several questions relevant to the present study. First, their constructions are not generally derived from a common geometric projection principle. In classical Gaussian linear regression, the coefficient of determination follows from orthogonal projection and the Pythagorean theorem \parencite{montgomery_introduction_1982}. Extending this interpretation to GLMMs requires identifying the statistical object being decomposed and specifying an appropriate inner product. Second, the definition and interpretation of explained variation depend on how the covariate design and underlying population target are treated. In particular, variation evaluated over an empirical covariate distribution need not represent the same estimand as a fixed-design quantity. Third, existing measures are not generally accompanied by a unified procedure for deriving influence functions, standard errors, and confidence intervals. These differences can complicate the interpretation and comparison of reported $R^2$ values, as well as the assessment of their estimation uncertainty.

We address these questions through a marginalized Quasi-GLM framework that separates the specification of the marginal statistical model from the geometric decomposition of explained variation. The construction proceeds in two steps. First, marginalization maps the joint distribution of responses and random effects onto the marginal response space. Although the latent random effects are integrated out, their associated uncertainty is retained in the marginal moment structure. These moments can be obtained from a fitted GLMM analytically in special cases or through numerical integration \parencite{liu_note_1994}. Second, a moment-based Quasi-GLM formulation equips this marginal space with a locally defined information metric. Orthogonal projection under the specified working metric yields an Pythagorean decomposition, extending the geometric interpretation of classical $R^2$ to non-Gaussian responses. Thus, the proposed measure assesses the explanatory contribution of auxiliary information within the marginalized geometry, without requiring a separate allocation of explained variation to fixed- and random-effect variance components. This construction defines a fixed-design population estimand and its corresponding sample estimator. We further develop a two-step estimation procedure, derive its influence function, and establish asymptotic normality as the number of independent clusters $M\to\infty$. Under the stated regularity conditions, we obtain cluster-robust sandwich standard errors and confidence intervals for the marginal quasi-$R^2$ estimator, accommodating working-covariance misspecification while allowing estimation uncertainty to be assessed alongside explained variation. The resulting framework connects marginalization, geometric interpretation, population estimands, and statistical inference across different link functions and response families.

The remainder of the paper is organized as follows. Sections 2--4 develop the marginalized Quasi-GLM framework, derive the proposed marginal quasi-$R^2$, and define its population and sample versions. Section 5 establishes the asymptotic theory under stated regularity conditions. Section 6 presents simulation studies, Section 7 provides two ecological applications, and Section 8 discusses interpretation and limitations.

\section{The Marginalized Quasi-GLM}
\subsection{Defining Fixed Effects on the Marginal Scale}
In standard generalized linear models (GLMs), the fixed-effect contribution to the response $y_i$ is $g^{-1}(\mathbf{x}_i^T \boldsymbol{\beta})$, representing the conditional mean given covariates $\mathbf{x}_i$ and the fixed-effect coefficient vector $\boldsymbol{\beta}$. In generalized linear mixed models (GLMMs), the population marginal mean is $m_i(\boldsymbol{\theta}_0) := \E_{\mathbf{u}_i} \left[ g^{-1}(\mathbf{x}_i^T \boldsymbol{\beta}_0 + \mathbf{z}_i^T \mathbf{u}_i) \right]$. Throughout the paper, subscript $0$ denotes the unknown population truth. $\boldsymbol{\beta}_0$ is the true fixed-effect parameter vector, $\mathbf{u}_i \sim \mathcal{N}(\mathbf{0}, \boldsymbol{\Sigma}_{\mathbf{u}}(\boldsymbol{\psi}_{0}))$ is the unobserved cluster-specific random effect with covariance matrix $\boldsymbol{\Sigma}_{\mathbf{u}}(\boldsymbol{\psi}_{0})$, $\mathbf{z}_i$ is the corresponding random-effect design vector, $\boldsymbol{\psi}_{0}$ parametrizes the true random effect covariance, and $\boldsymbol{\theta}_0$ denotes the full true parameter vector collecting both fixed effects and variance components. For nonlinear link functions, $m_i(\boldsymbol{\theta}_0) \neq g^{-1}(\mathbf{x}_i^T \boldsymbol{\beta}_0)$ due to Jensen's inequality. For example, in a Poisson-log model, evaluating the normal moment-generating function yields $m_i(\boldsymbol{\theta}_0) = \exp(\mathbf{x}_i^T \boldsymbol{\beta}_0) \exp(\frac{1}{2}\mathbf{z}_i^T \boldsymbol{\Sigma}_{\mathbf{u}} (\boldsymbol{\psi}_{0})\mathbf{z}_i)$, showing a discrepancy driven by the random-effect variance. Because the fixed-effect contribution is ambiguous on the marginal scale, we propose a Quasi-GLM framework to formalize it.

Let $y_i$ denote the observed response for the $i$-th observation. We define the marginalized moments for the $i$-th observation. Let $\mathbf{d}_i = (\mathbf{x}_i^T, \mathbf{z}_i^T)^T$ denote the complete covariate profile, where $\mathbf{x}_i$ is the fixed-effect covariate and $\mathbf{z}_i$ is the random effect design covariate. Let $\boldsymbol{\gamma} = (\boldsymbol{\psi}^T, \phi)^T$ collect the GLMM nuisance parameters, where $\boldsymbol{\psi}$ parametrizes the random-effect covariance $\boldsymbol{\Sigma}_\mathbf{u}(\boldsymbol{\psi})$ and $\phi$ is the dispersion parameter. The full parameter vector is $\boldsymbol{\theta} = (\boldsymbol{\beta}^T, \boldsymbol{\gamma}^T)^T$.

Let $\eta_i := \mathbf{x}_i^T \boldsymbol{\beta}$ denote the conditional linear predictor at $\mathbf{u} = \mathbf{0}$ and $\mathbf{u} \sim \mathcal{N}(\boldsymbol{0}, \Sigma_{\mathbf{u}}(\boldsymbol{\psi}))$. The marginal mean and its derivative with respect to $\eta_i$ are defined as:
$$
m_i(\boldsymbol{\theta}) := \E_{\mathbf{u}} \left[ g^{-1}(\eta_i + \mathbf{z}_i^T \mathbf{u}) \right], \quad
m_i^\prime(\boldsymbol{\theta}) := \E_{\mathbf{u} } \left[ \frac{\partial}{\partial \eta_i}g^{-1}(\eta_i + \mathbf{z}_i^T \mathbf{u}) \right]
$$

Applying a working independence assumption motivated by Generalized Estimating Equations (GEE) \parencite{liang_longitudinal_1986}, we formulate a quasi-score equation \parencite{wedderburn_quasi-likelihood_1974} with a working variance function $\V(m_i)$ evaluated at the marginal mean:
$$
\mathbf{U}_{\text{Quasi}}(\boldsymbol{\beta}) = \sum_{i=1}^{n} \mathbf{x}_i m_i^\prime \frac{y_i - m_i}{\V(m_i)} = \sum_{i=1}^{n} \mathbf{x}_i \underbrace{\left( \frac{(m_i^\prime)^2}{\V(m_i)} \right)}_{w_i^\star} \underbrace{\left( \frac{y_i - m_i}{m_i^\prime} \right)}_{R_i} = \mathbf{X}^T \mathbf{W}^\star \mathbf{R}
$$
where, $w_i^\star(\boldsymbol{\theta}) := [m_i^\prime(\boldsymbol{\theta})]^2 / \V(m_i(\boldsymbol{\theta}))$ is the Quasi-GLM observation-level working weight, and $\V(\cdot)$ is the variance function of the underlying exponential family (e.g., $\V(m) = m$ for Poisson, $\V(m) = m(1-m)$ for Binomial). Throughout this paper, the qualifier \emph{working} signals a quantity derived from the quasi-likelihood approximation to the marginal model rather than from the true joint distribution implied by the GLMM.
  
Indexed by the true GLMM parameter $\boldsymbol{\theta}_0 = (\boldsymbol{\beta}_0^T, \boldsymbol{\gamma}_0^T)^T$,  the Quasi-GLM is specified by the following working conditions for each observation $i$:
$$
\begin{aligned}
    \text{Mean function:} \quad & \E[y_i] = m_i(\boldsymbol{\theta}_0), \\
    \text{Working Variance:} \quad & \text{Var}_{\text{work}}(y_i) := \V(m_i(\boldsymbol{\theta}_0)) = \frac{[m_i^\prime(\boldsymbol{\theta}_0)]^2}{w_i^\star(\boldsymbol{\theta}_0)}, \\
    \text{Working Independence:} \quad & \text{Cov}_{\text{work}}(y_i, y_j) = 0 \quad \text{for } i \neq j.
\end{aligned}
$$
Because the true GLMM data-generating process introduces intra-cluster correlation (thus $\text{Cov}_{\text{true}}(y_i, y_j) \neq 0$) and overdispersion on the marginal scale arising from the unobserved between-cluster variance and the nonlinear distortion of the variance function via Jensen's inequality, the working moments differ from the true moments. We use the empirical sandwich estimator (Section \ref{section:asymptotic}) to account for this discrepancy. 

\textbf{Proposition 1 (Identification).} Under the GLMM sampling model, the population quasi-score equation $\E[\mathbf{U}_{\text{Quasi}} (\boldsymbol{\beta}; \boldsymbol{\gamma}_0)] = \boldsymbol{0}$ has a unique solution at $\boldsymbol{\beta} = \boldsymbol{\beta}_0$. That is, the Quasi-GLM identifies the same fixed-effect parameter as the GLMM. This follows directly from $\E[y_i] = m_i(\boldsymbol{\theta}_0)$ (by definition of $m_i$) and the strict monotonicity of the marginal mean map $\boldsymbol{\beta} \mapsto (m_1(\boldsymbol{\beta}, \boldsymbol{\gamma}_0),  \dots, m_n(\boldsymbol{\beta}, \boldsymbol{\gamma}_0))$ under full-rank $\mathbf{X}$ (Proof in Appendix A).

To operationalize the Quasi-GLM framework, evaluation of the marginal components is required. Table \ref{tab:population_marginal_mean} summarizes the closed-form expressions for the marginal mean $m_i(\boldsymbol{\theta})$ and its derivative $m_i^\prime(\boldsymbol{\theta})$ under standard exponential dispersion family distributions.
\begin{table}[H]
\centering
\caption{Population marginal mean $m_i(\boldsymbol{\theta})$ and derivative $m_i^{\prime}(\boldsymbol{\theta})$ across exponential-dispersion families. "CF" denotes closed form; "GH" denotes $K-\text{point}$ Gauss-Hermite quadrature.}
\label{tab:population_marginal_mean}
\setlength{\tabcolsep}{4pt}
\small
\renewcommand{\arraystretch}{1.4}
\begin{tabular}{@{} l l c c c @{}}
\toprule
\textbf{Family}& \textbf{Link} & $m_i(\boldsymbol{\theta})$& $m_i^{\prime} (\boldsymbol{\theta})$& \textbf{Method}\\ 
\midrule
Gaussian     & identity& $\eta_i$& $1$            & CF\\
Poisson& log (canonical)& $\exp(\eta_i + \frac{1}{2} \sigma^2_{\mathbf{z}_i})$& $\exp(\eta_i + \frac{1}{2} \sigma^2_{\mathbf{z}_i})$& CF\\
Gamma& log (non-canonical)& $\exp(\eta_i + \frac{1}{2} \sigma^2_{\mathbf{z}_i})$& $\exp(\eta_i + \frac{1}{2} \sigma^2_{\mathbf{z}_i})$& CF\\
Neg Binomial& log (non-canonical)  & $\exp(\eta_i + \frac{1}{2} \sigma^2_{\mathbf{z}_i})$& $\exp(\eta_i + \frac{1}{2} \sigma^2_{\mathbf{z}_i})$& CF\\
Bernoulli/Binomial& probit (non-canonical)& $\Phi\left( \frac{\eta_i}{\sqrt{1 + \sigma^2_{\mathbf{z}_i}}} \right)$& $\frac{1}{\sqrt{1 + \sigma_{\mathbf{z}_i}^2}} \varphi\left( \frac{\eta_i}{\sqrt{1 + \sigma_{\mathbf{z}_i}^2}} \right)$&CF\\
Bernoulli/Binomial& logit (canonical)& $\displaystyle \int \frac{\exp{(\eta_i + v)}}{1 + \exp{(\eta_i + v)}} \phi(v; 0 , \sigma^2_{\mathbf{z}_i}) \, dv$& $\displaystyle \int \frac{\exp{(\eta_i + v)}}{(1 + \exp{(\eta_i + v)})^2} \phi(v; 0, \sigma^2_{\mathbf{z}_i}) \, dv$& GH\\
Beta& logit& $\displaystyle \int \frac{\exp{(\eta_i + v)}}{1 + \exp{(\eta_i + v)}} \phi(v; 0 , \sigma^2_{\mathbf{z}_i}) \, dv$& $\displaystyle \int \frac{\exp{(\eta_i + v)}}{(1 + \exp{(\eta_i + v)})^2} \phi(v; 0, \sigma^2_{\mathbf{z}_i}) \, dv$& GH\\ 
Inverse Gaussian& log (non-canonical)& $\exp(\eta_i + \frac{1}{2} \sigma^2_{\mathbf{z}_i})$& $\exp(\eta_i + \frac{1}{2} \sigma^2_{\mathbf{z}_i})$&CF\\
Tweedie& log (non-canonical)& $\exp(\eta_i + \frac{1}{2} \sigma^2_{\mathbf{z}_i})$& $\exp(\eta_i + \frac{1}{2} \sigma^2_{\mathbf{z}_i})$&CF\\
\bottomrule
\end{tabular}
\vspace{1ex}

{\raggedright \footnotesize 
Note: $\sigma^2_{\mathbf{z}_i} := \mathbf{z}_i^T \boldsymbol{\Sigma}_{\mathbf{u}}(\boldsymbol{\psi}) \mathbf{z}_i$ is the random-effect contribution to the linear-predictor variance for observation $i$. $\eta_i := \mathbf{x}_i^T \boldsymbol{\beta}$ is the conditional linear predictor at $\mathbf{u} = \mathbf{0}$.  $\Phi$ is the cumulative distribution function of normal distribution. $\varphi$ is the probability density function of normal distribution. For Bernoulli and Binomial families, the response is modeled as a proportion. \par}
\end{table}

\section{\texorpdfstring{$R^2$}{R2} Calculation on the Quasi-GLM}
\subsection{The Marginal Working Response}
Building upon the Iteratively Reweighted Least Squares (IRLS) formulation of the quasi-score equation, we define the marginal working response for the $i$-th observation by linearizing around the population marginal mean:
$$
Y_{\text{Quasi}, i}^{\star}(\boldsymbol{\beta}) := \mathbf{x}_{i}^T \boldsymbol{\beta} + \frac{y_i - m_i(\boldsymbol{\theta})}{m_i^\prime(\boldsymbol{\theta})}
$$
By definition of the Quasi-GLM working independence structure, $\text{Cov}_{\text{work}}(Y_{\text{Quasi}, i}^{\star}, Y_{\text{Quasi}, j}^{\star}) = 0$ for $i \neq j$.

Evaluating the marginal components $m_i(\boldsymbol{\theta})$ and $m^{\prime}_{i}(\boldsymbol{\theta})$ requires integrating over the random-effect distribution. For the Gaussian-identity model, the expectations evaluate trivially because $g^{-1}$ is identical. For the log link and probit link, they admit closed-form expressions via the log-normal moment-generating function and probability density function of the normal distribution, respectively (as summarized in Table \ref{tab:population_marginal_mean}). For other link functions, these expectations reduce to low-dimensional Gauss-Hermite quadrature integrals.

When the random effect variance approaches zero  ($\boldsymbol{\Sigma}_{\mathbf{u}} \rightarrow \mathbf{0}$), the GLMM reduces to a standard GLM. Consequently, $m_i(\boldsymbol{\theta}) \rightarrow g^{-1}(\mathbf{x}_{i}^{T}\boldsymbol{\beta})$ and $m^{\prime}_{i} (\boldsymbol{\theta}) \rightarrow \frac{\partial}{\partial \eta_i} g^{-1} (\mathbf{X}_i^T \boldsymbol{\beta})$, and the marginal working response recovers the IRLS adjusted dependent variable of classical GLMs \parencite{mccullagh_generalized_1989}:
$$
Y_{\text{Quasi}, i}^\star \bigg|_{\boldsymbol{\Sigma}_{\mathbf{u}} = \mathbf{0}} = \mathbf{x}_i^T \boldsymbol{\beta} + \frac{y_i - g^{-1}(\mathbf{x}_i^T \boldsymbol{\beta})}{\frac{\partial}{\partial \eta_i}g^{-1}(\mathbf{x}_i^T \boldsymbol{\beta})}.
$$
Our construction naturally extends the standard GLM working response to the marginal space of GLMMs. 

\subsection{A Two-Step Estimation Procedure}
The underlying GLMM serves as the data-generating model, specifying the full conditional distribution $p(y_i \mid \mathbf{u}_{j(i)}; \boldsymbol{\beta}, \phi)$ and the random-effect distribution $p(\mathbf{u}_{j(i)} ; \boldsymbol{\psi})$ where $j$ indexes the $j$-th cluster and $i$ indexes the $i$-th observation in $j$-th cluster. The Quasi-GLM serves as the evaluation model, providing a working-scale variance decomposition necessary to formalize the variance explained by fixed effects.

The two frameworks are connected through a two-step estimation procedure \parencite{newey_chapter_1994}:

\textbf{Step 1 (GLMM fitting).} Fit the specified GLMM via maximum likelihood to obtain consistent estimates of the variance components $\hat{\boldsymbol{\psi}}$ and the dispersion parameter $\hat{\phi}$.

\textbf{Step 2 (Quasi-GLM evaluation).} Substitute $\hat{\boldsymbol{\psi}}$ and $\hat{\phi}$ into the Quasi-GLM conditions to construct the working weights $\hat{\mathbf{W}}^\star$ and the marginal working responses $\hat{\mathbf{Y}}_{\text{Quasi}}^\star$. Apply IRLS on this Quasi-GLM to obtain the quasi-likelihood fixed-effect estimate $\hat{\boldsymbol{\beta}}_{\text{Quasi}}$ and the associated variance decomposition. The marginal $R^2$ is then computed.

Under this paradigm, the GLMM estimates the nuisance parameters $(\boldsymbol{\psi}, \phi)$, while the Quasi-GLM provides inference on the fixed effects and the marginal $R^2$. This separation mirrors the generalized estimating equations (GEE) framework \parencite{liang_longitudinal_1986}, where a working correlation structure is evaluated using initial estimates and plugged into the estimating equations to provide inference on the primary regression parameters.

\subsection{Variance Decomposition on the Quasi-GLM}
The Quasi-GLM established above is fitted via iteratively reweighted least squares (IRLS). At convergence, the estimator $\hat{\boldsymbol{\beta}}_{\text{Quasi}}$ satisfies the quasi-score equation:
$$\mathbf{X}^T \mathbf{W}^\star(\hat{\boldsymbol{\beta}}_{\text{Quasi}}) \left[ \mathbf{Y}_{\text{Quasi}}^\star(\hat{\boldsymbol{\beta}}_{\text{Quasi}}) - \mathbf{X} \hat{\boldsymbol{\beta}}_{\text{Quasi}} \right] = \mathbf{0},$$
where $\mathbf{W}^\star = \text{diag}(w_1^\star, \dots, w_n^\star)$ is the working-weight matrix and $\mathbf{Y}^\star_{\text{Quasi}}$ is the marginal working response vector, both evaluated at the final estimates. This represents $\mathbf{W}^\star$-weighted normal equations, guaranteeing that the working residual vector is $\mathbf{W}^\star$-orthogonal to the column space of the design matrix $\mathbf{X}$. This geometric property directly underpins the variance decomposition that follows.

Let $\hat{\boldsymbol{\eta}} := \mathbf{X}\hat{\boldsymbol{\beta}}_{\text{Quasi}}$ denote the fitted linear predictors and $\hat{\mathbf{R}} := \mathbf{Y}^\star_{\text{Quasi}} - \hat{\boldsymbol{\eta}}$ the working residuals.

To construct the $R^2$ metric, we must measure the total variance of the working response relative to an intercept-only null model. Define the weighted working grand mean as $\bar{Y}_{w}^\star := (\mathbf{1}^T \mathbf{W}^\star \mathbf{1})^{-1} (\mathbf{1}^T \mathbf{W}^\star \mathbf{Y}_{\text{Quasi}}^\star)$, and similarly let $\bar{\eta}_w := (\mathbf{1}^T \mathbf{W}^\star \mathbf{1})^{-1} (\mathbf{1}^T \mathbf{W}^\star \hat{\boldsymbol{\eta}})$ be the weighted mean of the fitted values.

Algebraically, the centered total variance decomposes as: 
$$\mathbf{Y}_{\text{Quasi}}^\star - \bar{Y}_w^\star \mathbf{1} = \underbrace{(\hat{\boldsymbol{\eta}} - \bar{\eta}_w \mathbf{1})}_{\text{Explained variance}} + \underbrace{\hat{\mathbf{R}}}_{\text{Working residual}} + \underbrace{(\bar{\eta}_w - \bar{Y}_w^\star) \mathbf{1}}_{\text{Centering adjustment}}$$

Under the assumption that $\mathbf{X}$ includes an intercept column ($\mathbf{1} \in \text{col}(\mathbf{X})$), the normal equations ensure that $\mathbf{1}^T \mathbf{W}^\star \hat{\mathbf{R}} = 0$. This implies $\bar{Y}_w^\star = \bar{\eta}_w$, thereby vanishing the centering adjustment. The decomposition simplifies to:
$$\mathbf{Y}_{\text{Quasi}}^\star - \bar{Y}_w^\star \mathbf{1} = (\hat{\boldsymbol{\eta}} - \bar{\eta}_w \mathbf{1}) + \hat{\mathbf{R}}$$

Because the intercept is in the model and $\mathbf{X}^T \mathbf{W}^\star \hat{\mathbf{R}} = \mathbf{0}$, the explained variance is perfectly $\mathbf{W}^\star$-orthogonal to the working residual, yielding $(\hat{\boldsymbol{\eta}} - \bar{\eta}_w \mathbf{1})^T \mathbf{W}^\star \hat{\mathbf{R}} = 0$. This orthogonality leads immediately to the exact Pythagorean identity:
$$    \| \mathbf{Y}_{\text{Quasi}}^\star - \bar{Y}_{w}^{\star} \mathbf{1} \|^2_{\mathbf{W}^\star} = \|\hat{\boldsymbol{\eta}} - \bar{\eta}_w \mathbf{1} \|^2_{\mathbf{W}^\star} + \| \hat{\mathbf{R}} \|^2_{\mathbf{W}^\star}$$

We define these components as $\text{SST}_w$, $\text{SSR}_w$, and $\text{SSE}_w$ respectively, such that $\text{SST}_w = \text{SSR}_w + \text{SSE}_w$ holds exactly in finite samples. This decomposition applies to any IRLS convergence point for any positive diagonal $\mathbf{W}^\star$, remaining independent of distributional assumptions.

The generalized marginal $R^2$ is defined as the ratio of explained to total working-scale variance:
$$R^2 := \frac{\text{SSR}_w}{\text{SST}_w} = \frac{\|\hat{\boldsymbol{\eta}} - \bar{\eta}_w \mathbf{1} \|^2_{\mathbf{W}^\star}}{\| \mathbf{Y}_{\text{Quasi}}^\star - \bar{Y}_{w}^{\star} \mathbf{1}\|^2_{\mathbf{W}^\star}}$$
Because the decomposition is orthogonal under the $\mathbf{W}^\star$-inner product, the coefficient can be geometrically interpreted as $R^2 = \cos^2 \theta$, where $\theta \in [0, \pi / 2]$ is the $\mathbf{W}^\star$-angle between the centered working response $(\mathbf{Y}_{\text{Quasi}}^\star - \bar{Y}_w^\star \mathbf{1})$ and the centered fitted predictors $(\hat{\boldsymbol{\eta}} - \bar{\eta}_w \mathbf{1})$.

\section{Population Estimand and Empirical Estimator}
\subsection{Population Quantities under Fixed Design}
We establish the asymptotic properties under a fixed-design framework. Observations $(y_i, \mathbf{x}_i, \mathbf{z}_i)$ for $i = 1, \dots, n$ arise from $M$ clusters, where $j(i) \in \{1, \dots, M\}$ denotes the cluster index for observation $i$. The fixed-effect design vectors $\mathbf{x}_i \in \mathbb{R}^p$ and random-effect design vectors $\mathbf{z}_i \in \mathbb{R}^q$ are deterministic. The unobserved cluster-specific random effects are independent and identically distributed as $\mathbf{u}_j \sim \mathcal{N}(\mathbf{0}, \boldsymbol{\Sigma}_\mathbf{u}(\boldsymbol{\psi}_0))$. Conditional on $\mathbf{u}_{j(i)}$, the response follows an exponential dispersion family $y_i \mid \mathbf{u}_{j(i)} \stackrel{\text{ind}}{\sim} f(\cdot \mid \mu_i, \phi_0)$, where $\mu_i = g^{-1}(\mathbf{x}_i^T \boldsymbol{\beta}_0 + \mathbf{z}_i^T \mathbf{u}_{j(i)})$. 

The true parameter vector $\boldsymbol{\theta}_0 = (\boldsymbol{\beta}_0^T, \boldsymbol{\psi}_0^T, \phi_0)^T$ lies in the interior of a compact parameter space $\Theta$. We consider the asymptotic regime where the number of clusters $M \rightarrow \infty$, while the cluster sizes $n_j$ are uniformly bounded.

To define the population-level $R^2$ estimand, all Quasi-GLM working components are evaluated at the true parameter $\boldsymbol{\theta}_0$. Let $m_i := m_i(\boldsymbol{\theta}_0)$, $m_i^\prime := m_i^\prime(\boldsymbol{\theta}_0)$, and $w_i^\star := [m_i^\prime]^2 / \V(m_i)$ denote the population marginal mean, its derivative, and the working weight, respectively.

\textbf{Population Signal ($\mathcal{F}_0$).} The signal represents the total working-weighted variance of the fixed-effect linear predictor across the design space:
$$
\mathcal{F}_0 := \boldsymbol{\beta}_0^T \mathbf{C}_0 \boldsymbol{\beta}_0,
$$
where $\mathbf{C}_0 = \left[ n^{-1} \mathbf{X}^T \mathbf{W}^\star \mathbf{X} - \bar{\mathbf{x}}_w \bar{\mathbf{x}}_w^T / s_{n} \right]$, with $\bar{\mathbf{x}}_w := n^{-1} \mathbf{X}^T \mathbf{W}^\star \mathbf{1}$ and $s_{n} := n^{-1} \mathbf{1}^T \mathbf{W}^\star \mathbf{1}$. The matrix $\mathbf{C}_0$ is the weighted covariance of the covariates, quantifying the spread of the marginal mean predictions after removing the intercept.

\textbf{Population Noise ($\mathcal{N}_0$).} The population noise measures the working-weighted expectation of the marginal squared working residuals. Evaluating the expectation yields:
$$
\mathcal{N}_0 := \frac{1}{n} \sum_{i = 1}^{n} w^\star_i \cdot \frac{\E[(y_i - m_i)^2]}{(m_i^\prime)^2}
$$
By definition of the marginal mean, $\E[y_i] = m_i$. Thus, the central moment $\E[(y_i - m_i)^2]$ is the true marginal variance $\text{Var}(y_i)$. Substituting the definition of the working weight $w_i^\star = (m_i^\prime)^2 / \V(m_i)$, the derivatives cancel out, yielding:
$$
\mathcal{N}_0 = \frac{1}{n} \sum_{i = 1}^{n} \frac{\text{Var}(y_i)}{\V(m_i)}
$$
Here, the true marginal variance is $\text{Var}(y_i) = \E_{\mathbf{u}}[\text{Var}(y_i \mid \mathbf{u})] + \text{Var}_{\mathbf{u}}(\E[y_i \mid \mathbf{u}])$ by the Law of Total Variance. This cancellation demonstrates that $\mathcal{N}_0$ is the average ratio between the true data-generating variance and the Quasi-GLM working variance. It acts as a global dispersion factor, capturing all variance unexplainable by the fixed effects: conditional sampling noise, unobserved random-effect heterogeneity, and the nonlinear distortion induced by Jensen's inequality.

\textbf{Cross-term Vanishing and Additive Decomposition.} 
To justify the $R^2$ estimand, the total population variance must additively decompose into the signal and noise components. We define the true marginal working response as $Y_i^\star := \mathbf{x}_i^T \boldsymbol{\beta}_0 + R_i$, and let $\bar{\eta}_0 := (\boldsymbol{1}^T \mathbf{W}^\star \boldsymbol{1})^{-1} (\boldsymbol{1}^T \mathbf{W}^\star \mathbf{X}\boldsymbol{\beta}_0)$ denote the weighted intercept projection. The total population variance is formalized as the limit of the expected working-weighted dispersion:
$$
\frac{1}{n} \sum_{i=1}^n w_i^\star \cdot \E \left[ (Y_i^\star - \bar{\eta}_0)^2 \right].
$$
Expanding the squared quadratic separates the total variance into a signal component $(\mathbf{x}_i^T \boldsymbol{\beta}_0 - \bar{\eta}_0)^2$, a noise component $R_i^2$, and a cross-term $2(\mathbf{x}_i^T \boldsymbol{\beta}_0 - \bar{\eta}_0)R_i$. 

Under the fixed-design asymptotic framework, the covariates $\mathbf{x}_i$, the true parameters $\boldsymbol{\beta}_0$, and the weights $w_i^\star$ are deterministic constants. Therefore, the expectation operator $\E[\cdot]$ penetrates these terms and acts only  on the random residual $R_i$. By the construction of the Quasi-GLM marginal mean, $m_i = \E[y_i]$, the working residual satisfies $\E[R_i] = \E[y_i - m_i] / m_i^\prime = 0$. Consequently, the expected cross-term perfectly vanishes:
$$
\frac{1}{n} \sum_{i=1}^n w_i^\star (\mathbf{x}_i^T \boldsymbol{\beta}_0 - \bar{\eta}_0) \cdot \underbrace{\E[R_i]}_{=0} = 0.
$$

This probabilistic unbiasedness geometrically orthogonalizes the signal from the noise, guaranteeing that the total expected variance equals $\mathcal{F}_0 + \mathcal{N}_0$. This confirms that the orthogonal decomposition empirically enforced by the IRLS normal equations is tracking a theoretically valid and additive population estimand.

To operationalize the Quasi-GLM framework, evaluation of the population noise ($\mathcal{N}_0$) is required. Table \ref{tab:n_0_tab} summarizes the closed-form expressions for the noise term $\mathcal{N}_{0}$.

\begin{table}[H]
\centering
\caption{Closed forms of the population noise $\mathcal{N}_M = n^{-1}\sum_i \text{Var}(y_i) / \V(m_i)$. "CF" denotes closed form; "GH" denotes $K$-point Gauss-Hermite quadrature.}
\label{tab:n_0_tab}

\renewcommand{\arraystretch}{1.4}

\begin{tabular*}{\textwidth}{@{\extracolsep{\fill}} l l l c @{}}
\toprule
\textbf{Family} & \textbf{Link} & $\mathcal{N}_0$ & \textbf{Method} \\
\midrule
Gaussian & identity & $\displaystyle 1 + n^{-1}\sum_i \sigma^2_{\mathbf{z}_i} / \sigma^2_0$ & CF \\
Poisson & log & $\displaystyle 1 + n^{-1}\sum_i m_i(\exp{(\sigma^2_{\mathbf{z}_i})}-1)$ & CF \\
Gamma & log & $\displaystyle n^{-1}\sum_i [(\phi+1)\exp{(\sigma^2_{\mathbf{z}_i}})-1]/\phi$ & CF \\
Neg Binomial & log & $\displaystyle n^{-1}\sum_i \frac{1+m_i[\exp{(\sigma^2_{\mathbf{z}_i})}(1+1/k)-1]}{1+m_i/k}$ & CF \\
Bernoulli & logit / probit & $1$ & CF \\
Binomial($n_{i}^{\text{trial}} > 1$) & logit  & 
  $\displaystyle 1 + n^{-1}\sum_i 
  \frac{(n_i^{\text{trial}}-1)\,\text{Var}_u(\mu_i)}
  {m_i(1-m_i)}$ & GH \\
Binomial($n_{i}^{\text{trial}} > 1$) & probit &
$1 + n^{-1} \sum_i \frac{(n_i^\text{trial} - 1)[\Phi_2 (\tilde{\eta}_i, \tilde{\eta}_i; \sigma^2_{\mathbf{z}_i}/ (1 + \sigma^2_{\mathbf{z}_i})) - m_i^2]}{m_i (1 - m_i)}$ & CF \\
Beta & logit & $\displaystyle 1 + n^{-1} \sum_{i} \frac{\tau \cdot \text{Var}_{u} (\mu_i)}{m_i (1 - m_i)}$ & GH \\
Inv.\ Gaussian & $\text{log}$ & $\displaystyle n^{-1}\sum_i \left [ \exp{(3 \sigma^2_{\mathbf{z}_i})} + \frac{\exp{(\sigma^2_{\mathbf{z}_i})} - 1}{\phi m_i} \right ]$ & CF \\
Tweedie & log & $\displaystyle n^{-1}\sum_i \left[\exp{\left (\frac{p(p-1)}{2}\sigma^2_{\mathbf{z}_i} \right)} + \frac{m_i^{2-p}}{\phi}(\exp{(\sigma^2_{\mathbf{z}_i})}-1)\right]$ & CF \\
\bottomrule
\end{tabular*}
\vspace{1ex}

{\raggedright \footnotesize 
\textbf{Note:} Consistent with our Quasi-GLM formulation, $\V(m_i)$ denotes the \textit{full} working variance function evaluated at the marginal mean, absorbing the family-specific nuisance parameters. For example, $\V(m_i) = \sigma^2_0$ for Gaussian, $m_i$ for Poisson, $\phi m_i^2$ for Gamma, $m_i + m_i^2/k$ for Negative Binomial, and $\phi m_i^p$ for Tweedie. Here, $\sigma^2_0$ is the Gaussian residual variance; $\phi$ is the dispersion parameter; $k$ is the overdispersion parameter for the Negative Binomial; $\tau = \alpha + \beta$ is the precision parameter for the Beta distribution and $\text{Var}_u(\mu_i) = \E_{u} [\mu_{i}^{2}] - m_i^2$, where $\mu_{i} = g^{-1}(\eta_{i} + \mathbf{z}_i^T \mathbf{u})$; and $p \in (1, 2)$ is the Tweedie power parameter. The conditional linear predictor at $\mathbf{u}=\mathbf{0}$ is defined as $\eta_i := \mathbf{x}_i^T \boldsymbol{\beta}$, and the conditional mean is $\mu_i^c := g^{-1}(\eta_i + \mathbf{z}_i^T \mathbf{u})$. Furthermore, $\sigma^2_{\mathbf{z}_i} := \mathbf{z}_i^T \boldsymbol{\Sigma}_{\mathbf{u}}(\boldsymbol{\psi}) \mathbf{z}_i$ denotes the random-effect contribution to the linear-predictor variance, where $\mathbf{z}_i$ is the corresponding random-effect design vector and $\boldsymbol{\Sigma}_{\mathbf{u}}(\boldsymbol{\psi})$ is the random-effect covariance matrix parameterized by $\boldsymbol{\psi}$. $\Phi_2$ is the standard normal bivariate cumulative distribution function. $\Phi$ is the standard normal cumulative distribution function. $\tilde{\eta}_i = \eta_i /\sqrt{1 + \sigma_{\mathbf{z}_i}^2}$ \par}
\end{table}

\textbf{Remark on Asymptotic Noise and Binary Data:} The behavior of $\mathcal{N}_0 = n^{-1} \sum_i \text{Var}(y_i) / \V(m_i)$ varies across exponential families. For Poisson count data where $\V(m_i) = m_i$, latent random effects induce overdispersion such that $\text{Var}(y_i) > \V(m_i)$, yielding $\mathcal{N}_0 > 1$. For Bernoulli data, however, the marginal variance is constrained by the marginal mean such that $\text{Var}(y_i) = m_i(1-m_i)$. Because the Quasi-GLM working variance is identically $\V(m_i) = m_i(1-m_i)$, it follows that $\mathcal{N}_0 = 1$, regardless of the random-effect magnitude. Under this invariant noise term, any reduction in the marginal $R^2$ due to increasing latent heterogeneity is driven entirely by the signal component $\mathcal{F}$. Specifically, as the random-effect variance increases, the non-linear transformation via Jensen's inequality attenuates the marginal slope $m_i^\prime$ toward zero. This flattening reduces the quasi-information weights $w_i^\star$, thereby diminishing the signal $\mathcal{F}$. Consequently, for binary data, the framework captures the diminished explanatory power of fixed effects through the systematic attenuation of the marginal signal rather than an inflation of the residual variance.

\subsection{The Finite-Design Population \texorpdfstring{$R^2$}{R2} Estimand}
Under the fixed-design framework where the covariate profile $\mathbf{D} = (\mathbf{X}, \mathbf{Z})$ is treated as deterministic, the population generalized marginal $R^2$ is defined directly conditional on the realized sample of $M$ independent clusters as:
$$
\rho_M^2 := \frac{\mathcal{F}_M}{\mathcal{F}_M + \mathcal{N}_M},
$$
where $\mathcal{F}_M = \boldsymbol{\beta}_0^T \mathbf{C}_M \boldsymbol{\beta}_0$, $\mathbf{C}_M = n^{-1} \mathbf{X}^T \mathbf{W}^\star(\boldsymbol{\theta}_0) \mathbf{X} - \bar{\mathbf{x}}_w \bar{\mathbf{x}}_w^T / s_n$, and $\mathcal{N}_M =  \frac{1}{n} \sum_{i = 1}^{n} \frac{\text{Var}(y_i)}{\V(m_i)}$ represent the deterministic signal and noise components evaluated at the true parameter $\boldsymbol{\theta}_0$, respectively.

\subsection{Empirical Estimators}
Following the two-step estimation procedure, we construct the empirical estimators by evaluating the Quasi-GLM components at the joint parameter estimate $\hat{\boldsymbol{\theta}} = (\hat{\boldsymbol{\beta}}_{\text{Quasi}}^T, \hat{\boldsymbol{\gamma}}^T)^T$. Here, $\hat{\boldsymbol{\gamma}} = (\hat{\boldsymbol{\psi}}^T, \hat{\phi})^T$ collects the nuisance parameters from the initial GLMM fit, and $\hat{\boldsymbol{\beta}}_{\text{Quasi}}$ is the marginalized fixed-effect estimate derived from the Quasi-GLM IRLS convergence.

The empirical fixed-effect signal is quantified by the sample working-weighted variance of the fitted linear predictors:
$$
\hat{\mathcal{F}} := \hat{\boldsymbol{\beta}}_{\text{Quasi}}^T \hat{\mathbf{C}} \hat{\boldsymbol{\beta}}_{\text{Quasi}},
$$
where $\hat{\mathbf{C}} := n^{-1} \mathbf{X}^T \hat{\mathbf{W}}^\star \mathbf{X} - \hat{\bar{\mathbf{x}}}_w \hat{\bar{\mathbf{x}}}_w^T / \hat{s}_n$ serves as the empirical weighted covariance matrix of the fixed-effect design. The weight matrix $\hat{\mathbf{W}}^\star = \text{diag}(\hat{w}_1^\star, \dots, \hat{w}_n^\star)$, the weighted covariate mean $\hat{\bar{\mathbf{x}}}_w = n^{-1} \mathbf{X}^T \hat{\mathbf{W}}^\star \mathbf{1}$, and the mean sum of weights $\hat{s}_n = n^{-1} \mathbf{1}^T \hat{\mathbf{W}}^\star \mathbf{1}$ are all evaluated at $\hat{\boldsymbol{\theta}}$.

The empirical noise is calculated by aggregating the squared marginal working residuals:
$$
\hat{\mathcal{N}} := \frac{1}{n} \sum_{i = 1}^{n} \hat{w}^\star_i \cdot \left( \frac{y_i - m_i(\hat{\boldsymbol{\theta}})}{m_i^\prime(\hat{\boldsymbol{\theta}})} \right)^2.
$$
The empirical squared residuals $(y_i - m_i(\hat{\boldsymbol{\theta}}))^2$ absorb unmodeled overdispersion, unobserved random-effect heterogeneity, and nonlinear Jensen distortions from the data.

The sample generalized marginal $R^2$ estimator is thus defined as the plug-in ratio:
$$
\hat{R}^2 := \frac{\hat{\mathcal{F}}}{\hat{\mathcal{F}} + \hat{\mathcal{N}}}.
$$
Because the IRLS convergence guarantees the sample-level orthogonal decomposition $\text{SST}_w = \text{SSR}_w + \text{SSE}_w$, this estimator is bounded within $[0, 1]$.

\section{Asymptotic Theory}
\subsection{Regularity Conditions}
The asymptotic results require the following key conditions (full conditions stated formally in Appendix A): (i) Correct specification of the marginal mean $m_i(\boldsymbol{\beta})$ (ii) No single cluster dominates the total sample, (iii) Finite fourth moments of the quasi-working residuals, which governs the rate at which the sandwich central limit theorem takes effect, and (iv) A fixed covariate design. Conditions (i)---(iii) are standard in clustered estimation literature.

\subsection{Asymptotic Properties}
\label{section:asymptotic}
Building upon the regularity conditions and the $\sqrt{M}$-consistency of the empirical parameters, we now establish the asymptotic consistency and normality of the proposed $R^2$ estimator. For notational consistency, let $\rho^{2}_{M} := \rho^{2}_{M}(\boldsymbol{\theta}_0, \mathbf{X})$ denote the true population estimand, and let $D_M := \mathcal{F}_M + \mathcal{N}_M$ denote the total expected population variance.

\textbf{Theorem 1 (Consistency).} Under the regularity conditions, as the number of independent clusters $M \rightarrow \infty$, the sample generalized marginal $R^2$ converges in probability to the population estimand:
$$
\hat{R}^2 \xrightarrow{p} \rho_{M}^2.
$$

\textbf{Theorem 2 (Asymptotic Normality).} Under the regularity conditions, as $M \rightarrow \infty$, the standardized estimator is asymptotically normally distributed:
$$ 
\sqrt{M} \left( \hat{R}^2 - \rho_{M}^2 \right) \xrightarrow{d} \mathcal{N} \left(0, \sigma^2_{\rho, M} \right), 
$$
where the asymptotic variance $\sigma^2_{\rho, M}$ is determined by the limit of the average cluster-level influence functions:
$$ 
\sigma^2_{\rho, M} = \frac{1}{M} \sum_{j=1}^M \text{Var}\left( \psi_j^{(R^2)} \right).
$$
This cluster-robust sandwich variance structure relies on the empirical influence function framework and generalized estimating equations for correlated data \parencite{huber_behavior_1967, white_heteroskedasticity-consistent_1980, liang_longitudinal_1986}. By an application of the functional delta method, the overall cluster-$j$ influence function $\psi_j^{(R^2)}$ is a linear combination of the signal influence ($\psi_j^{(\mathcal{F})}$) and the noise influence ($\psi_j^{(\mathcal{N})}$):
$$
\psi_j^{(R^2)} := \frac{\mathcal{N}_M}{D_M^2}\psi_j^{(\mathcal{F})} - \frac{\mathcal{F}_M}{D_M^2}\psi_j^{(\mathcal{N})}.
$$
These specific influence components map the random cluster-level fluctuations via the model score equations, taking the exact asymptotic linear forms:
\begin{align*}
    \psi_j^{(\mathcal{F})} &= \left[ 2\boldsymbol{\beta}_0^T \mathbf{C}_0 + \mathbf{D}_{\boldsymbol{\beta}_0}^T \right] \bar{\mathbf{I}}_0^{-1} \mathbf{s}_j^\star + \mathbf{D}_{\boldsymbol{\gamma}_0}^T \boldsymbol{a}_j
    \, \\
    \psi_j^{(\mathcal{N})} &= \mathbf{g}_{\mathcal{N}, \boldsymbol{\beta}}^T \bar{\mathbf{I}}_0^{-1} \mathbf{s}_j^\star + \mathbf{g}_{\mathcal{N}, \boldsymbol{\gamma}}^T \boldsymbol{a}_{j} + \frac{M}{n} \sum_{i \in j} A_i(\boldsymbol{\theta}_0) \left[ (y_i - m_i(\boldsymbol{\theta}_0))^2 - \text{Var}(y_i) \right],
\end{align*}
where the constituent matrices and gradients are deterministically defined as follows: $\mathbf{s}_j^\star = \mathbf{s}_{j} + \bar{\mathbf{J}}_{\boldsymbol{\gamma}, 0} \boldsymbol{a}_j$ is the nuisance-adjusted quasi-score for cluster $j$, where $\boldsymbol{a}_{j}$ is the first-stage influence function (formally specified in Assumption (R5) in Appendix A). This adjusts for the uncertainty in the two-stage estimation process \parencite{newey_chapter_1994}. 
$\mathbf{D}_{\boldsymbol{\beta}_0}$ and $\mathbf{D}_{\boldsymbol{\gamma}_0}$ are the parameter gradients of the covariance matrix $\mathbf{C}(\boldsymbol{\theta})$ evaluated at $\boldsymbol{\theta}_0$, with their $k$-th elements analytically defined as $\boldsymbol{\beta}_0^T \left( \partial \mathbf{C}_0 / \partial \beta_k \right) \boldsymbol{\beta}_0$ and $\boldsymbol{\beta}_0^T \left( \partial \mathbf{C}_0 / \partial \gamma_k \right) \boldsymbol{\beta}_0$.
$\mathbf{g}_{\mathcal{N}, \boldsymbol{\beta}}$ and $\mathbf{g}_{\mathcal{N}, \boldsymbol{\gamma}}$ are the conformably partitioned parameter-specific limits of the overall noise gradient $\mathbf{g}_{\mathcal{N}, \boldsymbol{\theta}} := \lim_{n \to \infty} \frac{1}{n} \sum_{i=1}^n \nabla_{\boldsymbol{\theta}} A_i(\boldsymbol{\theta}_0) \cdot \text{Var}(y_i)$.
$A_i(\boldsymbol{\theta}) := w_i^\star(\boldsymbol{\theta}) / (m_i^\prime(\boldsymbol{\theta}))^2$ defines the deterministic multiplier applied to the squared marginal residuals.

In practice, the theoretical variance $\sigma^2_{\rho, M}$ is consistently estimated by the empirical variance of the sample plug-in influence functions across all $M$ clusters:
$$ 
\hat{\sigma}^{2}_{\rho, M} = \frac{1}{M - 1} \sum_{j = 1}^{M} \left( \hat{\psi}_j^{(R^2)} - \bar{\hat{\psi}}^{(R^2)} \right)^2.
$$
The derivations for both theorems are provided in Appendix A.

\textbf{Confidence Interval Construction.} 
Because the generalized marginal $R^2$ is bounded within $[0, 1]$, standard symmetric Wald intervals computed directly on the original scale may yield bounds outside this mathematical support, particularly when the true explanatory power is near the boundaries (e.g., $0$ or $1$). To enforce these domain restrictions and improve finite-sample coverage properties, we construct the confidence interval on the logit scale \parencite{agresti_categorical_2013}.

Let $\hat{\zeta}$ denote the logit-transformed estimator:
$$ \hat{\zeta} := \log \left( \frac{\hat{R}^2}{1 - \hat{R}^2} \right). $$
By an application of the Delta method \parencite{vaart_asymptotic_1998}, the standard error of this transformed estimator is derived as:
$$ \hat{\text{se}}_{\zeta} := \frac{\hat{\sigma}_{\rho,M} / \sqrt{M}}{\hat{R}^2 \left( 1 - \hat{R}^2 \right)}, $$
where $\hat{\sigma}_{\rho, M}$ is the empirical sandwich standard deviation defined above. 

The asymptotic $100(1 - \alpha)\%$ confidence interval for the true estimand $\rho^{2}_{M}$ is then obtained by mapping the symmetric Wald interval back to the original probability scale via the inverse logit (expit) function:
$$ \text{CI}_{1-\alpha} = \left[ \text{expit} \left(\hat{\zeta} - z_{1 - \alpha / 2} \cdot \hat{\text{se}}_{\zeta} \right), \; \text{expit} \left(\hat{\zeta} + z_{1 - \alpha / 2} \cdot \hat{\text{se}}_{\zeta} \right) \right], $$
where $\text{expit}(x) = \exp(x) / (1 + \exp(x))$, and $z_{1 - \alpha / 2}$ is the upper $(1-\alpha/2)$ quantile of the standard normal distribution. By construction, this back-transformed interval is confined to $(0, 1)$. This transformation prevents the boundary violations inherent to naive symmetric intervals, ensuring stable and theoretically valid inference across the entire parameter space.

In finite samples, two corrections are applied.  First, the HC1 correction of \textcite{mackinnon_heteroskedasticity-consistent_1985} replaces the sandwich meat with the rescaled $\frac{M}{M-p}\sum_{j=1}^{M}\hat{\psi}_{j}\hat{\psi}_{j}^{\!\top}$ where $p$ is the number of fixed effect covariates, compensating for the downward bias that arises when the number of clusters~$M$ is small relative to the number of estimated parameters. Second, the normal quantile $z_{1-\alpha/2}$ is replaced by the Student-$t$ quantile $t_{M-p,\,1-\alpha/2}$, which provides more conservative intervals that account for the additional uncertainty in estimating the cluster-level variance from $M$ independent units. The finite-sample interval is therefore
\[
  \mathrm{CI}_{1-\alpha}^{\mathrm{HC1}}
  =
  \Bigl[\,
    \operatorname{expit}\!\bigl(\hat\zeta
      - t_{M-p,\,1-\alpha/2}\;\hat{\mathrm{se}}_{\zeta}^{\,\mathrm{HC1}}\bigr),
    \;
    \operatorname{expit}\!\bigl(\hat\zeta
      + t_{M-p,\,1-\alpha/2}\;\hat{\mathrm{se}}_{\zeta}^{\,\mathrm{HC1}}\bigr)
  \,\Bigr].
\]
When $M$ is large both corrections vanish: $M/(M-p)\to 1$ and $t_{M-p}\to z$, recovering the asymptotic interval above.

\section{Simulation Studies}
\subsection{Design}
Following the simulation design in our pre-registration \parencite{hao_qin_new_2026}, which is adapted from \textcite{jiang_usable_2022}, we evaluate the finite-sample performance of the proposed marginal $R^2$ estimator and its sandwich-based confidence interval across eight exponential dispersion families: Gaussian (identity), Poisson (log), negative binomial (log), Gamma (log), Tweedie (log, $p = 1.5$), binomial (logit), binomial (probit), and beta (logit). Random intercept ($n_j = 10$ observations per cluster) model is considered. The number of independent clusters varies over $M \in \{30, 50, 100, 200, 300\}$. HC1 corrected confidence intervals are used in this simulation study.

We consider three parameter conditions: (i) A (strong signal, small RE): $\boldsymbol{\beta} = (0.5, 1.0, -0.8)^T$, small random-effect variance. True $\rho^2_M \approx 0.20$--$0.50$, (ii) B (moderate signal, moderate RE): $\boldsymbol{\beta} = (0.3, 0.5, -0.3)^T$, moderate random-effect variance. True $\rho^2_M \approx 0.04$--$0.15$, and (iii) C (weak signal, large RE): $\boldsymbol{\beta} = (0.8, 0.2, -0.1)^T$, large random-effect variance. True $\rho^2_M \approx 0.006$--$0.015$.

Covariates $x_1, x_2 \overset{\text{iid}}{\sim} \text{Uniform}(-0.8, 0.8)$ are generated once and held fixed across replications (fixed design). For each scenario, the true population estimand $\rho^2_M$ is computed from the known parameters via Gauss--Hermite quadrature. All results are based on 10000 Monte Carlo replications. Models are fitted using \texttt{glmmTMB}; the proposed $R^2$ and its confidence intervals are computed using the \texttt{quasiR2} package.

\subsection{Performance metrics}

For each scenario, we report: (i) relative bias $(\hat{R}^2 - \rho^2_M)/\rho^2_M$ (in percent), (ii) empirical coverage of the 95\% logit-delta confidence interval, (iii) the ratio ASE/ESE of the average sandwich standard error to the empirical standard deviation of $\hat{R}^2$ across replications, and (iv) model convergence rate. An ASE/ESE ratio near 1.0 confirms the consistency of the sandwich variance estimator.

\subsection{Results}
\subsubsection{Consistency and bias decay}

\begin{figure}[H]
    \centering
    \includegraphics[width=\textwidth]{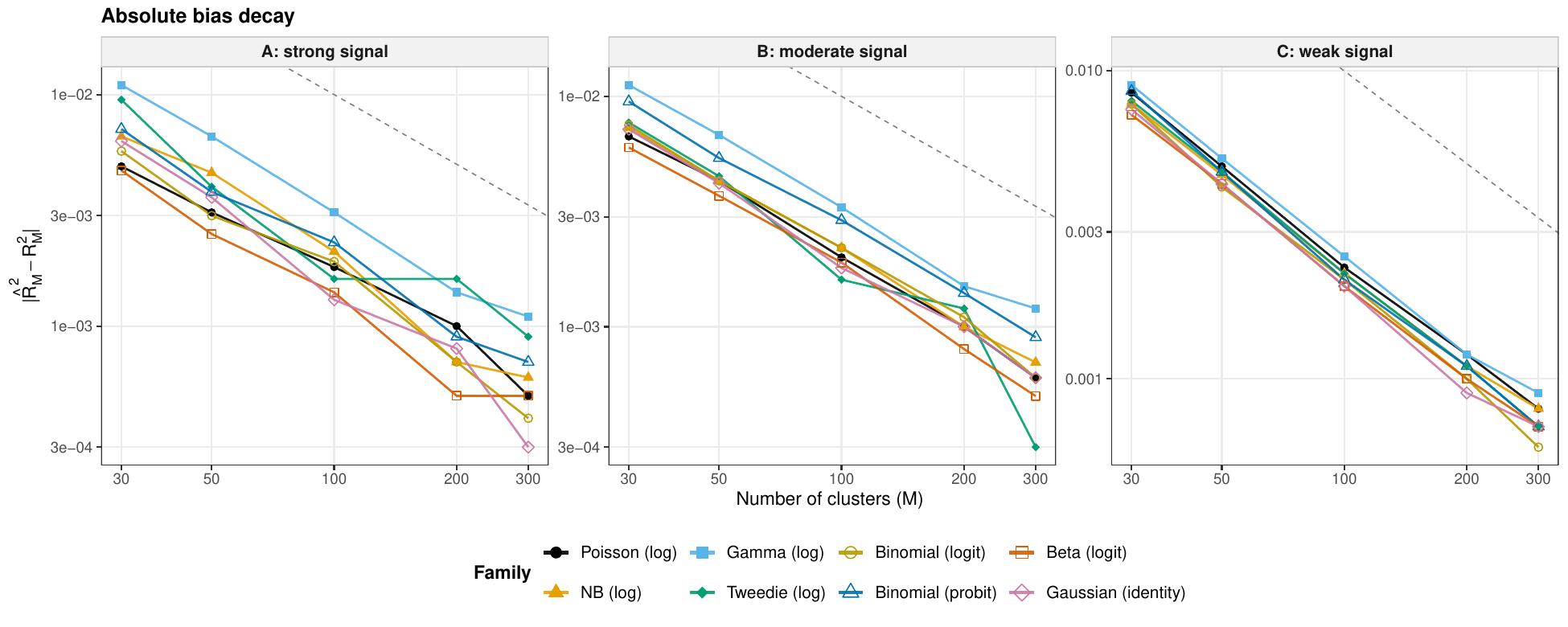}
    \caption{Absolute bias $\mid \hat{R}^{2} - \rho_{M}^{2} \mid$ as a function of the number of clusters $M$ (log--log scale) under the random intercept model ($n_{j} = 10$, $10000$ replications). The dashed line indicates the $O(1/M)$ reference rate. Bias decays at the theoretical rate across all families and configurations, confirming the $\sqrt{M}$-consistency established in Theorem 1.}
    \label{fig:log_log_deacy_int}
\end{figure}

The point estimator $\hat{R}^2$ is consistent across all families and specifications. Under A and B, the relative bias is below 4\% at $M = 30$ and below 1\% at $M \geq 200$ for all families (Figure~\ref{fig:log_log_deacy_int}). Under C, where $\rho^2_M < 0.02$, the relative bias is large in percentage terms at small $M$ (e.g., 70--130\% at $M = 30$), but the absolute bias $|\hat{R}^2 - \rho^2_M|$ remains below 0.01 and decays at the expected $O(1/M)$ rate. By $M = 300$, the relative bias is below 12\% even under C.

\subsubsection{Confidence interval coverage}

\begin{figure}[H]
    \centering
    \includegraphics[width=\textwidth]{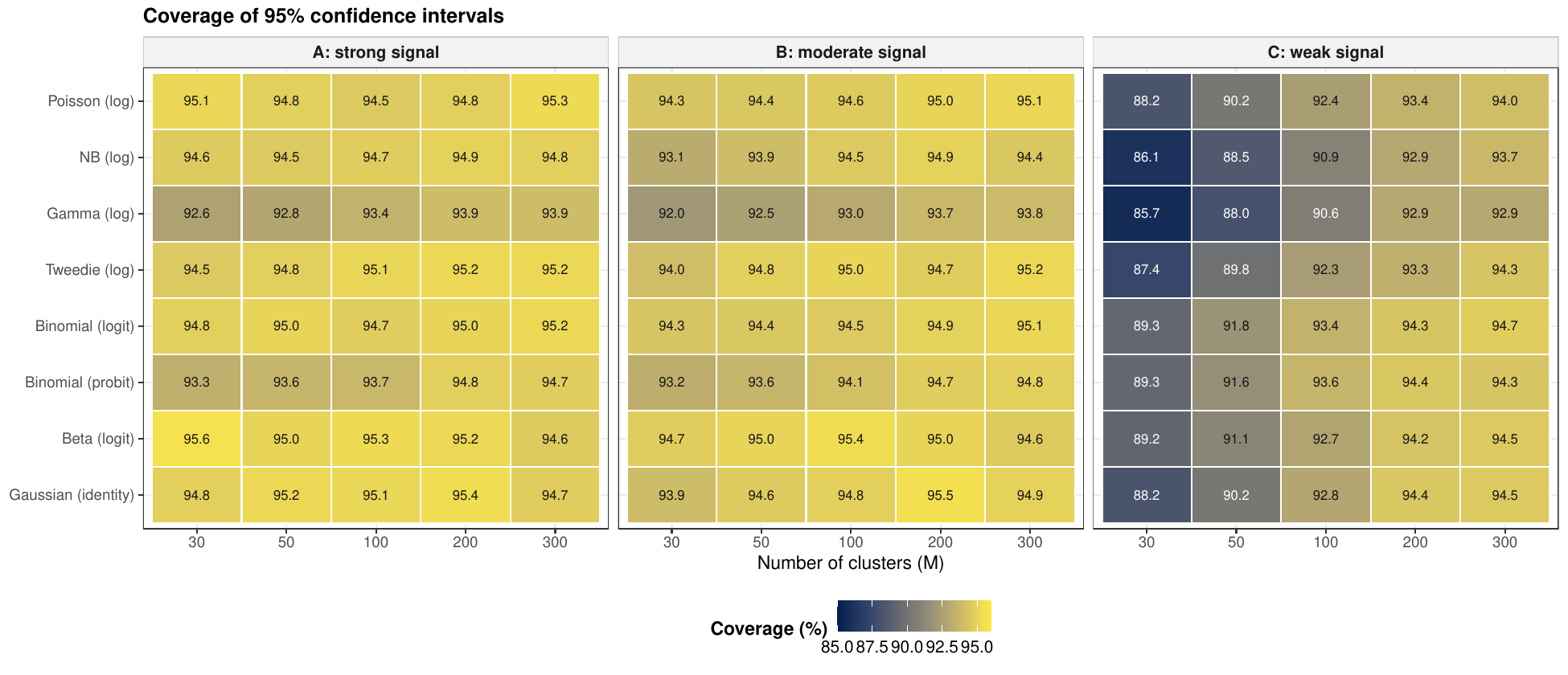}
    \caption{Empirical coverage (\%) of the 95\% log-delta confidence interval for $\rho_{M}^{2}$ under the random intercept model ($n_j = 10$, $10000$ replications). Rows correspond to eight exponential dispersion families; columns to three parameter specifications spanning strong ($\rho_{M}^{2} \approx 0.20 - 0.50$), moderate ($\rho_{M}^{2} \approx 0.04 - 0.15$), and weak ($\rho_{M}^{2} \le 0.02$) signal regimes. The nominal level (95\%) corresponds to light yellow. Coverage is at or above 93\% for all families when $M \ge 50$ under A and B, and recovers to nominal by $M \ge 200$ under C.}
    \label{fig:coverage}
\end{figure}

Figure~\ref{fig:coverage} displays the empirical coverage of the 95\% sandwich-based logit-delta confidence interval. Under A ($\rho^2_M \approx 0.20$--$0.50$), coverage is nominal ($\geq 92\%$) for all families and all $M$. Under B ($\rho^2_M \approx 0.04$--$0.15$), coverage is at or above 93\% for all families, with Binomial and Beta families achieving 94--96\%.

Under C ($\rho^2_M < 0.02$), coverage is below nominal at small $M$ (85--90\% at $M = 30$) but recovers to $\geq 93\%$ by $M = 200$--$300$, confirming the asymptotic validity of the estimator. The undercoverage at small $M$ arises from the logit-delta transformation: the variance inflation factor $[\hat{R}^2(1-\hat{R}^2)]^{-2}$ grows rapidly as $\hat{R}^2 \to 0$, amplifying small discrepancies between the sandwich standard error and the true standard deviation. This is a known limitation shared by Wald-type confidence intervals for near-boundary parameters.

\subsubsection{Sandwich variance consistency}

\begin{figure}[H]
    \centering
    \includegraphics[width=\textwidth]{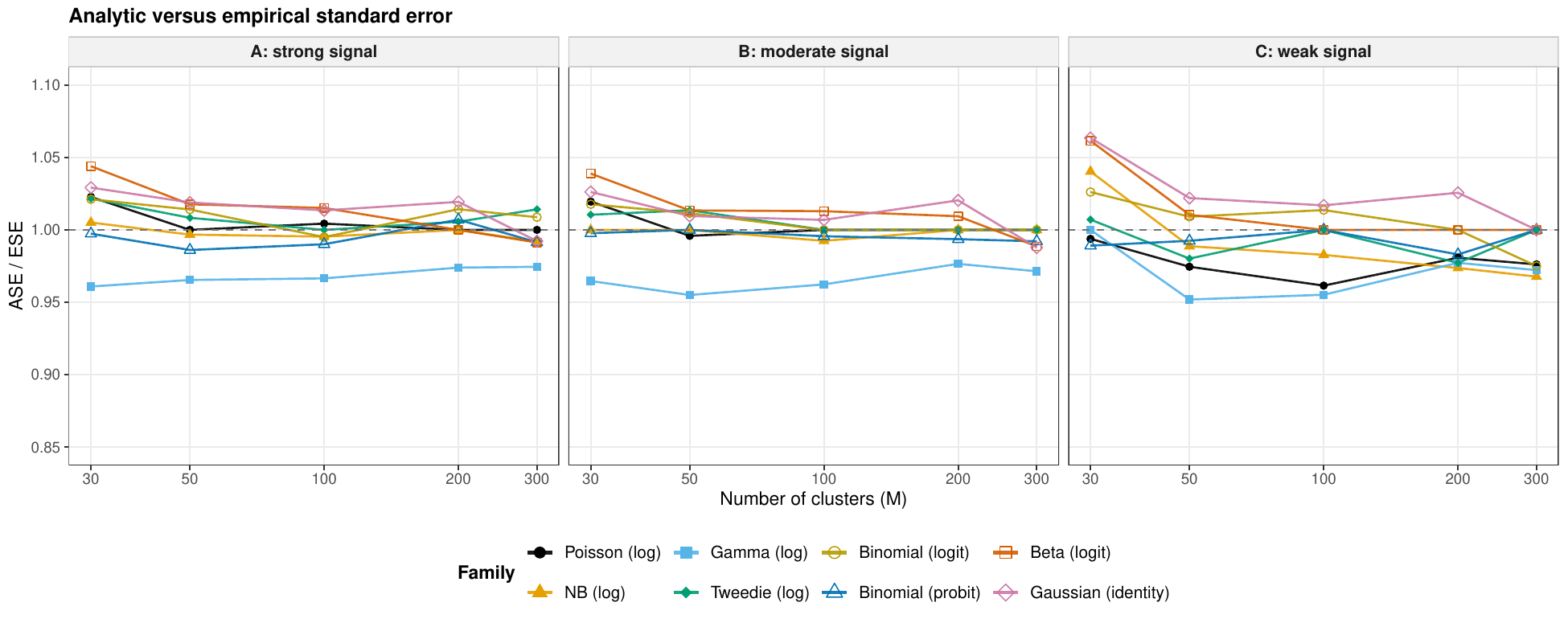}
    \caption{Ratio of the average sandwich standard error (ASE) to the empirical standard deviation of $\hat{R}^{2}$ (ESE) under the random intercept model ($n_j = 10$, $10000$ replications). A ratio of $1.0$ (dashed line) indicates agreement. The sandwich variance estimator is consistent across all families, with $\text{ASE/ESE} \in [0.90, 1.07]$ throughout.}
    \label{fig:ase_ese}
\end{figure}

The ASE/ESE ratio is uniformly close to 1.0 under all specifications for the random intercept model (Figure~\ref{fig:ase_ese}). For the random intercept and slope model under B and C, a mild underestimation (ASE/ESE $\approx 0.91$--$0.95$) is observed for log-link count families (Poisson, NB, Tweedie), reflecting residual sensitivity to the nuisance gradient $\nabla_{\boldsymbol{\gamma}}\rho^2$ in the two-dimensional random-effect integration. Binomial and Gaussian families, for which the nuisance gradient is smaller, maintain ASE/ESE $\approx 1.0$ throughout.

\subsection{Responsiveness to Signal Strength}
\begin{figure}[H]
    \centering
    \includegraphics[width=\textwidth]{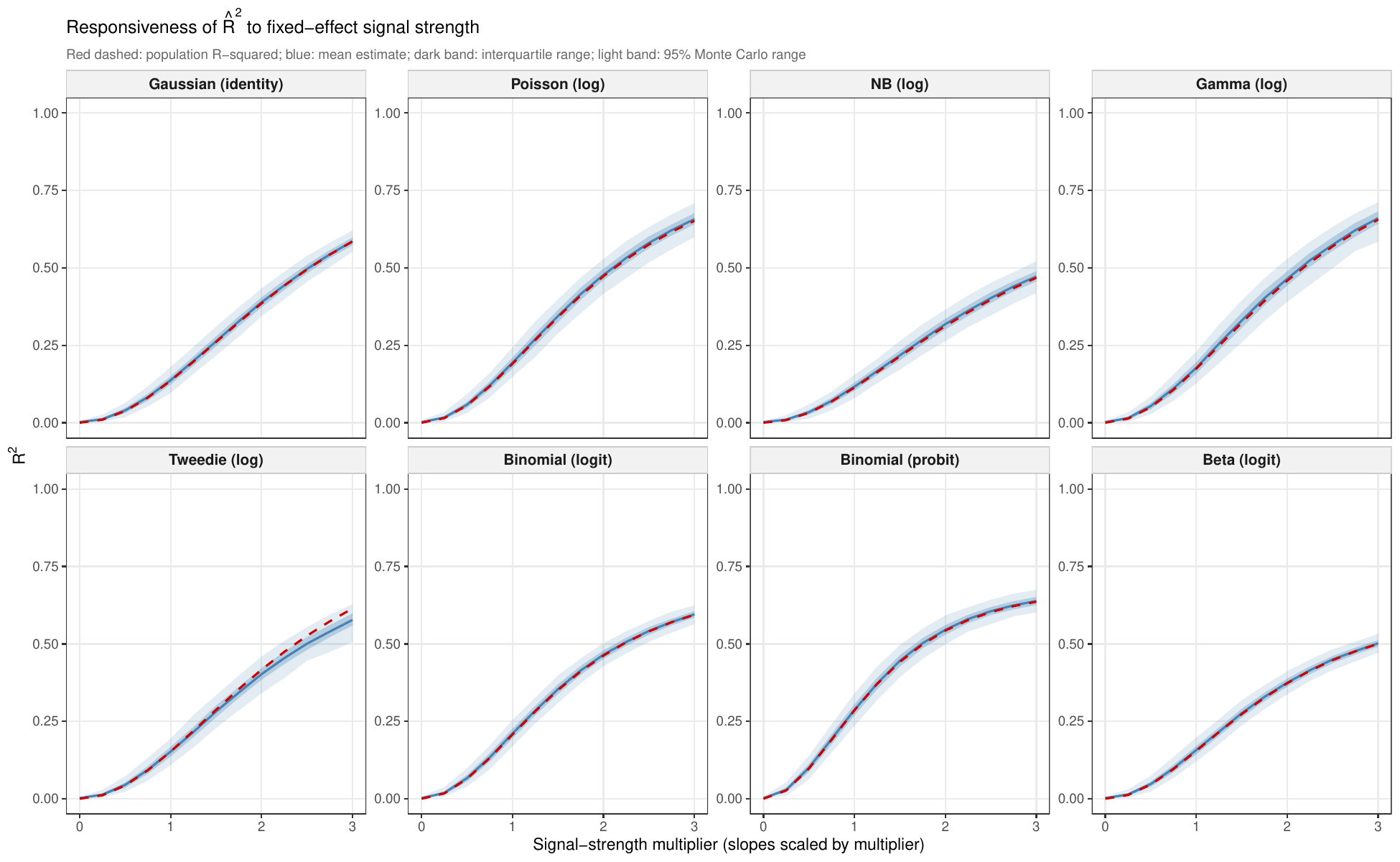}
    \caption{Responsiveness of $\hat{R}^{2}$ to fixed-effect strength $(M = 100, n_j = 10, \sigma_{u}^{2} = 0.25, 1000 \text{ replications})$. The fixed-effect slopes are scaled by a multiplier ranging from $0$ (null model) to $3$ (strong signal); the intercept is held constant. Red dashed line: true population estimand $\rho^{2}_{M}$. Blue solid line: mean $\hat{R}^2$. Dark band: interquartile range. Light band: 2.5th - 97.5th percentile range. The estimator tracks the true $\rho_{M}^{2}$ closely across all eight families, with the expected upward finite-sample bias diminishing as signal strength increases and $\rho^{2}_{M}$ moves away from the boundary.}
    \label{fig:signal}
\end{figure}

We evaluated the responsiveness of the proposed quasi-$R^2$ to the fixed-effects signal strength. Figure \ref{fig:signal} shows the mean estimated $R^2$ over 1000 replications as the fixed-effect coefficients are scaled by a multiplier ranging from 0 to 3. As expected, the estimated $R^2$ exhibits a strict monotonic increase across all eight distributions. Under the null model (multiplier = 0), the estimates approach zero. 

We restrict this sensitivity analysis to the empirically plausible domain (multiplier $\le 3$) to preclude theoretical boundary behaviors at extreme signal strengths, which violate the foundational regularity conditions established in Section 5. For bounded distributions such as the binomial, extreme signals induce complete separation. The predicted probabilities converge to 0 or 1 (violating Assumption R1 in Appendix A), yielding the data-generating mechanism deterministic and causing the Fisher information matrix to degenerate (violating Assumption R4 in Appendix A). In this deterministic limit, the concept of marginal stochastic explained variance loses its statistical interpretation. Conversely, for unbounded count distributions with a log-link (e.g., Poisson), extreme coefficients trigger an exponential growth. The $\exp(\eta)$ mapping generates massive outliers, heavily inflating estimation noise and destroying the uniform boundedness of the moments required by Assumption R3. Within the practically meaningful parameter space, the proposed quasi-$R^2$ reasonably quantifies the proportion of explained variance under standard asymptotic regularity.

Additional simulations involving random-slope GLMMs, as specified in the preregistered simulation protocol, are reported in the Supporting Information.

\section{Case studies}
We apply the proposed marginal quasi-$R^2$ using two ecological longitudinal data sets. The first application is fitted using a Gaussian mixed model and the second application analyzes parasite counts with a GLMM.

\subsection{Density-dependent size variation in macroalgal recruits}
We reanalyzed the macroalgal restoration data collected by
\parencite{mulla_density-driven_2026} and archived in the Dryad Digital Repository \parencite{mulla_data_2026}. The experiment investigated the effects of initial settlement density on the development of \textit{Cystoseira compressa}, a canopy-forming Mediterranean brown alga. Twenty experimental units were assign to four initial densities (10, 100, 250, and 500 individuals per unit) and monitored
at three-month intervals. At each census, the longest axis length (cm) of five randomly selected surviving recruits was recorded for each unit. Our response was individual algal body size. The density-time mean estimand was defined as
$$
\mu_{dt}
=
E\left(Y_{ijt}\mid D_i=d,T=t,S_{ijt}=1\right)
$$
where $Y_{ijt}$ denotes the size of sampled recruit $j$ from unit $i$ at census time $T = t$, $D_i$ is the initial density treatment, and $S_{ijt}=1$ indicates membership in the surviving population. Thus, the estimand describes the mean size of surviving recruits at each census, without attempting to infer the size distribution of recruits that did not survive. The baseline census ($t=0$) was excluded because all initial sizes were recorded as 0.1 cm by experimental design. The resulting analysis contained 400 individual measurements from 20 units across four post-baseline censuses ($t=3,6,9,12$ months).

We specified a Gaussian identity-link linear mixed model with categorical density, categorical census time, and their interaction as fixed effects:
$$
Y_{ijt}
=
\beta_0+\alpha_{D_i}+\tau_t+\delta_{D_i,t} +b_i t+e_{ijt},
$$
where $\beta_0$ is the reference-group intercept, $\alpha_{D_i}$ represents the main effect of initial density, $\tau_t$ represents the categorical census-time effect, and $\delta_{D_i,t}$ represents their interaction. Using density 10 and month 3 as reference levels, we impose $\alpha_{10}=\tau_3=0$ and $\delta_{10,t}=\delta_{d,3}=0$. The unit-specific random slope $b_i$ captures deviations from the treatment-specific mean trajectory, with $t$ entering the random-effect component as numeric time in months. We assume
$$
b_i\sim N(0,A),
\qquad
e_{ijt}\overset{\mathrm{working}}{\sim}N(0,\sigma_e^2).
$$
The categorical interaction allowed each density treatment to have its own temporal mean trajectory without imposing linearity on the fixed mean structure. A unit-specific random slope was included to represent persistent differences in temporal trajectories among experimental units. The corresponding model formula was
$$
\text{individual size}
\sim
\text{density}\times\text{time}
+
(0+\text{numeric time}\mid\text{unit}).
$$

The fitted model had an AIC of 543.65, with an estimated random-slope standard deviation of 0.00911 cm per month and a residual standard deviation of 0.45143 cm. The random-slope variance was small, and the AIC difference relative to the fixed-effects-only Gaussian model was negligible ($543.65$ versus $543.73$). Residual diagnostics revealed increasing dispersion across successive censuses, together with departures from Gaussian tail behavior. Consequently, the homoscedastic Gaussian residual variance was treated as a working assumption rather than an empirically established property of the observations. The estimated marginal quasi-$R^2$ was 0.9752, indicating a strong contribution of the fitted density-specific temporal mean structure under the proposed quasi-$R^2$ decomposition. The HC1 cluster-sandwich standard error was 0.00564, giving a 95\% confidence interval of $[0.9536,\,0.9870]$. The corresponding parametric-bootstrap standard error was 0.00165, with a logit-scale 95\% confidence interval of $[0.9715,\,0.9783]$ (Table~\ref{tab:seaweed-uncertainty}). The parametric-bootstrap standard error was smaller than the HC1 sandwich standard error (0.00165 versus 0.00564). This difference should be interpreted in light of both the working variance specification and the limited number of independent clusters. The observed residual dispersion varied substantially across census times, whereas the parametric bootstrap generated responses under the fitted homoscedastic Gaussian model. In addition, the HC1 procedure involved 20 independent unit clusters and 16 fixed-effect columns, resulting in only four residual cluster degrees of freedom and a substantial finite-sample adjustment. These features may contribute to the difference between the two uncertainty estimates, but the observed discrepancy does not identify a unique source of error or establish the coverage performance of either interval. The bootstrap was therefore used as a complementary model-based numerical comparison rather than as a validation of bootstrap inference under the actual hierarchical sampling mechanism. This example provides a relatively simple Gaussian mixed-model illustration of marginal quasi-$R^2$ estimation and uncertainty.

\begin{table}[tbp]
\centering
\small
\caption{Marginal quasi-$R^2$ uncertainty for the Gaussian
random-slope model of macroalgal recruit size. Parametric-bootstrap
results are based on 2000 unconditional model simulations with
complete model refitting.}
\label{tab:seaweed-uncertainty}
\begin{tabular}{lrrrr}
\toprule
Method
& $\widehat R_M^2$
& SE
& 95\% CI lower
& 95\% CI upper\\
\midrule
HC1 cluster sandwich
& 0.9752 & 0.0056 & 0.9536 & 0.9870\\
Parametric bootstrap
& 0.9752 & 0.0016 & 0.9715 & 0.9783\\
\bottomrule
\end{tabular}

\begin{minipage}{0.96\textwidth}
\footnotesize
\textit{Note.}
$\widehat R_M^2$ denotes the estimated marginal quasi-$R^2$. Sandwich standard errors were clustered by experimental unit. The HC1 procedure used a finite-sample correction and a $t$ reference distribution with four degrees of freedom. The bootstrap standard error was calculated from the standard deviation of 2000 bootstrap estimates. Bootstrap confidence intervals were constructed using a normal approximation on the logit scale and transformed back to the unit interval. Unconditional simulations regenerated unit-level random slopes and individual responses from the fitted Gaussian working model.
\end{minipage}
\end{table}

\subsection{Parasite burden in noise-exposed guppies}

We reanalyzed the guppy parasite-count data from the study of \textcite{masud_noise_2020}, archived in the Dryad Digital Repository \parencite{masud_data_2020}. The experiment included individually housed female guppies (\textit{Poecilia reticulata}) infected with \textit{Gyrodactylus turnbulli}, a skin parasite, and exposed to no-noise, acute-noise, or chronic-noise treatments. Parasite counts were recorded repeatedly over 17 days. The original article reported 52 control fish, whereas the archived spreadsheet contained 54 control identifiers. The records for C53 and C54 were identical to those for C18 and C19, respectively, for every variable other than fish ID. We therefore excluded C53 and C54 as duplicate records. After excluding these duplicates and observations with missing parasite counts, the analysis contained 899 observations from 102 fish.

Following the original study, we specified a log-link GLMM with a random intercept for fish identity and excluded standard length from the conditional mean structure. We included treatment, categorical observation day, and their interaction:
$$
\text{parasite count}
\sim
\text{treatment}\times\text{observation day}
+
(1\mid\text{Fish ID}).
$$
The categorical interaction allowed treatment-specific infection trajectories without imposing a particular functional form over time, reducing the risk that misspecification of the temporal mean structure would be attributed to overdispersion. We fitted the same mean and random-effect structure using Poisson and Negative Binomial (NB2) working families, respectively, with conditional variance functions
$$
\operatorname{Var}(Y_{it}\mid b_i)=\mu_{it}
\qquad\text{and}\qquad
\operatorname{Var}(Y_{it}\mid b_i)
=
\mu_{it}+\frac{\mu_{it}^{2}}{\theta}.
$$

The Negative Binomial GLMM had an AIC of 7507.14, compared with 13951.58 for the Poisson GLMM, giving an AIC difference of 6444.44. Both models used the same categorical treatment-by-day mean specification, log link, and random-effects structure. The substantial AIC difference favors the Negative Binomial working family, consistent with its ability to accommodate extra-Poisson dispersion. Nevertheless, this comparison concerns the complete working likelihoods rather than an isolated variance-function effect. Under the Negative Binomial working model, the estimated marginal quasi-$R^2$ was 0.7914, compared with 0.5343 under the Poisson working model. The corresponding HC1 cluster-sandwich standard errors were 0.0381 and 0.1829, respectively, with the Poisson standard error approximately 4.8 times larger. These differences reflect not only the fitted model parameters but also the working-family-dependent estimating equations and precision geometries. In particular, the two quasi-$R^2$ estimates do not generally represent the same population estimand, because each working family induces its own variance decomposition. Consequently, their numerical difference should not be interpreted as incremental explained variation under a common estimand.

\begin{table}[tbp]
\centering
\small
\caption{Marginal quasi-$R^2$ uncertainty for the guppy
parasite-count models. Parametric-bootstrap results are based
on 2000 unconditional model simulations and complete model
refitting.}
\label{tab:guppy-uncertainty}
\begin{tabular}{llrrrr}
\toprule
Model
& Method
& $\widehat R_M^2$
& SE
& 95\% CI lower
& 95\% CI upper\\
\midrule
Poisson
& HC1 cluster sandwich
& 0.5343
& 0.1829
& 0.2095
& 0.8324\\
Poisson
& Parametric bootstrap
& 0.5343
& 0.1063
& 0.2844
& 0.7681\\
\addlinespace
Negative Binomial
& HC1 cluster sandwich
& 0.7914
& 0.0381
& 0.7055
& 0.8573\\
Negative Binomial
& Parametric bootstrap
& 0.7914
& 0.0544
& 0.7108
& 0.8542\\
\bottomrule
\end{tabular}

\begin{minipage}{0.96\textwidth}
\footnotesize
\textit{Note.}
$\widehat R_M^2$ denotes the estimated marginal quasi-$R^2$. Both models used the same categorical treatment-by-day mean structure, log link, and fish-level random intercept. Sandwich standard errors were clustered by fish. The HC1 procedure applies a finite-sample correction and uses a $t$ reference distribution. The bootstrap SE is the standard deviation of the 2000 bootstrap estimates. Bootstrap confidence intervals were constructed using a normal approximation on the logit scale, with the standard error estimated from the logit-transformed bootstrap replicates. The resulting bounds were transformed back to the unit interval. Unconditional simulations regenerated fish-level random effects and parasite counts from each fitted working model.
\end{minipage}
\end{table}

For the Poisson working family, the parametric-bootstrap standard error was approximately 58\% of the HC1 sandwich standard error (0.106 versus 0.183). The corresponding bootstrap confidence interval $[0.284,\;0.768]$ was narrower than the HC1 interval $[0.210,\;0.832]$. The parametric bootstrap generated responses under the fitted Poisson conditional distribution, whose variance equals its conditional mean. It therefore did not reproduce the extra-Poisson dispersion suggested by the observed data and the substantially better Negative Binomial likelihood fit. This discrepancy in the working variance specification may contribute to the smaller bootstrap standard error. By contrast, the sandwich variance is constructed from empirical fish-cluster influence functions and is designed to accommodate variance misspecification under the stated regularity conditions. Under the Negative Binomial working family, the ordering of the standard errors reversed: the bootstrap standard error exceeded the HC1 sandwich standard error by approximately 43\% (0.054 versus 0.038). However, the corresponding bootstrap confidence interval $[0.711,\;0.854]$ was slightly narrower than the HC1 interval $[0.706,\;0.857]$, with substantial overlap between the two intervals. This reflects the distinction between standard errors reported on the original $R^2$ scale and confidence intervals constructed using different reference distributions and transformations. Each bootstrap replicate re-estimated the model parameters and recomputed the marginal quasi-$R^2$, whereas the analytic sandwich procedure used a first-order influence-function approximation. The observed differences may reflect working model specification, finite-sample behavior, and nonlinear parameter estimation, but do not identify a unique source of discrepancy. These comparisons illustrate that parametric-bootstrap and sandwich uncertainty estimates need not agree in finite samples. The parametric bootstrap describes repeated estimation under the fitted working model, whereas the sandwich procedure targets robustness to variance misspecification through empirical cluster-level contributions. Neither comparison establishes the finite-sample coverage of the resulting intervals under the actual data-generating mechanism. Accordingly, the bootstrap results are presented as complementary model-based numerical comparisons rather than as independent validation of the proposed analytic inference.

\section{Discussion}
In this paper, we defined a marginal $R^2$ for GLMMs from a geometric perspective and developed an inference procedure for its estimator. The central construction consists of two steps that serve distinct statistical purposes. First, marginalization maps the joint distribution of responses and random effects onto the marginal response space through a pushforward operation, thereby specifying the statistical model in which explanatory strength is assessed. Integrating out the latent random effects does not discard their associated uncertainty; rather, latent heterogeneity is retained in the marginal moment structure, potentially affecting both the mean and variance of the response. Second, the Quasi-GLM formulation provides a moment-based statistical structure from which a local information metric can be constructed. Within this geometry, orthogonal projection yields a Pythagorean decomposition under the specified working metric, preserving the geometric interpretation of classical $R^2$ \parencite{amari_information_2016}. Thus, marginalization determines the statistical setting for explanatory assessment, while geometric projection determines how explained and unexplained variation are distinguished. Building on this construction, we defined the population estimand, developed a two-step estimation procedure, derived the influence function and asymptotic distribution of the estimator, and constructed confidence intervals. Under the stated regularity conditions, the resulting inference does not require full specification of the joint distribution of responses and random effects beyond the relevant moment structure. The proposed approach accommodates different link functions and response families within a common geometric and inferential formulation.

The Quasi-GLM construction offers a different perspective on explained variation from existing $R^2$ measures for GLMMs \parencite{nakagawa_general_2013, jaeger_r2_2019, piepho_coefficient_2019}. These approaches provide useful summaries through variance-component decompositions or covariance-based formulations, whereas our construction begins by specifying the marginal statistical space in which explanatory strength is assessed. The distinction is not whether variation is decomposed, but the statistical object and principle underlying the decomposition. Rather than directly attributing explained variation to separate fixed- and random-effect variance components, we first marginalize over the latent random effects and then construct an orthogonal decomposition within the resulting local information geometry. The uncertainty associated with random effects therefore informs the assessment of explanatory strength through the marginal moment structure, without being assigned a separate additive contribution to explained variation. In classical Gaussian linear regression, orthogonal projection provides the geometric basis for the coefficient of determination through the Pythagorean theorem \parencite{montgomery_introduction_1982}. Our construction extends this principle by specifying a local metric appropriate to the marginal quasi-model. Furthermore, some existing measures quantify fixed-effect variation through the empirical distribution of covariates, without explicitly distinguishing a fixed-design estimand from a random-design population target. Our fixed-design formulation instead evaluates the explanatory contribution of auxiliary information within the specified marginal response structure and relative to a defined reference model.

Defining the population estimand does not eliminate the inferential challenges of GLMMs \parencite{jiang_linear_2007}. The precision of the marginal quasi-$R^2$ estimator depends on the model parameters, working family, random-effects structure, and study design, particularly the number of independent clusters \parencite{jiang_usable_2022}. In our construction, a diagonal working covariance provides a tractable basis for quasi-estimation and geometric projection. Although marginalization incorporates random-effect uncertainty into the marginal moments, the working covariance does not explicitly represent the full within-cluster dependence structure. Cluster-robust sandwich inference accommodates working-covariance misspecification asymptotically under appropriate regularity conditions, but does not alter the working geometry defining quasi-$R^2$. Our simulations suggest that more complex settings may require more clusters for first-order confidence intervals to attain nominal coverage. Higher-order corrections may improve coverage in small samples, but are beyond the scope of this paper. Consequently, the magnitude of explained marginal variation and the precision of its estimate are distinct inferential quantities, motivating the joint reporting of marginal quasi-$R^2$ and its confidence interval. Interpretation also depends on the working information geometry. Unlike the homoscedastic Gaussian identity-link case, information weights in non-Gaussian models may depend on the fitted means, so the signs as well as the magnitudes of regression coefficients can affect the resulting quasi-$R^2$. Nevertheless, quasi-$R^2$ is not a signed measure of individual effects. Furthermore, different working families may induce different geometric decompositions and population estimands; their quasi-$R^2$ values should therefore not be interpreted as incremental explained variation without establishing a common inferential target.

For applied researchers, we recommend using marginal quasi-$R^2$ as an inferential summary of the explanatory contribution of auxiliary information, rather than as a standalone criterion for model selection. The proposed measure evaluates explained variation within the marginal response structure, where latent heterogeneity is reflected in the marginal moments underlying the working geometry. Its interpretation therefore depends on the specified auxiliary information, marginal moment structure, working metric, and reference model. The working family, link function, and random-effects structure should be selected according to the scientific question and the characteristics of the data, with particular attention to the adequacy of the mean and variance specifications \parencite{rao_small_2015}. We recommend reporting the marginal quasi-$R^2$ estimate together with its confidence interval, the number of independent clusters, and scientifically relevant regression coefficients. When the number of clusters is limited, confidence intervals based on first-order asymptotic approximations should be interpreted with caution. Comparisons across different working models should also account for possible changes in the underlying estimand. In this way, marginal quasi-$R^2$ provides a complementary measure of explained variation and its estimation uncertainty, while substantive interpretation remains grounded in the scientific question and the fitted model.

\newpage

\appendix
\section{Derivation}
\section*{Notation and Parameter Dependencies:} Throughout this appendix, the quasi-score function $\mathbf{U}$ and the working weight matrix $\mathbf{W}^\star$ depend on both the mean parameters $\boldsymbol{\beta}$ and the GLMM nuisance parameters (the random-effects variance components $\boldsymbol{\psi}$ and the dispersion parameter $\phi$). Let $\boldsymbol{\gamma} = (\boldsymbol{\psi}^T, \phi)^T$ denote the combined nuisance vector, with its true oracle value $\boldsymbol{\gamma}_0 = (\boldsymbol{\psi}^T_0, \phi_0)^T$ and maximum likelihood estimator $\hat{\boldsymbol{\gamma}} = (\hat{\boldsymbol{\psi}}^T, \hat{\phi})^T$. Under the regularity conditions stated below, the nuisance estimators are $\sqrt{M}$-consistent, satisfying $\|\hat{\boldsymbol{\gamma}} - \boldsymbol{\gamma}_0 \| = O_p(M^{-1/2})$. Recognizing that the marginalization depends on both the fixed-effect covariates $\mathbf{x}_i$ and the random-effect design vector $\mathbf{z}_i$, we bundle the complete covariate profile as $\mathbf{d}_i = (\mathbf{x}_i^T, \mathbf{z}_i^T)^T$ and the full parameter vector as $\boldsymbol{\theta} = (\boldsymbol{\beta}^T, \boldsymbol{\gamma}^T)^T$ for notational conciseness. The true parameter is denoted as $\boldsymbol{\theta}_0 = (\boldsymbol{\beta}_0^T, \boldsymbol{\gamma}_0^T)^T$ We denote the quasi-score equation and weight matrix as $\mathbf{U}(\boldsymbol{\theta})$ and $\mathbf{W}^\star(\boldsymbol{\theta})$.

Population quantities $\mathcal{F}_0$, $\mathcal{N}_0$, $\mathbf{C}_0$ defined as in Section 3 (deterministic functions of $\boldsymbol{\theta}_0$ and design).

Under the fixed-design framework, the population quantities $\mathcal{F}_0$, $\mathcal{N}_0$, $\mathbf{C}_0$, $\rho^2$ coincide with their finite-design counterparts $\mathcal{F}_M$, $\mathcal{N}_M$, $\mathbf{C}_M$, $\rho^{2}_{M}$ defined in Section 3, as both are deterministic functions of $\boldsymbol{\theta}_0$ and the realized design. We use the subscript 0 throughout the proofs to emphasize evaluation at the true parameter.

\section*{Regularity Conditions}
To establish the asymptotic properties of the proposed estimation framework, we impose the following foundational regularity conditions.

\textbf{Assumption (R1) (Compactness and Bounded Domain).} Under a fixed design $(\mathbf{X}, \mathbf{Z})$, the true parameter vector $\boldsymbol{\theta}_0 = (\boldsymbol{\beta}_0, \boldsymbol{\gamma}_0)$ lies in the interior of a compact parameter space $\Theta$. The covariates $\mathbf{x}_i$ and $\mathbf{z}_i$ reside in a bounded domain $\mathcal{X}$. Consequently, the linear predictors remain bounded, preventing the marginal means from approaching the boundaries of the parameter space or the singularities of the link functions (e.g., $0$ or $\infty$ for log and inverse-square links).

\textbf{Assumption (R2) (Asymptotic Design and Bounded Clusters).} Under the asymptotic regime where the number of clusters $M \rightarrow \infty$, the individual cluster sizes are bounded by a finite maximum size $n_{\text{max}}$ (i.e., $\max_j n_j \le n_{\text{max}}$). Consequently, the total sample size grows proportionally to the number of clusters ($n \asymp M$), and the maximum relative cluster contribution naturally vanishes: $\max_j n_j / n \le n_{\text{max}} / M \rightarrow 0$. Furthermore, the bounded domain of the covariates (Assumption R1) automatically satisfies the standard Noether condition ($\max_i \mathbf{x}_i^T \mathbf{x}_i / n \rightarrow 0$), preventing any single observation from dominating the asymptotic variance.

\textbf{Assumption (R3) (Integrability and Bounded Moments).} Standard link functions (such as log, probit, logit, and inverse-square) are smoothly differentiable. By the Extreme Value Theorem, the marginalized moments, mean derivatives $m_i^\prime$, variance components, and Quasi-GLM weights $w^\star_i$ are continuously differentiable and bounded over $\Theta \times \mathcal{X}$. Furthermore, the conditional log-likelihood and its derivatives up to the third order with respect to $\boldsymbol{\theta}$ admit integrable envelopes against the Gaussian random-effect measure. Consequently, the independent cluster-level aggregated random vectors (including the cluster-level GLMM marginal scores, the cluster-level quasi-scores $\mathbf{s}_j$, and the quadratic residual forms composing the dispersion $\hat{\mathcal{N}}$) possess uniformly bounded absolute moments of order $4+\delta$ for some $\delta > 0$.

\textbf{Assumption (R4) (Information Limits and Positive Definiteness).} The asymptotic matrices evaluated at the true parameters satisfy:
\begin{enumerate}
    \item Information Limits and Positive Definiteness: For any finite number of clusters $M$, the expected information matrices for both the GLMM nuisance parameters and the Quasi-GLM estimators are positive definite, ensuring parameter identifiability. Furthermore, as $M \to \infty$, the asymptotic average Fisher information matrix $\bar{\mathbf{I}}_{\boldsymbol{\gamma}, 0}$ for the GLMM nuisance parameters and the Quasi-GLM average information matrix $\bar{\mathbf{I}}_0$ both exist and remain positive definite.
    $$\bar{\mathbf{I}}_{\boldsymbol{\gamma}, 0} := \lim_{M \to \infty} M^{-1} \sum_{j=1}^M \E\left[-\frac{\partial^2 \ell_j^{\text{marg}}(\boldsymbol{\gamma}_0)}{\partial \boldsymbol{\gamma} \partial \boldsymbol{\gamma}^T}\right], \quad \bar{\mathbf{I}}_0 := \lim_{M \to \infty} M^{-1} \mathbf{X}^T \mathbf{W}^\star(\boldsymbol{\theta}_0) \mathbf{X}$$
where $\ell_{j}^{\text{marg}} (\boldsymbol{\gamma}) = \log \int p(y_{j} \mid \boldsymbol{u}_j, \boldsymbol{\beta}_0, \phi) p(\boldsymbol{u}_j; \boldsymbol{\psi}) \, d\boldsymbol{u}_j$ is the cluster-$j$ marginal log-likelihood function.
    
    \item Cross-Jacobian Limit: The limit of the expected cross-derivative matrix $\bar{\mathbf{J}}_{\boldsymbol{\gamma}, 0}$ exists and is uniformly bounded.
    $$ \bar{\mathbf{J}}_{\boldsymbol{\gamma}, 0} := \lim_{M\to\infty} \frac{1}{M}\E\left[\frac{\partial \mathbf{U}(\boldsymbol{\beta}_0; \boldsymbol{\gamma}_0)}{\partial \boldsymbol{\gamma}^T}\right]$$
    
    \item Covariance Limit: The average covariance matrix of the augmented parameter scores $\mathbf{s}_j^\star := \mathbf{s}_j + \bar{\mathbf{J}}_{\boldsymbol{\gamma}, 0} \boldsymbol{a}_j$ converges to a positive definite limit $\mathbf{V}^\star$, where $\boldsymbol{a}_j$ is the first-stage influence function from Assumption (R5).
\end{enumerate}

\textbf{Assumption (R5) (First-Stage Regularity).} The first-stage nuisance estimator $\hat{\boldsymbol{\gamma}}$  admits a cluster-level asymptotic linear representation:$$\hat{\boldsymbol{\gamma}} - \boldsymbol{\gamma}_0 = \frac{1}{M}\sum_{j=1}^M \mathbf{a}_j + o_p(M^{-1/2})$$
where $\{ \boldsymbol{a}_j \}_{j = 1}^{M}$ are independent, mean-zero random vectors depending only on cluster $j$'s data, with uniformly bounded $(2 + \delta)$-th absolute moments. For GLMM maximum likelihood estimation, this representation is guaranteed under (R1) - (R4) by \parencite[Lemma 2]{nie_convergence_2007} and \parencite[Part II]{bradley_asymptotic_1962}.

\subsection*{Theoretical Consequences}
These five conditions establish the joint asymptotic validity of the two-stage estimation procedure.

\textbf{$\sqrt{M}$-Consistency of GLMM Nuisance Estimators.} In generalized linear mixed models, the cluster-level data constitute independent but not identically distributed (i.n.i.d.) observations. Under Assumptions (R1) and (R3), the integrability of the conditional derivatives against the Gaussian measure rigorously satisfies Conditions I(i) and I(ii) of \textcite{bradley_asymptotic_1962}. Unbounded-growth links (identity, log, inverse squared) combine the conditional density's natural decay with the dominant squared-exponential tail decay of the Gaussian distribution, permitting the interchange of differentiation and integration via the Dominated Convergence Theorem. Coupled with the positive definite Fisher information (Assumption R4) and bounded cluster sizes (Assumption R2), the uniformly bounded $(2+\delta)$-th moments satisfy Lyapunov's condition. This validates the multivariate Lindeberg condition, extending the i.n.i.d. maximum likelihood regularity \parencite{nie_convergence_2007} to guarantee the $\sqrt{M}$-consistency and asymptotic normality of the nuisance parameters $\hat{\boldsymbol{\psi}}$ and $\hat{\phi}$:
$$ \|\hat{\boldsymbol{\psi}} - \boldsymbol{\psi}_0\| = O_p(M^{-1/2}) \quad \text{and} \quad |\hat{\phi} - \phi_0| = O_p(M^{-1/2}). $$

\textbf{Quasi-GLM Regularity and Joint Asymptotic Normality.} The preceding regularity conditions enable the rigorous derivation of the $\sqrt{M}$-consistency, uniqueness, and asymptotic normality of the Quasi-GLM root in Appendix A, accurately accounting for the nuisance parameter plug-in effect. The augmented cluster scores $\mathbf{s}_j^\star = \mathbf{s}_j + \bar{\mathbf{J}}_{\gamma, 0} \boldsymbol{a}_{j}$ inherit uniformly bounded $(2+\delta)$-th moments from $\mathbf{s}_j$ (R3) and $\boldsymbol{a}_j$ (R5) through the bounded linear transformation coefficients (R4), so Lyapunov's condition is satisfied and the multivariate Lindeberg-Feller theorem yields asymptotic normality of the cluster-summed score. Following the cluster-level M-estimation framework which traces back to \textcite{liang_longitudinal_1986} for GEE and \textcite{heyde_quasi-likelihood_1997} for quasi-likelihood, we map these independent cluster contributions to establish joint asymptotic normality of $\hat{\boldsymbol{\beta}}_{\text{Quasi}}$, with the asymptotic covariance consistently estimated via the sandwich form (Theorem 2). This unified framework operates through analytical moment properties only, rendering it applicable to standard exponential dispersion families with arbitrary continuously differentiable link functions, independently of canonical-link restrictions.

\subsubsection*{Remark}
Assumption (R3) imposes a uniformly bounded $(4+\delta)$-th moment condition on the marginal responses. We emphasize that this specific order is exactly the minimal sufficient requirement to simultaneously establish the Lyapunov Central Limit Theorem for the quasi-scores (which requires $>2$ moments) and the empirical variance components $\hat{\mathcal{N}}$ (which involves squared residuals and hence requires $>4$ moments for its own Lyapunov condition). Given our motivating framework where the conditional distribution belongs to the standard exponential dispersion family (possessing infinitely differentiable moment generating functions on the compact space), this $(4+\delta)$ requirement is trivially and naturally satisfied.

\section*{Proposition 1} Under the GLMM sampling model, the population quasi-score equation $\E[\mathbf{U}_{\text{Quasi}} (\boldsymbol{\beta}; \boldsymbol{\gamma}_0)] = \mathbf{0}$ has a unique solution at $\boldsymbol{\beta} = \boldsymbol{\beta}_0$.

\textbf{Proof.}
The proof consists of two parts: establishing that $\boldsymbol{\beta}_0$ is a root (unbiasedness), and proving its global uniqueness.

\textbf{Part 1: Root Existence (Unbiasedness).} 
Let the true data-generating parameter be $\boldsymbol{\theta}_0 = (\boldsymbol{\beta}_0, \boldsymbol{\gamma}_0)$. By the law of total expectations over the random effects, the unconditional expectation of the response is the marginal mean evaluated at the true parameters: $\E[y_i] = \E_{\mathbf{u}}[\E[y_i \mid \mathbf{u}]] = m_i(\boldsymbol{\beta}_0, \boldsymbol{\gamma}_0)$. 

Taking the expectation of the quasi-score evaluated at $\boldsymbol{\beta} = \boldsymbol{\beta}_0$ yields:
$$
\E[\mathbf{U}_{\text{Quasi}}(\boldsymbol{\beta}_0; \boldsymbol{\gamma}_0)] = \sum_{i=1}^n \mathbf{x}_i \frac{m_i^\prime(\boldsymbol{\beta}_0)}{\V(m_i(\boldsymbol{\beta}_0))} \underbrace{\E[y_i - m_i(\boldsymbol{\beta}_0)]}_{= 0} = \mathbf{0}.
$$
Thus, the true parameter $\boldsymbol{\beta}_0$ satisfies the population quasi-score equation.

\textbf{Part 2: Uniqueness.}
To prove uniqueness, suppose there exists an arbitrary parameter $\boldsymbol{\beta} \in \mathbb{R}^p$ such that $\E[\mathbf{U}_{\text{Quasi}}(\boldsymbol{\beta}; \boldsymbol{\gamma}_0)] = \mathbf{0}$. We must show that $\boldsymbol{\beta} = \boldsymbol{\beta}_0$.

Since $\E[y_i] = m_i(\boldsymbol{\beta}_0)$, the expected score evaluated at $\boldsymbol{\beta}$ is:
$$
\E[\mathbf{U}_{\text{Quasi}}(\boldsymbol{\beta})] = \sum_{i=1}^n \mathbf{x}_i \frac{m_i^\prime(\boldsymbol{\beta})}{\V(m_i(\boldsymbol{\beta}))} \left[ m_i(\boldsymbol{\beta}_0) - m_i(\boldsymbol{\beta}) \right] = \mathbf{0}.
$$
Multiplying this equation by the transposed difference $(\boldsymbol{\beta}_0 - \boldsymbol{\beta})^T$, we obtain the scalar equation:
$$
\sum_{i=1}^n \underbrace{(\mathbf{x}_i^T \boldsymbol{\beta}_0 - \mathbf{x}_i^T \boldsymbol{\beta})}_{\eta_{0i} - \eta_i} \left[ \frac{m_i^\prime(\boldsymbol{\beta})}{\V(m_i(\boldsymbol{\beta}))} \right] \left[ m_i(\boldsymbol{\beta}_0) - m_i(\boldsymbol{\beta}) \right] = 0.
$$
For standard exponential dispersion families with the links considered in this paper, the conditional inverse link function $g^{-1}(\cdot)$ is strictly monotone (either strictly increasing for identity, log, logit, probit, or strictly decreasing for other cases). Because the marginal mean $m_i(\boldsymbol{\beta})$ is the expectation of $g^{-1}(\cdot)$ over the Gaussian measure, the map $\eta_i \mapsto m_i$ is also strictly monotone, ensuring $m_i^\prime(\boldsymbol{\beta}) \ne 0$. Furthermore, the variance function $\V(m_i) > 0$. 

Due to strict monotonicity of $m_i$ in $\eta_i$, the sign of $[m_{i} (\boldsymbol{\beta_0}) - m_{i}(\boldsymbol{\beta})]$ equals to the sign of $(\eta_{0i} - \eta_i)$ multiplied by the sign of $m_i^{\prime}$. Consequently, the term $(\eta_{0i} - \eta_i) \cdot m_{i}^{\prime} (\boldsymbol{\beta}) \cdot [m_{i} (\boldsymbol{\beta_0}) - m_i(\boldsymbol{\beta})]$ is always non-negative:
$$
(\eta_{0i} - \eta_i) \left[ \frac{m_i^\prime(\boldsymbol{\beta})}{\V(m_i(\boldsymbol{\beta}))} \right] \left[ m_i(\boldsymbol{\beta}_0) - m_i(\boldsymbol{\beta}) \right] \ge 0,
$$
with equality holding if and only if $\eta_{0i} = \eta_i$.

Since every term in the sum is guaranteed to be non-negative, for the sum to equal zero, every individual term must be zero. This implies $\mathbf{x}_i^T \boldsymbol{\beta}_0 = \mathbf{x}_i^T \boldsymbol{\beta}$ for all $i = 1, \dots, n$. 

In matrix form, this requires $\mathbf{X} \boldsymbol{\beta}_0 = \mathbf{X} \boldsymbol{\beta}$, or equivalently $\mathbf{X}(\boldsymbol{\beta}_0 - \boldsymbol{\beta}) = \mathbf{0}$. Under the standard assumption that the design matrix $\mathbf{X}$ is of full column rank, the null space of $\mathbf{X}$ is trivial. This implies $\boldsymbol{\beta} = \boldsymbol{\beta}_0$, concluding the proof of global uniqueness.

\section*{Lemma A.1 Quasi-score properties} 
Under Assumptions (R1), (R2), (R3), and (R4), the total Quasi-GLM quasi-score function $\boldsymbol{U}(\boldsymbol{\theta})$ is aggregated at the cluster level as:
$$ \mathbf{U}(\boldsymbol{\theta}) = \sum_{j=1}^M \mathbf{U}_j(\boldsymbol{\theta}) $$
where for each cluster $j = 1, \dots, M$, the cluster-level quasi-score vector $\mathbf{U}_j(\boldsymbol{\theta})$ is defined as:
$$ \mathbf{U}_j(\boldsymbol{\theta}) := \sum_{k=1}^{n_j} \mathbf{x}_{jk} w_{jk}^\star(\boldsymbol{\theta}) R_{jk}(\boldsymbol{\theta}) $$
Here, $j$ indexes the cluster, $k$ indexes the $k$-th observation within cluster $j$, and $n_j$ denotes the cluster size. The marginal residual $R_{jk}(\boldsymbol{\theta})$ is given by:
$$ R_{jk}(\boldsymbol{\theta}) = \frac{y_{jk} - m_{jk}(\boldsymbol{\theta})}{m_{jk}^\prime(\boldsymbol{\theta})} $$
satisfies:
\begin{enumerate}
    \item Unbiasedness at $\boldsymbol{\theta}_0$: $\E[\mathbf{U}(\boldsymbol{\theta}_0)] = \mathbf{0}$.
    \item Positive definiteness of the Information Matrix: The expected information matrix of the quasi-score, $\mathbf{I}(\boldsymbol{\theta}_0) := \E\left[-\frac{\partial \mathbf{U}(\boldsymbol{\theta}_0)}{\partial \boldsymbol{\beta}^T}\right] = \mathbf{X}^T \mathbf{W}^\star(\boldsymbol{\theta}_0) \mathbf{X}$, is positive definite (by Assumption (R4)).
    \item Lipschitz continuity: $\|\mathbf{U}(\boldsymbol{\beta}_1; \boldsymbol{\gamma}_0) - \mathbf{U}(\boldsymbol{\beta}_2; \boldsymbol{\gamma}_0) \| \le L_M \|\boldsymbol{\beta}_1 - \boldsymbol{\beta}_2 \|$ uniformly over $\boldsymbol{\beta}_1, \boldsymbol{\beta}_2 \in \Theta$, with $L_M = O_p(M)$.
\end{enumerate}

\textbf{Proof.} \\
\textbf{Lemma A.1.1 Unbiasedness at the true parameter $\boldsymbol{\theta}_0$: $\E[\mathbf{U}(\boldsymbol{\theta}_0)] = \mathbf{0}$}. \\
Recall that the total quasi-score is $\mathbf{U}(\boldsymbol{\theta}_0) = \sum_{j=1}^M \sum_{k=1}^{n_j} \mathbf{x}_{jk} w_{jk}^\star(\boldsymbol{\theta}_0) R_{jk}(\boldsymbol{\theta}_0)$, where the working residual is $R_{jk}(\boldsymbol{\theta}_0) = (y_{jk} - m_{jk}(\boldsymbol{\theta}_0))/m_{jk}^\prime(\boldsymbol{\theta}_0)$ 

To evaluate the expectation of the quasi-score, we must account for the GLMM data-generating process. By the Law of Iterated Expectations, the marginal expectation of the response $y_{jk}$ given the fixed-effects covariates $\boldsymbol{x}_{jk}$ is obtained by integrating out the cluster-specific random effects $\boldsymbol{u}_j$:
$$\E[y_{jk} \mid \mathbf{x}_{jk}] = \E_{\mathbf{u}_j} \Big[ \E[y_{jk} \mid \mathbf{x}_{jk}, \mathbf{u}_j] \Big] = \E_{\mathbf{u}_j} \big[ g^{-1}(\mathbf{x}_{jk}^T \boldsymbol{\beta}_0 + \mathbf{z}_{jk}^T \mathbf{u}_j) \big]$$
By the definition of the population-averaged mean in the Quasi-GLM framework, this integral is $m_{jk}(\boldsymbol{\theta}_0)$. 

Consequently, the marginal expectation of the working residual evaluates to zero:
$$\E[R_{jk}(\boldsymbol{\theta}_0) \mid \mathbf{x}_{jk}] = \frac{\E[y_{jk} \mid \mathbf{x}_{jk}] - m_{jk}(\boldsymbol{\theta}_0)}{m_{jk}^\prime(\boldsymbol{\theta}_0)} = 0$$
Since the fixed-effects design matrix and the population weights $w_{jk}^\star(\boldsymbol{\theta}_0)$ are deterministic under a fixed-design framework, it follows that $\E[\mathbf{x}_{jk} w_{jk}^\star(\boldsymbol{\theta}_0) R_{jk}(\boldsymbol{\theta}_0)] = \mathbf{0}$. Therefore, the overall expected quasi-score is zero: $\E[\mathbf{U}(\boldsymbol{\theta}_0)] = \mathbf{0}$. \hfill $\blacksquare$

\textbf{Lemma A.1.2 Positive definiteness of the Quasi Information Matrix} \\
Let $\mathbf{I}(\boldsymbol{\theta}_0)$ denote the expected information matrix of the total quasi-score vector $\mathbf{U}(\boldsymbol{\theta})$ evaluated at the true parameter $\boldsymbol{\theta}_0$. By defining the cluster-level sensitivity matrix as $\mathbf{D}_j(\boldsymbol{\theta}) := \E\left[-\frac{\partial \mathbf{U}_j(\boldsymbol{\theta})}{\partial \boldsymbol{\beta}^T}\right]$, the total expected information matrix is aggregated as:
$$ \mathbf{I}(\boldsymbol{\theta}_0) := \E\left[-\frac{\partial \mathbf{U}(\boldsymbol{\theta}_0)}{\partial \boldsymbol{\beta}^T}\right] = \sum_{j=1}^M \mathbf{D}_j(\boldsymbol{\theta}_0) = \mathbf{X}^T \mathbf{W}^\star(\boldsymbol{\theta}_0) \mathbf{X} $$
By Assumption (R4), this finite-sample expected information matrix is assumed to be positive definite, which guarantees the invertibility required for the Quasi-GLM estimation and its subsequent Taylor expansions. \hfill $\blacksquare$

\textbf{Lemma A.1.3 Lipschitz continuity} We express the total quasi-score as $\mathbf{U}(\boldsymbol{\beta}; \boldsymbol{\gamma}_0) = \sum_{j=1}^M \mathbf{U}_j(\boldsymbol{\beta}; \boldsymbol{\gamma}_0)$,
where each cluster contribution is $\boldsymbol{U}_j(\boldsymbol{\beta}; \boldsymbol{\gamma}_0) = \sum_{k=1}^{n_j} \boldsymbol{x}_{jk} w_{jk}^\star(\boldsymbol{\beta}; \boldsymbol{\gamma}_0) R_{jk}(\boldsymbol{\beta}; \boldsymbol{\gamma}_0)$.
Applying the product and chain rules with $\partial \eta_{jk} / \partial \boldsymbol{\beta}^T = \mathbf{x}_{jk}^T$, each per-observation Jacobian decomposes as:
$$
\frac{\partial \mathbf{U}_{jk}(\boldsymbol{\beta}, \boldsymbol{\gamma}_0)}{\partial \boldsymbol{\beta}^T} = \underbrace{\left[ (y_{jk} - m_{jk}(\boldsymbol{\beta}, \boldsymbol{\gamma}_0)) \frac{\partial}{\partial \eta_{jk}}\left( \frac{m_{jk}^\prime(\boldsymbol{\beta}, \boldsymbol{\gamma}_0)}{\V(m_{jk}(\boldsymbol{\beta}, \boldsymbol{\gamma}_0))} \right) - w_{jk}^\star(\boldsymbol{\beta}, \boldsymbol{\gamma}_0) \right]}_{\alpha_{jk}(\boldsymbol{\beta}, \boldsymbol{\gamma}_0)} \mathbf{x}_{jk} \mathbf{x}_{jk}^T,
$$
where $w_{jk}^\star(\boldsymbol{\beta}, \boldsymbol{\gamma}_0) = [m_{jk}^\prime(\boldsymbol{\beta}, \boldsymbol{\gamma}_0)]^2 / \V(m_{jk}(\boldsymbol{\beta}, \boldsymbol{\gamma}_0))$. The decomposition into $\mathbf{x}_{jk} \mathbf{x}_{jk}^T$ isolates the random observation $y_{jk}$ within the scalar multiplier $\alpha_{jk}(\boldsymbol{\beta}, \boldsymbol{\gamma}_0)$.

\textbf{Per-observation bound.} By Assumption (R1), the parameter space $\Theta$ is compact. By Assumption (R3), the weights $w_{jk}^\star (\boldsymbol{\beta}, \boldsymbol{\gamma}_0)$, marginal means $m_{jk}(\boldsymbol{\beta}, \boldsymbol{\gamma}_0)$, and their derivatives are continuously differentiable and uniformly bounded over $\Theta$. Therefore, the scalar multiplier $\alpha_{jk} (\boldsymbol{\beta}; \boldsymbol{\gamma}_0)$, which is a smooth transformation of these bounded deterministic envelopes and linear in the response $y_{jk}$, admits a uniformly bounded first moment. Thus, there exists a universal constant $K < \infty$ such that $\E[\sup_{\boldsymbol{\beta} \in \Theta} | \alpha_{jk} (\boldsymbol{\beta}; \gamma_0) |] \le K$. We directly obtain:
$$\E \left[ \sup_{\boldsymbol{\beta} \in \Theta} \left\| \frac{\partial \mathbf{U}_{jk} (\boldsymbol{\beta}, \boldsymbol{\gamma}_0)}{\partial \boldsymbol{\beta}^T} \right\| \right] \le \E\left[ \sup_{\boldsymbol{\beta} \in \Theta} |\alpha_{jk}(\boldsymbol{\beta}; \boldsymbol{\gamma}_0)| \right] \|\mathbf{x}_{jk}\|^2 \le K \|\mathbf{x}_{jk}\|^2$$
\textbf{Global Jacobian bound.} Summing the per-observation uniform bounds across all $M$ clusters and their respective observations via the triangle inequality yields:
$$ \E \left[ \sup_{\boldsymbol{\beta} \in \Theta} \left\| \frac{\partial \mathbf{U}(\boldsymbol{\beta}; \boldsymbol{\gamma}_0)}{\partial \boldsymbol{\beta}^T} \right\| \right] \le \sum_{j=1}^M \sum_{k=1}^{n_j} \E\left[ \sup_{\boldsymbol{\beta} \in \Theta} \left\| \frac{\partial \mathbf{U}_{jk}(\boldsymbol{\beta}, \boldsymbol{\gamma}_0)}{\partial \boldsymbol{\beta}^T} \right\| \right] \le K \sum_{j=1}^M \sum_{k=1}^{n_j} \|\mathbf{x}_{jk}\|^2. $$

Under the fixed design, the double sum $\sum_{j=1}^M \sum_{k=1}^{n_j} \|\mathbf{x}_{jk}\|^2$ is deterministic. By Assumption (R1), the covariates reside in a bounded domain $\mathcal{X}$, implying there exists a finite constant $C_x$ such that $\sup_{j,k} \|\mathbf{x}_{jk}\|^2 \le C_x$. Furthermore, Assumption (R2) enforces that the maximum cluster size is bounded ($\max_j n_j \le p_0$), which guarantees that the total sample size grows proportionally to the number of clusters, i.e., $\sum_{j=1}^M n_j \le p_0 M$. Consequently, the deterministic sum is bounded by:
$$ \sum_{j=1}^M \sum_{k=1}^{n_j} \|\mathbf{x}_{jk}\|^2 \le \sum_{j=1}^M n_j C_x \le (p_0 C_x) M. $$

Combining this bound with the global Jacobian expectation gives:
$$ \E\left[ \sup_{\boldsymbol{\beta} \in \Theta} \left\| \frac{\partial \mathbf{U}(\boldsymbol{\beta}; \boldsymbol{\gamma}_0)}{\partial \boldsymbol{\beta}^T} \right\| \right] \le K \cdot (p_0 C_x) M. $$
By Markov's inequality, the supremum of the Jacobian norm satisfies $\sup_{\boldsymbol{\beta} \in \Theta} \| \partial \mathbf{U} / \partial \boldsymbol{\beta}^T \| = O_p(M)$ uniformly over the compact parameter space.

\textbf{Lipschitz conclusion.} By the Mean Value Theorem for vector-valued functions, at some intermediate point $\tilde{\boldsymbol{\beta}}$ on this segment:
$$ \| \mathbf{U}(\boldsymbol{\beta}_1; \boldsymbol{\gamma}_0) - \mathbf{U}(\boldsymbol{\beta}_2; \boldsymbol{\gamma}_0) \| \le \left\| \frac{\partial \mathbf{U}(\tilde{\boldsymbol{\beta}}; \boldsymbol{\gamma}_0)}{\partial \boldsymbol{\beta}^T} \right\| \cdot \| \boldsymbol{\beta}_1 - \boldsymbol{\beta}_2 \|. $$
Because $\tilde{\boldsymbol{\beta}} \in \Theta$, the Jacobian norm at this specific intermediate point is upper-bounded by its global supremum over the entire parameter space. Combining this with our previously established global bound gives:
$$ \left\| \frac{\partial \mathbf{U}(\tilde{\boldsymbol{\beta}}; \boldsymbol{\gamma}_0)}{\partial \boldsymbol{\beta}^T} \right\| \le \sup_{\boldsymbol{\beta} \in \Theta} \left\| \frac{\partial \mathbf{U}(\boldsymbol{\beta}; \boldsymbol{\gamma}_0)}{\partial \boldsymbol{\beta}^T} \right\| = O_p(M). $$
Thus, the total quasi-score equation $\mathbf{U}(\boldsymbol{\beta}; \boldsymbol{\gamma}_0)$ satisfies a stochastic Lipschitz condition with coefficient $L_M = \sup_{\boldsymbol{\beta} \in \Theta} \| \partial \mathbf{U} / \partial \boldsymbol{\beta}^T \| = O_p(M)$. \hfill $\blacksquare$

\subsection{Corollary A.1.4 Quasi-score perturbation under nuisance plug-in}
Under Assumptions (R5), the perturbation introduced by plug-in of 
$\hat{\boldsymbol{\gamma}}$ in place of $\boldsymbol{\gamma}_0$ satisfies
$$\mathbf{U}(\boldsymbol{\beta}_0; \hat{\boldsymbol{\gamma}}) - \mathbf{U}(\boldsymbol{\beta}_0; \boldsymbol{\gamma}_0) = O_p(M^{1/2}).$$

\textbf{Proof.} By the Mean Value Theorem applied to $\mathbf{U}(\boldsymbol{\beta}_0; \cdot)$ along the segment between $\boldsymbol{\gamma}_0$ and $\hat{\boldsymbol{\gamma}}$,
$$\mathbf{U}(\boldsymbol{\beta}_0; \hat{\boldsymbol{\gamma}}) - \mathbf{U}(\boldsymbol{\beta}_0; \boldsymbol{\gamma}_0) = \frac{\partial \mathbf{U}(\boldsymbol{\beta}_0; \tilde{\boldsymbol{\gamma}})}{\partial \boldsymbol{\gamma}^T}(\hat{\boldsymbol{\gamma}} - \boldsymbol{\gamma}_0)$$
for some $\tilde{\boldsymbol{\gamma}}$ on the segment. By the analogous Lipschitz argument applied to differentiation in $\boldsymbol{\gamma}$ (which holds under the same R1-R5 conditions via the smooth dependence of $w^\star, m, m'$ on $\boldsymbol{\gamma}$), $\|\partial\mathbf{U}/\partial\boldsymbol{\gamma}^T\| = O_p(M)$. Combined with $\|\hat{\boldsymbol{\gamma}} - \boldsymbol{\gamma}_0\| = O_p(M^{-1/2})$ from the Theoretical Consequences derived under (R5),
$$\|\mathbf{U}(\boldsymbol{\beta}_0; \hat{\boldsymbol{\gamma}}) - \mathbf{U}(\boldsymbol{\beta}_0; \boldsymbol{\gamma}_0)\| \le \left\| \frac{\partial \mathbf{U}}{\partial \boldsymbol{\gamma}^T} \right\| \|\hat{\boldsymbol{\gamma}} - \boldsymbol{\gamma}_0\| = O_p(M^{1/2}).$$ \hfill $\blacksquare$

\section*{Lemma A.2 Existence, uniqueness, consistency of Quasi-score root} Under (R1) -- (R5) with $M \rightarrow \infty$:

(a) There exists a root $\hat{\boldsymbol{\beta}}_{\text{Quasi}}$ of $\mathbf{U}(\boldsymbol{\beta}; \hat{\boldsymbol{\gamma}}) = \mathbf{0}$ in the ball $\| \boldsymbol{\beta} - \boldsymbol{\beta}_0 \| \le \delta M^{-1/2}$ for some finite constant $\delta > 0$. Consequently, the estimator is $\sqrt{M}$-consistent: $\|\hat{\boldsymbol{\beta}}_{\text{Quasi}} - \boldsymbol{\beta}_0\| = O_p(M^{-1/2})$.

\textbf{Proof of (a).} 
To prove the existence of a root in a shrinking $\sqrt{M}$-neighborhood of $\boldsymbol{\beta}_0$, we study the behavior of the quasi-score vector field on the boundary of a $p$-dimensional sphere. Let $B_{M}(\delta) = \{\boldsymbol{\beta}: \|\boldsymbol{\beta} - \boldsymbol{\beta}_0 \| \le \delta M^{-1/2} \}$ be a closed ball centered at the true parameter $\boldsymbol{\beta}_0$ with radius $\delta M^{-1/2}$, where $\delta > 0$ is a constant to be determined. We denote its boundary as $\partial B_M(\delta)$.

By standard topology, the closed ball $B_M(\delta) \subset \mathbb{R}^{p}$ is homeomorphic to the closed unit hypercube $[0, 1]^p$. Applying the homeomorphism preserves the boundary behavior required by Poincaré-Miranda. We invoke the Poincaré-Miranda existence theorem \parencite{miranda_osservazione_1940}. If the continuous vector field $\mathbf{U}(\boldsymbol{\beta}; \hat{\boldsymbol{\gamma}})$ points inward everywhere on the boundary $\partial B_M(\delta)$, then there must exist at least one point inside the ball where $\mathbf{U}(\boldsymbol{\beta}; \hat{\boldsymbol{\gamma}}) = \mathbf{0}$. It suffices to show that the inner product is negative on the boundary with probability approaching one:
$$ \Prob \left( \sup_{\boldsymbol{\beta} \in \partial B_M(\delta)} (\boldsymbol{\beta} - \boldsymbol{\beta}_0)^T \mathbf{U}(\boldsymbol{\beta}; \hat{\boldsymbol{\gamma}}) < 0 \right) \rightarrow 1 \quad \text{as} \quad M \rightarrow \infty. $$

To evaluate this inner product, we define the scalar function $g(t) = (\boldsymbol{\beta} - \boldsymbol{\beta}_0)^T \mathbf{U}(\boldsymbol{\beta}_0 + t(\boldsymbol{\beta} - \boldsymbol{\beta}_0); \hat{\boldsymbol{\gamma}})$ for $t \in [0, 1]$. Applying the standard scalar Mean Value Theorem yields $g(1) - g(0) = g^\prime(\tilde{t})$ for some $\tilde{t} \in (0,1)$. Expanding this derivative gives the orthogonal decomposition:
$$ (\boldsymbol{\beta} - \boldsymbol{\beta}_0)^T \mathbf{U}(\boldsymbol{\beta}; \hat{\boldsymbol{\gamma}}) = \underbrace{(\boldsymbol{\beta} - \boldsymbol{\beta}_0)^T \mathbf{U}(\boldsymbol{\beta}_0; \hat{\boldsymbol{\gamma}})}_{\text{Stochastic term (I)}} - \underbrace{(\boldsymbol{\beta} - \boldsymbol{\beta}_0)^T \mathbf{I}_M(\tilde{\boldsymbol{\beta}}; \hat{\boldsymbol{\gamma}}) (\boldsymbol{\beta} - \boldsymbol{\beta}_0)}_{\text{Deterministic penalty (II)}}, $$
where $\mathbf{I}_M(\boldsymbol{\beta}; \boldsymbol{\gamma}) = -\partial \mathbf{U}(\boldsymbol{\beta}; \boldsymbol{\gamma}) / \partial \boldsymbol{\beta}^T$ is the observed quasi-information matrix, and $\tilde{\boldsymbol{\beta}}= \boldsymbol{\beta}_0 + \tilde{t}(\boldsymbol{\beta} - \boldsymbol{\beta}_0)$ is a specific point on the line segment between $\boldsymbol{\beta}_0$ and $\boldsymbol{\beta}$.

\textbf{1. Bounding the stochastic term (I):}
By Lemma A.1.1, the oracle quasi-score is unbiased: $\E[\mathbf{U}(\boldsymbol{\beta}_0; \boldsymbol{\gamma}_0)] = \mathbf{0}$. Under the uniform moment bounds established in Assumption (R3), the $(2+\delta)$-th absolute moments of the working residuals are uniformly bounded. Coupled with the bounded covariates (R1) and bounded cluster sizes (R2), the cluster-level quasi-scores satisfy the Lyapunov condition, $\sum_{j=1}^M \E[\|\mathbf{U}_j(\boldsymbol{\beta}_0; \boldsymbol{\gamma}_0)\|^{2+\delta}] \le CM$ for some constant $C$. Consequently, by the Lyapunov Central Limit Theorem, Cramér–Wold theorem, and assumption (R4), we have the stochastic order $\| M^{-1/2} \mathbf{U}(\boldsymbol{\beta}_0; \boldsymbol{\gamma}_0) \| = O_p(1)$. Furthermore, Corollary A.1.4 establishes the perturbation bound for the plug-in nuisance estimators as $\|\mathbf{U}(\boldsymbol{\beta}_0; \hat{\boldsymbol{\gamma}}) - \mathbf{U}(\boldsymbol{\beta}_0; \boldsymbol{\gamma}_0)\| = O_p(M^{1/2})$. By the triangle inequality, the evaluated score maintains the root-$M$ order: $\| M^{-1/2} \mathbf{U}(\boldsymbol{\beta}_0; \hat{\boldsymbol{\gamma}}) \| = O_p(1)$.

For any $\boldsymbol{\beta} \in \partial B_M(\delta)$, we have $\| \boldsymbol{\beta} - \boldsymbol{\beta}_0 \| = \delta M^{-1/2}$. Thus, term (I) is bounded via the Cauchy-Schwarz inequality:
$$ | (\boldsymbol{\beta} - \boldsymbol{\beta}_0)^T \mathbf{U}(\boldsymbol{\beta}_0; \hat{\boldsymbol{\gamma}}) | \le \| \boldsymbol{\beta} - \boldsymbol{\beta}_0 \| \cdot  \|\mathbf{U}(\boldsymbol{\beta}_0; \hat{\boldsymbol{\gamma}}) \| = (\delta M^{-1/2}) \cdot (M^{1/2} Z_M) = \delta \cdot Z_M, $$
where $Z_M := \| M^{-1/2} \mathbf{U}(\boldsymbol{\beta}_0; \hat{\boldsymbol{\gamma}}) \| = O_p(1)$ is a stochastically bounded random variable.

\textbf{2. Bounding the deterministic penalty (II):}
We analyze the sample information matrix by decomposing it into three tractable components:
$$ M^{-1} \mathbf{I}_M(\tilde{\boldsymbol{\beta}}; \hat{\boldsymbol{\gamma}}) = \underbrace{M^{-1} \E[\mathbf{I}_M(\boldsymbol{\beta}_0; \boldsymbol{\gamma}_0)]}_{\text{Limit}} + \underbrace{M^{-1} \big( \mathbf{I}_M(\boldsymbol{\beta}_0; \boldsymbol{\gamma}_0) - \E[\mathbf{I}_M(\boldsymbol{\beta}_0; \boldsymbol{\gamma}_0)] \big)}_{\text{Chebyshev's inequality}} + \underbrace{M^{-1} \big( \mathbf{I}_M(\tilde{\boldsymbol{\beta}}; \hat{\boldsymbol{\gamma}}) - \mathbf{I}_M(\boldsymbol{\beta}_0; \boldsymbol{\gamma}_0) \big)}_{\text{Local Smoothness}}. $$

First, we evaluate the deterministic limit of the normalized expected information matrix. Recall from Lemma A.1.2 that the expectation of the observed quasi-information matrix, when evaluated at the true parameters yields:
$$ \E[\mathbf{I}_M(\boldsymbol{\beta}_0; \boldsymbol{\gamma}_0)] = \E\left[ -\frac{\partial \mathbf{U}(\boldsymbol{\beta}_0; \boldsymbol{\gamma}_0)}{\partial \boldsymbol{\beta}^T} \right] = \mathbf{X}^T \mathbf{W}^\star(\boldsymbol{\theta}_0) \mathbf{X}. $$
Applying the $M^{-1}$ normalization, we have $M^{-1} \E[\mathbf{I}_M(\boldsymbol{\beta}_0; \boldsymbol{\gamma}_0)] = M^{-1} \mathbf{X}^T \mathbf{W}^\star(\boldsymbol{\theta}_0) \mathbf{X}$. By the design regularity in Assumption (R4), this normalized deterministic sequence converges to a finite, positive definite asymptotic limit:
$$ \lim_{M \to \infty} M^{-1} \E[\mathbf{I}_M(\boldsymbol{\beta}_0; \boldsymbol{\gamma}_0)] = \lim_{M \to \infty} M^{-1} \mathbf{X}^T \mathbf{W}^\star(\boldsymbol{\theta}_0) \mathbf{X} = \bar{\mathbf{I}}_0. $$
Because the limit matrix $\bar{\mathbf{I}}_0$ is positive definite, its minimal eigenvalue is bounded away from zero, denoted as $\lambda_{\min}(\bar{\mathbf{I}}_0) > 0$. By the continuity of eigenvalues, for sufficiently large $M$, the minimal eigenvalue of the finite-sample expected matrix $M^{-1} \E[\mathbf{I}_M(\boldsymbol{\beta}_0; \boldsymbol{\gamma}_0)]$ is similarly bounded below by some constant arbitrarily close to $\lambda_{\min}(\bar{\mathbf{I}}_0)$.

Second, evaluating the matrix at the fixed true parameter $\boldsymbol{\theta}_0$, the clusters are independent. Because Assumption (R3) bounds the moments of the derivatives, the variance of the sample average diminishes at the rate of $O(M^{-1})$. By Chebyshev's inequality (or the WLLN for independent, non-identically distributed variables), the pointwise deviation vanishes: $M^{-1} \big( \mathbf{I}_M(\boldsymbol{\beta}_0; \boldsymbol{\gamma}_0) - \E[\mathbf{I}_M(\boldsymbol{\beta}_0; \boldsymbol{\gamma}_0)] \big) = o_p(1)$.

Third, Assumption (R3) ensures the third-order derivatives of the quasi-likelihood are uniformly bounded on the compact space $\Theta$ (Assumption R1). By an argument analogous to Lemma A.1.3, bounding the per-observation third-derivative scalar coefficient and summing over observations using the deterministic $\sum_{j, k} \| \mathbf{x}_{jk} \|^3 \le C M$ structure for some constant $C$, the information matrix satisfies a uniform stochastic Lipschitz condition: $ \|\mathbf{I}_M(\tilde\beta; \hat\gamma) - \mathbf{I}_M(\beta_0; \gamma_0)\| \le O_p(M) \cdot (\|\tilde\beta - \beta_0\| + \|\hat\gamma - \gamma_0\|)$. Since $\tilde{\boldsymbol{\beta}} \in B_M(\delta)$ implies $\|\tilde{\boldsymbol{\beta}} - \boldsymbol{\beta}_0\| \le \delta M^{-1/2}$, and the nuisance estimators satisfy $\|\hat{\boldsymbol{\gamma}} - \boldsymbol{\gamma}_0\| = O_p(M^{-1/2})$ by (R5), this uniform smoothness restricts the difference:
$$ M^{-1} \left\| \mathbf{I}_M(\tilde{\boldsymbol{\beta}}; \hat{\boldsymbol{\gamma}}) - \mathbf{I}_M(\boldsymbol{\beta}_0; \boldsymbol{\gamma}_0) \right\| \le M^{-1} \cdot O_p(M) \cdot \left( \|\tilde{\boldsymbol{\beta}} - \boldsymbol{\beta}_0\| + \|\hat{\boldsymbol{\gamma}} - \boldsymbol{\gamma}_0\| \right) = O_p(M^{-1/2}) = o_p(1). $$

Consequently, the sample information matrix converges locally to the positive definite limit: $M^{-1}\mathbf{I}_M(\tilde{\boldsymbol{\beta}}; \hat{\boldsymbol{\gamma}}) = \bar{\mathbf{I}}_0 + o_p(1)$. For sufficiently large $M$, its minimum eigenvalue is bounded below by $c \lambda_{\min}$ for some constant $c \in (0,1)$ (e.g., $c=1/2$) with probability approaching 1. This provides a lower bound for the deterministic quadratic penalty:
\begin{align*}
    (\boldsymbol{\beta} - \boldsymbol{\beta}_0)^T \mathbf{I}_M(\tilde{\boldsymbol{\beta}}; \hat{\boldsymbol{\gamma}}) (\boldsymbol{\beta} - \boldsymbol{\beta}_0) 
    &\ge M c \lambda_{\min} \|\boldsymbol{\beta} - \boldsymbol{\beta}_0\|^2 \\
    &= M c \lambda_{\min} (\delta M^{-1/2})^2 \\
    &= c \lambda_{\min} \delta^2.
\end{align*}

\textbf{3. Establishing the dominance of the penalty and root existence:} 
Combining the bounds from the stochastic term (I) and the deterministic penalty (II), we evaluate the overall inner product uniformly on the boundary $\partial B_M(\delta)$:
\begin{align*}
    (\boldsymbol{\beta} - \boldsymbol{\beta}_0)^T \mathbf{U}(\boldsymbol{\beta}; \hat{\boldsymbol{\gamma}}) 
    &\le \delta Z_M - c \lambda_{\min} \delta^2 \\
    &= \delta (Z_M - c \lambda_{\min} \delta).
\end{align*}

Because $Z_M = O_p(1)$ is stochastically bounded, for any arbitrarily small $\epsilon > 0$, we can specify a sufficiently large, fixed constant $\delta > 0$ such that $\mathbb{P}(Z_M < c \lambda_{\min} \delta) > 1 - \epsilon$ for all sufficiently large $M$. 

When this probabilistic event holds, $Z_M - c \lambda_{\min} \delta < 0$, forcing the inner product to be negative simultaneously for all $\boldsymbol{\beta} \in \partial B_M(\delta)$. By the Poincaré-Miranda theorem, this ensures the existence of at least one root $\hat{\boldsymbol{\beta}}_{\text{Quasi}}$ inside the ball where $\mathbf{U}(\hat{\boldsymbol{\beta}}_{\text{Quasi}}; \hat{\boldsymbol{\gamma}}) = \mathbf{0}$. 

Since $\epsilon$ is arbitrary, the probability of root existence inside $B_M(\delta)$ converges to 1 as $M \rightarrow \infty$. Because this root lies within the defined boundary, it satisfies $\|\hat{\boldsymbol{\beta}}_{\text{Quasi}} - \boldsymbol{\beta}_0\| \le \delta M^{-1/2}$, thereby completing the proof of the $\sqrt{M}$-consistency rate. \hfill $\blacksquare$

(b) This root is unique within the shrinking ball $B_M(\delta)$ with probability approaching one.

To establish the uniqueness of the root $\hat{\boldsymbol{\beta}}_{\text{Quasi}}$ within the compact shrinking neighborhood $B_M(\delta)$, we demonstrate that the quasi-score mapping $\mathbf{U}(\boldsymbol{\beta}; \hat{\boldsymbol{\gamma}})$ is injective in this region. This is algebraically guaranteed if its Jacobian matrix, the negative observed information matrix $-\mathbf{I}_M(\boldsymbol{\beta}; \hat{\boldsymbol{\gamma}})$, is negative definite everywhere inside $B_M(\delta)$.

The stochastic Lipschitz bound derived in Step 2 of part (a) holds uniformly for all points within the defined boundary. Thus, the maximal deviation inside the ball is bounded:
$$ \sup_{\boldsymbol{\beta} \in B_M(\delta)} \left\| M^{-1}\mathbf{I}_M(\boldsymbol{\beta}; \hat{\boldsymbol{\gamma}}) - \bar{\mathbf{I}}_0 \right\| \le \underbrace{\left\| M^{-1}\mathbf{I}_M(\boldsymbol{\beta}_0; \boldsymbol{\gamma}_0) - \bar{\mathbf{I}}_0 \right\|}_{o_p(1) \text{ via Chebyshev's inequality}} + \underbrace{\sup_{\boldsymbol{\beta} \in B_M(\delta)} M^{-1} \left\| \mathbf{I}_M(\boldsymbol{\beta}; \hat{\boldsymbol{\gamma}}) - \mathbf{I}_M(\boldsymbol{\beta}_0; \boldsymbol{\gamma}_0) \right\|}_{o_p(1) \text{ via Lipschitz}}. $$
Because the asymptotic expected information matrix $\bar{\mathbf{I}}_0$ is positive definite with a positive minimum eigenvalue $\lambda_{\min} > 0$, it follows that with probability approaching $1$ as $M \rightarrow \infty$, the observed information matrix $M^{-1}\mathbf{I}_M(\boldsymbol{\beta}; \hat{\boldsymbol{\gamma}})$ remains positive definite uniformly over the entire convex ball $B_M(\delta)$.

Suppose, for the sake of contradiction, that there exist two distinct roots $\hat{\boldsymbol{\beta}}_1 \neq \hat{\boldsymbol{\beta}}_2$ inside $B_M(\delta)$. By the Fundamental Theorem of Calculus, the difference can be expressed as:
$$ \mathbf{0} = \mathbf{U}(\hat{\boldsymbol{\beta}}_1; \hat{\boldsymbol{\gamma}}) - \mathbf{U}(\hat{\boldsymbol{\beta}}_2; \hat{\boldsymbol{\gamma}}) = - \left[ \int_0^1 \mathbf{I}_M \big(\hat{\boldsymbol{\beta}}_2 + t(\hat{\boldsymbol{\beta}}_1 - \hat{\boldsymbol{\beta}}_2); \hat{\boldsymbol{\gamma}} \big) \, dt \right] (\hat{\boldsymbol{\beta}}_1 - \hat{\boldsymbol{\beta}}_2). $$
Because the ball $B_M(\delta)$ is convex, the line segment parameterized by $t \in [0,1]$ lies within the neighborhood. Consequently, the matrix integral represents a convex combination of positive definite matrices, yielding the integrated matrix positive definite itself. Since a positive definite matrix is invertible, multiplying both sides by its inverse definitively yields $\hat{\boldsymbol{\beta}}_1 - \hat{\boldsymbol{\beta}}_2 = \mathbf{0}$, establishing a direct contradiction. 

Therefore, coupled with the existence established in (a), the $\sqrt{M}$-consistent root $\hat{\boldsymbol{\beta}}_{\text{Quasi}}$ is unique within $B_M(\delta)$ with probability approaching 1. \hfill $\blacksquare$

(c) $\hat{\boldsymbol{\beta}}_{\text{Quasi}} \xrightarrow{p} \boldsymbol{\beta}_0$

The weak consistency of the estimator is a direct topological consequence of the existence bounds derived in part(a). By definition, the unique root $\hat{\boldsymbol{\beta}}_{\text{Quasi}}$ is bounded within the closed ball $B_M = \{\boldsymbol{\beta} : \|\boldsymbol{\beta} - \boldsymbol{\beta}_0 \| \le \delta M^{-1/2} \}$ with probability approaching $1$.

Taking the limit as the number of independent clusters $M \rightarrow \infty$, the radius of this ball shrinks to zero:
$$\underset{M  \rightarrow \infty}{\lim} \delta M^{-1/2} = 0$$
Therefore, for any arbitrarily small tolerance $\epsilon > 0$, the probability that the estimator lies outside the $\epsilon$-neighborhood of the true parameter vanishes: $\mathbb{P}(\| \hat{\boldsymbol{\beta}}_{\text{Quasi}} - \boldsymbol{\beta}_0 \| > \epsilon) \rightarrow 0$. This establishes that $\hat{\boldsymbol{\beta}}_\text{Quasi} \xrightarrow{p} \boldsymbol{\beta}_0$, confirming the consistency of the Quasi-GLM estimator. $\blacksquare$

(d) IRLS initialized at any $\boldsymbol{\beta}^{(0)}$ with $\| \boldsymbol{\beta}^{(0)} - \boldsymbol{\beta}_0 \| = O_p(M^{-1/2})$ converges to $\hat{\boldsymbol{\beta}}_{\text{Quasi}}$

The IRLS algorithm (equivalent to Fisher scoring in this context) at iteration $t$ replaces the observed information matrix with the positive definite working information matrix $\mathbf{I}_W(\boldsymbol{\beta}; \hat{\boldsymbol{\gamma}}) = \mathbf{X}^T \mathbf{W}^\star(\boldsymbol{\beta}; \hat{\boldsymbol{\gamma}}) \mathbf{X}$, yielding the update rule:
$$ \boldsymbol{\beta}^{(t + 1)} = \boldsymbol{\beta}^{(t)} + \mathbf{I}_W(\boldsymbol{\beta}^{(t)}; \hat{\boldsymbol{\gamma}})^{-1} \mathbf{U}(\boldsymbol{\beta}^{(t)}; \hat{\boldsymbol{\gamma}}). $$

By definition, the Quasi-GLM estimator satisfies $\mathbf{U}(\hat{\boldsymbol{\beta}}_{\text{Quasi}}; \hat{\boldsymbol{\gamma}}) = \mathbf{0}$. To expand the quasi-score $\mathbf{U}(\boldsymbol{\beta}^{(t)}; \hat{\boldsymbol{\gamma}})$ we apply the Fundamental Theorem of Calculus along the convex line segment parameterized by $s \in [0, 1]$:
$$ \mathbf{U}(\boldsymbol{\beta}^{(t)}; \hat{\boldsymbol{\gamma}}) = \mathbf{U}(\boldsymbol{\beta}^{(t)}; \hat{\boldsymbol{\gamma}}) - \mathbf{U}(\hat{\boldsymbol{\beta}}_{\text{Quasi}}; \hat{\boldsymbol{\gamma}}) = - \underbrace{\left[ \int_0^1 \mathbf{I}_M \big(\hat{\boldsymbol{\beta}}_{\text{Quasi}} + s(\boldsymbol{\beta}^{(t)} - \hat{\boldsymbol{\beta}}_{\text{Quasi}}); \hat{\boldsymbol{\gamma}}\big) \, ds \right]}_{\mathbf{I}_M^{\ast(t)}} (\boldsymbol{\beta}^{(t)} - \hat{\boldsymbol{\beta}}_{\text{Quasi}}), $$
where $\mathbf{I}_M = -\partial \mathbf{U} / \partial \boldsymbol{\beta}^T$ is the observed quasi-information matrix, and $\mathbf{I}_M^{\ast(t)}$ represents the integrated observed information matrix over the path.

Substituting this integral expansion into the IRLS update step yields:
\begin{align*}
    \boldsymbol{\beta}^{(t + 1)} - \hat{\boldsymbol{\beta}}_{\text{Quasi}} &= (\boldsymbol{\beta}^{(t)} - \hat{\boldsymbol{\beta}}_{\text{Quasi}}) - \mathbf{I}_W(\boldsymbol{\beta}^{(t)}; \hat{\boldsymbol{\gamma}})^{-1} \mathbf{I}_M^{\ast(t)} (\boldsymbol{\beta}^{(t)} - \hat{\boldsymbol{\beta}}_{\text{Quasi}}) \\
    &= \left[ \mathbf{I} - \mathbf{I}_W(\boldsymbol{\beta}^{(t)}; \hat{\boldsymbol{\gamma}})^{-1} \mathbf{I}_M^{\ast(t)} \right] (\boldsymbol{\beta}^{(t)} - \hat{\boldsymbol{\beta}}_{\text{Quasi}}).
\end{align*}

To evaluate the contraction mapping matrix inside the brackets, we bound the matrix difference by isolating the integral:
$$ \mathbf{I}_W(\boldsymbol{\beta}^{(t)}; \hat{\boldsymbol{\gamma}}) - \mathbf{I}_M^{\ast(t)} = \int_0^1 \left[ \mathbf{I}_W(\boldsymbol{\beta}^{(t)}; \hat{\boldsymbol{\gamma}}) - \mathbf{I}_M(\boldsymbol{\beta}(s); \hat{\boldsymbol{\gamma}}) \right] ds. $$
For any point $\boldsymbol{\beta}(s)$ on the integration segment, we decompose the integrand into two tractable deviations:
$$ \mathbf{I}_W(\boldsymbol{\beta}^{(t)}; \hat{\boldsymbol{\gamma}}) - \mathbf{I}_M(\boldsymbol{\beta}(s); \hat{\boldsymbol{\gamma}}) = \underbrace{\left[ \mathbf{I}_W(\boldsymbol{\beta}^{(t)}; \hat{\boldsymbol{\gamma}}) - \mathbf{I}_W(\boldsymbol{\beta}(s); \hat{\boldsymbol{\gamma}}) \right]}_{\text{Term A (Working Lipschitz)}} + \underbrace{\left[ \mathbf{I}_W(\boldsymbol{\beta}(s); \hat{\boldsymbol{\gamma}}) - \mathbf{I}_M(\boldsymbol{\beta}(s); \hat{\boldsymbol{\gamma}}) \right]}_{\text{Term B (Residual Perturbation)}}. $$

\textbf{1. Bounding Term A:} Assuming the iterate $\boldsymbol{\beta}^{(t)}$ remains in the shrinking ball $B_M(\delta)$ (which holds initially by the consistency of $\boldsymbol{\beta}^{0}$ and which we verify propagates via the contraction below), both $\boldsymbol{\beta}^{(t)}$ and $\hat{\boldsymbol{\beta}}_{\text{Quasi}}$ reside in the $O_p(M^{-1/2})$-neighborhood of $\boldsymbol{\beta}_0$, implying the intermediate point $\boldsymbol{\beta}(s)$ satisfies $\|\boldsymbol{\beta}^{(t)} - \boldsymbol{\beta}(s)\| \le \|\boldsymbol{\beta}^{(t)} - \hat{\boldsymbol{\beta}}_{\text{Quasi}}\| = O_p(M^{-1/2})$. Because the working weights $w^\star(\boldsymbol{\beta})$ are continuously differentiable over the compact space (Assumption R3), the working information matrix satisfies a deterministic Lipschitz condition scaling at $L_M = O(M)$. Thus, the difference is bounded: 
$$\| \text{Term A} \| \le L_M \|\boldsymbol{\beta}^{(t)} - \boldsymbol{\beta}(s)\| = O_p(M^{1/2}).$$

\textbf{2. Bounding Term B:} The difference between the working and observed information matrices is driven by the residual-dependent terms. We evaluate this difference by pivoting through the true parameter $\boldsymbol{\beta}_0$:
$$ \text{Term B} = \left[ \mathbf{I}_W(\boldsymbol{\beta}_0; \boldsymbol{\gamma}_0) - \mathbf{I}_M(\boldsymbol{\beta}_0; \boldsymbol{\gamma}_0) \right] + \left[ \text{Local Lipschitz perturbation to } \boldsymbol{\beta}(s) \text{ and plug-in error } \hat{\boldsymbol{\gamma}} \right]. $$
At the true parameter $\boldsymbol{\beta}_0$, the residuals have a mean of zero. Because the moments are uniformly bounded by Assumption (R3) and covariates are bounded by (R1), the sum of independent, zero-mean cluster-level matrices has a variance of $O(M)$. By Chebyshev's inequality, this baseline difference scales as $O_p(M^{1/2})$. Furthermore, the uniform Lipschitz property of both matrices ensures that shifting the parameters from $\boldsymbol{\beta}_0$ to $\boldsymbol{\beta}(s)$ adds a perturbation bounded by $O_p(M^{1/2})$. Hence, uniformly across the integral, $\|\text{Term B}\| = O_p(M^{1/2})$.

Combining these bounds, the integrand is uniformly $O_p(M^{1/2})$, implying the integrated deviation evaluates to $\left\|\mathbf{I}_W(\boldsymbol{\beta}^{(t)}; \hat{\boldsymbol{\gamma}}) - \mathbf{I}_M^{\ast(t)}\right\| = O_p(M^{1/2})$. Multiplying by the inverse working matrix $\mathbf{I}_W(\boldsymbol{\beta}^{(t)}; \hat{\boldsymbol{\gamma}})^{-1}$,  where $\left\| \mathbf{I}_W(\boldsymbol{\beta}^{(t)}; \hat{\boldsymbol{\gamma}})^{-1} \right\| = O_p(M^{-1})$ due to its verified $O(M)$ positive definiteness, gives the final contraction rate:
$$\left\| \mathbf{I}_W(\boldsymbol{\beta}^{(t)}; \hat{\boldsymbol{\gamma}})^{-1} \left( \mathbf{I}_W(\boldsymbol{\beta}^{(t)}; \hat{\boldsymbol{\gamma}}) - \mathbf{I}_M^{\ast(t)} \right) \right\| \le \left\| \mathbf{I}_W(\boldsymbol{\beta}^{(t)}; \hat{\boldsymbol{\gamma}})^{-1} \right\| \cdot \left\| \mathbf{I}_W(\boldsymbol{\beta}^{(t)}; \hat{\boldsymbol{\gamma}}) - \mathbf{I}_M^{\ast(t)} \right\| = O_p(M^{-1/2}).$$
Therefore, the contraction matrix simplifies to:
$$ \left\| \mathbf{I} - \mathbf{I}_W(\boldsymbol{\beta}^{(t)}; \hat{\boldsymbol{\gamma}})^{-1} \mathbf{I}_M^{\ast(t)} \right\| = O_p(M^{-1/2}). $$

Applying this scaling to the iterative Euclidean distance yields:
$$ \| \boldsymbol{\beta}^{(t + 1)} - \hat{\boldsymbol{\beta}}_{\text{Quasi}} \| \le O_p(M^{-1/2}) \cdot \|\boldsymbol{\beta}^{(t)} - \hat{\boldsymbol{\beta}}_{\text{Quasi}} \|. $$
Because the contraction factor $O_p(M^{-1/2}) \rightarrow 0$ stochastically as $M \rightarrow \infty$, starting from any properly initialized $\sqrt{M}$-consistent estimate $\|\boldsymbol{\beta}^{(0)} - \hat{\boldsymbol{\beta}}_{\text{Quasi}}\| = O_p(M^{-1/2})$, the algorithm acts as a stochastic contraction mapping. \hfill $\blacksquare$

\section*{Lemma A.3 Consistency of sample moments}
Under Assumption (R1)--(R5) with $M \rightarrow \infty$:

(a) $\hat{\mathbf{C}} \xrightarrow{p} \mathbf{C}_0$

(b) $\hat{\mathcal{F}} \xrightarrow{p} \mathcal{F}_0$

(c) $\hat{\mathcal{N}} \xrightarrow{p} \mathcal{N}_0$

\noindent where $\hat{\mathbf{C}}, \hat{\mathcal{F}}, \hat{\mathcal{N}}$ are evaluated at the IRLS convergence point $\hat{\boldsymbol{\beta}}_{\text{Quasi}}$ with plug-in $\hat{\boldsymbol{\gamma}}$.

(a) Consistency of the sample working-weighted covariance matrix $\hat{\mathbf{C}} \xrightarrow{p} \mathbf{C}_0$.

\textbf{Proof of (a).} 
The sample working-weighted covariance matrix is defined as:
$$ \hat{\mathbf{C}} = n^{-1} \mathbf{X}^T \hat{\mathbf{W}}^\star \mathbf{X} - \frac{\hat{\bar{\mathbf{x}}}_w \hat{\bar{\mathbf{x}}}_w^T}{\hat{s}_n}, $$
where $\hat{\mathbf{W}}^\star = \text{diag}\big(w^\star (\mathbf{x}_i; \hat{\boldsymbol{\beta}}_{\text{Quasi}}, \hat{\boldsymbol{\gamma}})\big)$, $\hat{\bar{\mathbf{x}}}_w = n^{-1} \mathbf{X}^T \hat{\mathbf{W}}^\star \mathbf{1}$, and $\hat{s}_n = n^{-1} \mathbf{1}^T \hat{\mathbf{W}}^\star \mathbf{1}$.

Under the fixed design framework, the design matrix $\mathbf{X}$ is deterministic. The randomness in $\hat{\mathbf{C}}$ originates from the estimated working weights $\hat{\mathbf{W}}^\star$, which depend on the joint parameter estimators $\hat{\boldsymbol{\theta}} = (\hat{\boldsymbol{\beta}}_{\text{Quasi}}^T, \hat{\boldsymbol{\gamma}}^T)^T$. 

By Assumption (R3), the weight function $\boldsymbol{\theta} \mapsto w^\star(\mathbf{x}_i; \boldsymbol{\theta})$ is continuously differentiable with bounded derivatives on the compact parameter space. By the Mean Value Theorem, the maximum absolute deviation in the weights is uniformly bounded by the parameter estimation error:
$$ \max_{1 \le i \le n} \big| \hat{w}^\star_i - w^\star_i \big| \le C \sup_{\mathbf{x} \in \mathcal{X}} \left\| \frac{\partial w^\star}{\partial \boldsymbol{\theta}} \right\| \cdot \|\hat{\boldsymbol{\theta}} - \boldsymbol{\theta}_0\| $$
for some finite constant $C > 0$. By Lemma A.2(a) for $\hat{\boldsymbol{\beta}}$ and Assumption (R5) for $\hat{\boldsymbol{\gamma}}$ the $\sqrt{M}$-Consistency of GLMM nuisance estimators, the parameter estimators are $\sqrt{M}$-consistent, i.e., $\|\hat{\boldsymbol{\theta}} - \boldsymbol{\theta}_0\| = O_p(M^{-1/2})$. Because the covariates are bounded (Assumption R1) and the derivative is uniformly bounded on a compact set, the maximal weight deviation is $O_p(M^{-1/2}) = o_p(1)$.

Using this uniform bound, we evaluate the stochastic perturbation of the leading matrix term:
\begin{align*}
    \left\| n^{-1} \mathbf{X}^T \hat{\mathbf{W}}^\star \mathbf{X} - n^{-1} \mathbf{X}^T \mathbf{W}^\star \mathbf{X} \right\| 
    &\le \left( \max_{1 \le i \le n} \big| \hat{w}^\star_i - w^\star_i \big| \right) \cdot \left\| n^{-1} \sum_{i=1}^n \mathbf{x}_i \mathbf{x}_i^T \right\| \\
    &= O_p(M^{-1/2}).
\end{align*}
Because $O_p(M^{-1/2})$ implies $o_p(1)$, and the deterministic baseline sequence converges by design limits (e.g., $n^{-1} \mathbf{X}^T \mathbf{W}^\star \mathbf{X} \rightarrow \mathbf{C}_{XX}$), it follows that $n^{-1} \mathbf{X}^T \hat{\mathbf{W}}^\star \mathbf{X} \xrightarrow{p} \mathbf{C}_{XX}$, where $\mathbf{C}_{XX} := \lim_n n^{-1} \mathbf{X}^T \mathbf{W}^\star (\boldsymbol{\theta}_0) \mathbf{X}$.

By an identical uniform bounding argument applied to the vector $\hat{\bar{\mathbf{x}}}_w$ and the scalar $\hat{s}$, we obtain $\hat{\bar{\mathbf{x}}}_w \xrightarrow{p} \mathbf{c}_x$ and $\hat{s}_n \xrightarrow{p} s_{n, 0} > 0$, where $\mathbf{c}_x := \lim_n n^{-1} \mathbf{X}^T \mathbf{W}^\star(\boldsymbol{\theta}_0) \boldsymbol{1}$, $s_{n,0} := \lim_n n^{-1} \boldsymbol{1}^T \mathbf{W}^\star (\boldsymbol{\theta}_0) \boldsymbol{1}$ are deterministic asymptotic limits, with $\mathbf{C}_0 = \mathbf{C}_{\mathbf{X} \mathbf{X}} - \mathbf{c}_\mathbf{x} \mathbf{c}_{\mathbf{x}}^T / s_{n, 0}$. Because $\hat{\mathbf{C}}$ is constructed from these three components via finite-dimensional fixed algebraic operations (addition, multiplication, and scalar division by a non-zero positive limit), the standard Slutsky's Theorem and Continuous Mapping Theorem for fixed-dimensional limits apply validly to their combination. Consequently, $\hat{\mathbf{C}} \xrightarrow{p} \mathbf{C}_0$. \hfill $\blacksquare$

(b) $\hat{\mathcal{F}} \xrightarrow{p} \mathcal{F}_0$

$\hat{\mathcal{F}} = \hat{\boldsymbol{\beta}}_{\text{Quasi}}^T \hat{\mathbf{C}} \hat{\boldsymbol{\beta}}_{\text{Quasi}}$.  By part (a), $\hat{\mathbf{C}} \xrightarrow{p} \mathbf{C}_0$; by Lemma A.2(a), $\hat{\boldsymbol{\beta}}_{\text{Quasi}} \xrightarrow{p} \boldsymbol{\beta}_0$. The map $(\boldsymbol{\beta}, \mathbf{C}) \mapsto \boldsymbol{\beta}^T \mathbf{C} \boldsymbol{\beta}$ is continuous. By the continuous mapping theorem, $\hat{\mathcal{F}} \xrightarrow{p} \boldsymbol{\beta}^T_0 \mathbf{C}_0 \boldsymbol{\beta}_0 = \mathcal{F}_0$. $\blacksquare$

(c) $\hat{\mathcal{N}} \xrightarrow{p} \mathcal{N}_0$

Let $\hat{\boldsymbol{\theta}} = (\hat{\boldsymbol{\beta}}_{\text{Quasi}}^T, \hat{\boldsymbol{\gamma}}^T)^T$ be the combined vector of all estimated parameters, and $\boldsymbol{\theta}_0 = (\boldsymbol{\beta}_0^T, \boldsymbol{\gamma}_0^T)^T$ denote their true oracle values. The estimator is defined as $\hat{\mathcal{N}} = n^{-1} \sum_{i=1}^n w_i^\star(\hat{\boldsymbol{\theta}}) R_i(\hat{\boldsymbol{\theta}})^2$, where the generalized working residual is $R_i(\boldsymbol{\theta}) = (y_i - m_i(\boldsymbol{\theta})) / m_i^\prime (\boldsymbol{\theta})$.

By grouping the observations into the $M$ independent clusters, we can express the estimator as a weighted average over clusters:
$$ \hat{\mathcal{N}} = \frac{1}{M} \sum_{j=1}^M \frac{M n_j}{n} \bar{h}_j(\hat{\boldsymbol{\theta}}), $$
where the cluster-level average is defined as $\bar{h}_j(\boldsymbol{\theta}) = n_j^{-1} \sum_{i \in j} h_i(\boldsymbol{\theta})$ with $h_i(\boldsymbol{\theta}) := w_i^\star(\boldsymbol{\theta}) R_i(\boldsymbol{\theta})^2$. 

To isolate the effect of parameter estimation, we decompose this empirical estimator into an oracle term and a plug-in penalty term:
$$ \hat{\mathcal{N}} = \underbrace{ \frac{1}{M} \sum_{j=1}^M \frac{M n_j}{n} \bar{h}_j(\boldsymbol{\theta}_0) }_{\text{Oracle Term } (\mathcal{N}^\ast)} + \underbrace{ \frac{1}{M} \sum_{j=1}^M \frac{M n_j}{n} \left[ \bar{h}_j(\hat{\boldsymbol{\theta}}) - \bar{h}_j(\boldsymbol{\theta}_0) \right] }_{\text{Plug-in Penalty } (\Delta_{\mathcal{N}})}. $$

\textbf{1. Convergence of the Oracle Term (${\mathcal{N}}^\ast$):}
Evaluated at the true parameter $\boldsymbol{\theta}_0$, the cluster averages $\bar{h}_j(\boldsymbol{\theta}_0)$ are independent random variables. Under Assumption (R2), the cluster sizes $n_j$ are uniformly bounded, meaning the weights $(M n_j)/n = O(1)$. By Assumption (R3), the marginal $(4+\delta)$-th moments of $y_i$ are uniformly bounded, which implies the variance components $\E[h_i(\boldsymbol{\theta}_0)]$ and their variances exist and are uniformly bounded. By Chebyshev's inequality, the oracle term converges in probability to its expected limit: 
$$ {\mathcal{N}}^\ast \xrightarrow{p} \lim_{M \to \infty} \frac{1}{M} \sum_{j=1}^M \frac{M n_j}{n} \E[\bar{h}_j(\boldsymbol{\theta}_0)] = \mathcal{N}_0. $$

\textbf{2. Bounding the Plug-in Penalty ($\Delta_{\mathcal{N}}$):}
We define the aggregated scalar function $H_n(\boldsymbol{\theta}) = n^{-1} \sum_{i=1}^n h_i(\boldsymbol{\theta})$. Applying the Mean Value Theorem for scalar functions, we expand $H_n(\hat{\boldsymbol{\theta}})$ around $\boldsymbol{\theta}_0$:
$$ \Delta_{\mathcal{N}} = H_n(\hat{\boldsymbol{\theta}}) - H_n(\boldsymbol{\theta}_0) = \left[ \nabla_{\boldsymbol{\theta}} H_n(\tilde{\boldsymbol{\theta}}) \right]^T (\hat{\boldsymbol{\theta}} - \boldsymbol{\theta}_0) = \left[ \frac{1}{n} \sum_{i=1}^n \nabla_{\boldsymbol{\theta}} h_i(\tilde{\boldsymbol{\theta}}) \right]^T (\hat{\boldsymbol{\theta}} - \boldsymbol{\theta}_0), $$
where $\tilde{\boldsymbol{\theta}}$ lies on the line segment between $\hat{\boldsymbol{\theta}}$ and $\boldsymbol{\theta}_0$.

By the chain rule, $\nabla_{\boldsymbol{\theta}} h_i(\boldsymbol{\theta}) = \nabla_{\boldsymbol{\theta}} w_i^\star \cdot R_i^2 + 2 w_i^\star R_i \cdot \nabla_{\boldsymbol{\theta}} R_i$. Because $R_i(\boldsymbol{\theta})$ is linear in $y_i$, its derivative $\nabla_{\boldsymbol{\theta}} R_i$ also contains terms linear in $y_i$. Consequently, the gradient $\nabla_{\boldsymbol{\theta}} h_i(\boldsymbol{\theta})$ is a polynomial of at most degree 2 in $y_i$. 

Because the covariates are bounded on a compact space (Assumption R1) and the functions $w_i^\star, m_i, m_i^\prime$ are smoothly differentiable (Assumption R3), the coefficients of this polynomial are uniformly bounded on the parameter neighborhood. Coupled with the uniformly bounded second moments of $y_i$ (Assumption R3), the expected absolute gradient is bounded: $ \E[\sup_{\boldsymbol{\theta} \in B_M(\delta)}\|\nabla_{\boldsymbol{\theta}} h_i(\boldsymbol{\theta})\|] < \infty$. By Markov's inequality, the sample average of the gradient is stochastically bounded: $\| \nabla_{\boldsymbol{\theta}} H_n(\tilde{\boldsymbol{\theta}}) \| = O_p(1)$.

Combined with the established root-$M$ consistency of the parameter estimators, $\|\hat{\boldsymbol{\theta}} - \boldsymbol{\theta}_0\| = O_p(M^{-1/2})$, we obtain the rigorous asymptotic bound for the plug-in penalty:
$$ |\Delta_{\mathcal{N}}| \le \| \nabla_{\boldsymbol{\theta}} H_n(\tilde{\boldsymbol{\theta}}) \| \cdot \|\hat{\boldsymbol{\theta}} - \boldsymbol{\theta}_0\| = O_p(M^{-1/2}). $$

Because $O_p(M^{-1/2}) = o_p(1)$, the penalty term vanishes asymptotically. Therefore, $\hat{\mathcal{N}} = {\mathcal{N}}^\ast + \Delta_{\mathcal{N}} = \mathcal{N}_0 + o_p(1)$, establishing consistency. \hfill $\blacksquare$

\section*{Lemma A.4 Convergence of the Integrated Cross-Jacobian}
Define the asymptotic Cross-Jacobian limit:
$$ \bar{\mathbf{J}}_{\boldsymbol{\gamma}, 0} := \lim_{M\to\infty} \frac{1}{M}\E\left[\frac{\partial \mathbf{U}(\boldsymbol{\beta}_0; \boldsymbol{\gamma}_0)}{\partial \boldsymbol{\gamma}^T}\right]. $$
Let the integrated sample cross-Jacobian be defined as $M^{-1}\mathbf{J}_M^\ast = \int_0^1 M^{-1} \mathbf{J}_M(\boldsymbol{\gamma}(s)) \, ds$ where $\boldsymbol{\gamma}(s) = \boldsymbol{\gamma}_0 + s(\hat{\boldsymbol{\gamma}} - \boldsymbol{\gamma}_0)$ for $s \in [0, 1]$, and $\mathbf{J}_M(\boldsymbol{\gamma}) = \sum_{j=1}^M \frac{\partial \mathbf{U}_j(\boldsymbol{\beta}_0; \boldsymbol{\gamma})}{\partial \boldsymbol{\gamma}^T}$. Under Assumptions (R3), (R4), and (R5), it holds that:
$$M^{-1}\mathbf{J}_M^\ast = \bar{\mathbf{J}}_{\boldsymbol{\gamma}, 0} + o_p(1).$$
\textbf{Proof.}
We evaluate the integrand $M^{-1}\mathbf{J}_M(\boldsymbol{\gamma}(s))$ at an arbitrary path parameter $s \in [0, 1]$. To establish its uniform convergence, we utilize a three-term decomposition:
$$\frac{1}{M}\mathbf{J}_M(\boldsymbol{\gamma}(s)) = \underbrace{\frac{1}{M}\E[\mathbf{J}_M(\boldsymbol{\gamma}_0)]}_{\text{Limit}} + \underbrace{\frac{1}{M}\left( \mathbf{J}_M(\boldsymbol{\gamma}_0) - \E[\mathbf{J}_M(\boldsymbol{\gamma}_0)] \right)}_{\text{Chebyshev's inequality}} + \underbrace{\frac{1}{M}\left( \mathbf{J}_M(\boldsymbol{\gamma}(s)) - \mathbf{J}_M(\boldsymbol{\gamma}_0) \right)}_{\text{Local Smoothness}}.$$
First, we evaluate the limit term, by Assumption (R4), the deterministic sequence $M^{-1}\E[\mathbf{J}_{M}(\boldsymbol{\gamma}_{0})]$ converges to the true limit as $M \rightarrow \infty$:
$$\frac{1}{M} \E [\mathbf{J}_{M}(\boldsymbol{\gamma}_{0})] = \bar{\mathbf{J}}_{\boldsymbol{\gamma}, 0} + o(1).$$
Second, we evaluate the sample mean term. Under Assumption (R3), the second-order derivatives of the quasi-likelihood have uniformly bounded moments, so $\text{Var}(\mathbf{J}_{M} (\boldsymbol{\gamma}_{0}) / M) = O(M^{-1})$. By Chebyshev's inequality: 
$$\frac{1}{M}\left( \mathbf{J}_M(\boldsymbol{\gamma}_0) - \E[\mathbf{J}_M(\boldsymbol{\gamma}_0)] \right) = O_p(M^{-1/2}) = o_p(1).$$
Third, we evaluate the local smoothness term. To bound the perturbation caused by replacing the true parameter $\boldsymbol{\gamma}_0$ with the intermediate estimate $\boldsymbol{\gamma}(s)$, we apply the Mean Value Theorem. For some $\tilde{\boldsymbol{\gamma}}$ on the segment connecting $\boldsymbol{\gamma}_0$ and $\boldsymbol{\gamma}(s)$, we have:
$$\left \| \frac{1}{M}\left( \mathbf{J}_M(\boldsymbol{\gamma}(s)) - \mathbf{J}_M(\boldsymbol{\gamma}_0) \right) \right \| \le \left( \frac{1}{M} \sum_{j=1}^M \sup_{\boldsymbol{\gamma} \in \Theta} \left\| \frac{\partial^2 \mathbf{U}_j(\boldsymbol{\beta}_0; \boldsymbol{\gamma})}{\partial \boldsymbol{\gamma} \partial \boldsymbol{\gamma}^T} \right\| \right) \|\boldsymbol{\gamma}(s) - \boldsymbol{\gamma}_0\|.$$

By Assumption (R3) which bounds the third-order derivatives of the quasi-likelihood uniformly, the second derivative $\partial^2 \mathbf{U}_{j} / \partial \boldsymbol{\gamma} \partial \boldsymbol{\gamma}^{T}$ (i.e., the third derivative of the quasi-likelihood) is uniformly bounded. Hence the bracketed supremum term is $O_p(1)$. Furthermore, because $\boldsymbol{\gamma}(s)$ lies on the convex combination path between $\hat{\boldsymbol{\gamma}}$ and $\boldsymbol{\gamma}_0$, its distance to the truth is bounded by the distance of the first-stage estimator:
$$\|\boldsymbol{\gamma}(s) - \boldsymbol{\gamma}_0\| \le \|\hat{\boldsymbol{\gamma}} - \boldsymbol{\gamma}_0\|.$$
By Assumption (R5), the first-stage estimator achieves $\sqrt{M}$-consistency, hence $\|\hat{\boldsymbol{\gamma}} - \boldsymbol{\gamma}_0\| = O_p(M^{-1/2})$. The product yields:
$$\left \| \frac{1}{M}\left( \mathbf{J}_M(\boldsymbol{\gamma}(s)) - \mathbf{J}_M(\boldsymbol{\gamma}_0) \right) \right \| = o_p(1).$$
Substituting the three terms back into the original path evaluation yields:
$$\frac{1}{M}\mathbf{J}_M(\boldsymbol{\gamma}(s)) = \bar{\mathbf{J}}_{\boldsymbol{\gamma}, 0} + o_p(1) + o_p(1) = \bar{\mathbf{J}}_{\boldsymbol{\gamma}, 0} + o_p(1).$$
Because the integrand is uniformly bounded and the integration interval $[0, 1]$ is finite, substituting this back into the fundamental theorem integral yields:
$$M^{-1}\mathbf{J}_M^\ast = \int_0^1 \left( \bar{\mathbf{J}}_{\boldsymbol{\gamma}, 0} + o_p(1) \right) \, ds = \bar{\mathbf{J}}_{\boldsymbol{\gamma}, 0} + o_p(1).$$
This concludes the proof. \hfill $\blacksquare$

\section*{Theorem 1 Consistency of \texorpdfstring{$R^2$}{R2}}
Under Assumptions (R1)--(R5) with $M \rightarrow \infty$,
$$\hat{R}^2 \xrightarrow{p} \rho^2 (\boldsymbol{\theta}_0, \mathbf{X})$$
\textbf{Proof.} By Lemma A.3, $\hat{\mathcal{F}} \xrightarrow{p} \mathcal{F}_0$ and $\hat{\mathcal{N}} \xrightarrow{p} \mathcal{N}_0$ with $\mathcal{N}_0 > 0$ (non-degenerate model). The map $(a, b) \mapsto a / (a + b)$ is continuous at $(\mathcal{F}_0, \mathcal{N}_0)$ with $\mathcal{F}_0 + \mathcal{N}_0 > 0$. By the continuous mapping theorem:
$$\hat{R}^2 = \frac{\hat{\mathcal{F}}}{\hat{\mathcal{F}} + \hat{\mathcal{N}}} \xrightarrow{p} \frac{\mathcal{F}_0}{\mathcal{F}_0 + \mathcal{N}_0} = \rho^2$$ $\blacksquare$

\section*{Theorem 2 Asymptotic normality of $\hat R^2$}

Under Assumptions (R1)--(R5), as $M \to \infty$,
$$ \sqrt{M}(\hat R^2 - \rho^2) \xrightarrow{d} \mathcal{N}(0, \sigma^2_\rho) $$
where the asymptotic variance is defined as
$$ \sigma^2_\rho = \frac{1}{M}\sum_{j=1}^M \text{Var}(\psi_j^{(R^2)}) $$
with $\psi_j^{(R^2)}$ being the cluster-level influence function defined in Section 5. This variance is consistently estimated by the sample variance of the plug-in influence functions $\hat{\psi}_j^{(R^2)}$.

\textbf{Proof.} The proof proceeds in three stages: 
(i) we establish the asymptotic normality and the linear representation of $\hat{\boldsymbol{\beta}}_{\text{Quasi}}$ via the Quasi-GLM score and GLMM marginal score integral expansions, accounting for the nuisance plug-in effect;
(ii) we expand $\hat R^2 - \rho^2$ via the delta method and reduce it to a sum of cluster-level influence functions;
(iii) we invoke the multivariate Lyapunov Central Limit Theorem.

For notational brevity, we write $\hat{\boldsymbol{\beta}} = \hat{\boldsymbol{\beta}}_{\text{Quasi}}$ throughout.

%%% Stage (i) %%%
\textbf{Stage (i): Asymptotic normality and linear representation of $\hat{\boldsymbol{\beta}} - \boldsymbol{\beta}_0$.}

At the Quasi-GLM root, $\mathbf{U}(\hat{\boldsymbol{\beta}}; \hat{\boldsymbol{\gamma}}) = \mathbf{0}$, apply the Fundamental Theorem of Calculus along the convex line segment $\boldsymbol{\beta}(s) = \boldsymbol{\beta}_0 + s(\hat{\boldsymbol{\beta}} - \boldsymbol{\beta}_0)$ for $s \in [0, 1]$:
$$ \mathbf{0} = \mathbf{U}(\hat{\boldsymbol{\beta}}; \hat{\boldsymbol{\gamma}}) = \mathbf{U}(\boldsymbol{\beta}_0; \hat{\boldsymbol{\gamma}}) - \underbrace{\left[ \int_0^1 \mathbf{I}_M(\boldsymbol{\beta}(s); \hat{\boldsymbol{\gamma}}) \, ds \right]}_{=: \mathbf{I}_M^\ast} (\hat{\boldsymbol{\beta}} - \boldsymbol{\beta}_0). $$
Rearranging yields:
\begin{equation}
\sqrt{M}(\hat{\boldsymbol{\beta}} - \boldsymbol{\beta}_0) = \underbrace{\left[M^{-1}\mathbf{I}_M^\ast \right]^{-1}}_{=: \mathbf{A}_M} \cdot \underbrace{M^{-1/2}\,\mathbf{U}(\boldsymbol{\beta}_0; \hat{\boldsymbol{\gamma}})}_{=: \mathbf{B}_M}.
\label{eq:beta_expansion}
\end{equation}
 
\textbf{Convergence of $\mathbf{A}_M$.} By Lemma A.2(a), $\|\hat{\boldsymbol{\beta}} - \boldsymbol{\beta}_0\| = O_p(M^{-1/2})$, implying the entire integration path $\boldsymbol{\beta}(s)$ resides in an $o_p(1)$ neighborhood of $\boldsymbol{\beta}_0$. By Assumption (R5), $\|\hat{\boldsymbol{\gamma}} - \boldsymbol{\gamma}_0\| = O_p(M^{-1/2}) = o_p(1)$. Because the scaled information matrix $M^{-1}\mathbf{I}_M(\boldsymbol{\beta}, \boldsymbol{\gamma})$ is uniformly stochastically continuous on a compact parameter space $\Theta$ (Assumption R1) via the smoothness of $w^\star$ in $(\boldsymbol{\beta}, \boldsymbol{\gamma})$ (Assumption R3) and converges such that $M^{-1}\mathbf{I}_M(\boldsymbol{\beta}_0, \boldsymbol{\gamma}_0) \xrightarrow{p} \bar{\mathbf{I}}_0$, the integrated matrix converges to the same limit:
$$ \mathbf{A}_M = \left[M^{-1}\mathbf{I}_M^\ast \right]^{-1} \xrightarrow{p} \bar{\mathbf{I}}_0^{-1}. $$

\textbf{Decomposition of $\mathbf{B}_M$ via nuisance plug-in.} Similarly, applying the integral expansion to the nuisance parameter path $\boldsymbol{\gamma}(s) = \boldsymbol{\gamma}_0 + s(\hat{\boldsymbol{\gamma}} - \boldsymbol{\gamma}_0)$:
$$ \mathbf{U}(\boldsymbol{\beta}_0; \hat{\boldsymbol{\gamma}}) = \mathbf{U}(\boldsymbol{\beta}_0; \boldsymbol{\gamma}_0) + \underbrace{\left[ \int_0^1 \frac{\partial \mathbf{U}(\boldsymbol{\beta}_0; \boldsymbol{\gamma}(s))}{\partial \boldsymbol{\gamma}^T} ds \right]}_{=: \mathbf{J}_M^\ast} (\hat{\boldsymbol{\gamma}} - \boldsymbol{\gamma}_0). $$
Dividing by $\sqrt{M}$ gives:
\begin{equation}
\mathbf{B}_M = \frac{1}{\sqrt{M}}\mathbf{U}(\boldsymbol{\beta}_0; \boldsymbol{\gamma}_0) + \left[\frac{1}{M} \mathbf{J}_M^\ast \right]\sqrt{M}(\hat{\boldsymbol{\gamma}} - \boldsymbol{\gamma}_0).
\label{eq:B_M_decomp}
\end{equation}

By Lemma A.4, $M^{-1}\mathbf{J}_M^\ast = \bar{\mathbf{J}}_{\boldsymbol{\gamma}, 0} + o_p(1)$. Substituting into \eqref{eq:B_M_decomp} and using $\sqrt{M} (\hat{\boldsymbol{\gamma}} - \boldsymbol{\gamma}_{0}) = O_p(1)$ by Assumption R5,
\begin{equation}
\mathbf{B}_M = \frac{1}{\sqrt{M}}\mathbf{U}(\boldsymbol{\beta}_0; \boldsymbol{\gamma}_0) + \bar{\mathbf{J}}_{\boldsymbol{\gamma}, 0}\sqrt{M}(\hat{\boldsymbol{\gamma}} - \boldsymbol{\gamma}_0) + o_p(1).
\label{eq:B_M_simplified}
\end{equation}

\textbf{Linearization of $\sqrt{M}(\hat{\boldsymbol{\gamma}} - \boldsymbol{\gamma}_0)$.} By Assumption (R5), the first-stage nuisance estimator admits the cluster-level asymptotic linear representation \begin{equation}
\hat{\boldsymbol{\gamma}} - \boldsymbol{\gamma}_0 = M^{-1} \sum_{j = 1}^{M} \boldsymbol{a}_j + o_{p}(M^{-1/2})
\label{eq:gamma_linearization}
\end{equation} 
where $\{ \boldsymbol{a}_j \}$ are independent, mean-zero random vectors with uniformly bounded $(2 + \delta)$-th moments.

\textbf{Combining: $\mathbf{B}_M$ as a sum of cluster-level scores.} 
Recall $\mathbf{U}(\boldsymbol{\beta}_0; \boldsymbol{\gamma}_0) = \sum_j \mathbf{s}_j$, where $\mathbf{s}_j := \sum_{i \in j} w_i^\star \mathbf{x}_i R_i$ is the cluster-$j$ quasi-score (independent across $j$, and zero-mean by Lemma A.1). Substituting \eqref{eq:gamma_linearization} into \eqref{eq:B_M_simplified}:
\begin{align*}
\mathbf{B}_M &= \frac{1}{\sqrt{M}}\sum_{j=1}^M \mathbf{s}_j + \bar{\mathbf{J}}_{\boldsymbol{\gamma}, 0} \cdot \frac{1}{\sqrt{M}}\sum_{j=1}^M \boldsymbol{a}_j + o_p(1) \\
&= \frac{1}{\sqrt{M}}\sum_{j=1}^M \mathbf{s}_j^\star + o_p(1),
\end{align*}
where the \emph{plug-in-corrected cluster score} is defined as:
$$\mathbf{s}_j^\star := \mathbf{s}_j + \bar{\mathbf{J}}_{\boldsymbol{\gamma}, 0} \boldsymbol{a}_j$$
The correction term dictates how cluster $j$'s information regarding the nuisance parameter $\boldsymbol{\gamma}$ propagates into the quasi-score evaluated at $\boldsymbol{\beta}_0$.

\textbf{Lyapunov CLT and Slutsky.} The corrected scores $\{\mathbf{s}_j^\star\}_{j=1}^M$ are independent (each depending solely on cluster $j$'s data) and mean zero. By Assumption (R3), the marginal responses possess uniformly bounded $(4+\delta)$-th moments. Coupled with the bounded covariate space (Assumption R1) and the compact parameter domain the $(2+\delta)$-th absolute moments of the corrected cluster scores are bounded: $\sup_j \E[\|\mathbf{s}_j^\star\|^{2+\delta}] \le C < \infty$. 

This uniform higher-order moment bound ensures that the sequence satisfies the Lyapunov condition (which implies the necessary Lindeberg condition). Therefore, applying the multivariate Lyapunov Central Limit Theorem, we obtain:
$$ \mathbf{B}_M \xrightarrow{d} \mathcal{N}(\mathbf{0}, \mathbf{V}^\star), \quad \mathbf{V}^\star := \lim_{M\to\infty} \frac{1}{M}\sum_{j=1}^M \text{Var}(\mathbf{s}_j^\star). $$
Combined with $\mathbf{A}_M \xrightarrow{p} \bar{\mathbf{I}}_0^{-1}$ via Slutsky's theorem, the asymptotic normality of the estimator is established:
\begin{equation*}
\sqrt{M}(\hat{\boldsymbol{\beta}} - \boldsymbol{\beta}_0) \xrightarrow{d} \mathcal{N}(\mathbf{0}, \bar{\mathbf{I}}_0^{-1}\mathbf{V}^\star\bar{\mathbf{I}}_0^{-1}).
\label{eq:beta_normality}
\end{equation*}

\textbf{Asymptotic Linear Representation of $\hat{\boldsymbol{\beta}}$.} 
Before advancing to the functional delta method, we extract the precise linear expansion of our estimator, which serves as the fundamental building block for the subsequent $R^2$ derivation. Recall from \eqref{eq:beta_expansion} that $\sqrt{M}(\hat{\boldsymbol{\beta}} - \boldsymbol{\beta}_0) = \mathbf{A}_M \mathbf{B}_M$. 
Given that $\mathbf{A}_M = \bar{\mathbf{I}}_0^{-1} + o_p(1)$ and $\mathbf{B}_M = \frac{1}{\sqrt{M}}\sum_{j=1}^M \mathbf{s}_j^\star + o_p(1)$ (where the leading sum is bounded as $O_p(1)$ by the CLT), their product yields:
$$ \sqrt{M}(\hat{\boldsymbol{\beta}} - \boldsymbol{\beta}_0) = \bar{\mathbf{I}}_0^{-1} \left( \frac{1}{\sqrt{M}}\sum_{j=1}^M \mathbf{s}_j^\star \right) + o_p(1). $$
Dividing both sides by $\sqrt{M}$ provides the influence function representation of the parameter estimation error:
\begin{equation*}
\hat{\boldsymbol{\beta}} - \boldsymbol{\beta}_0 = \frac{1}{M} \sum_{j=1}^M \bar{\mathbf{I}}_0^{-1} \mathbf{s}_j^\star + o_p(M^{-1/2}).
\label{eq:beta_linear}
\end{equation*}
This establishes that the estimation error can be decomposed into an average of independent cluster-level contributions, bounded precisely to a high-order remainder.

%%% Stage (ii) %%%
\textbf{Stage (ii): Asymptotic Linearization via Mean Value Theorem.}

We now linearize the components $\hat{\mathcal{F}}$ and $\hat{\mathcal{N}}$. Let $\hat{\boldsymbol{\theta}} = (\hat{\boldsymbol{\beta}}^T, \hat{\boldsymbol{\gamma}}^T)^T$ be the combined estimator and $\boldsymbol{\theta}_0$ be the true parameter. We denote the finite-sample empirical estimators as functions of the parameters: $\hat{\mathcal{F}} = \mathcal{F}(\hat{\boldsymbol{\theta}})$ and $\hat{\mathcal{N}} = \mathcal{N}(\hat{\boldsymbol{\theta}})$. 

\textbf{Linearization of $\hat{\mathcal{F}} - \mathcal{F}_0$.}
$$\hat{\mathcal{F}} - \mathcal{F}_0 = \underbrace{\left[ \mathcal{F}(\hat{\boldsymbol{\theta}}) - \mathcal{F}(\boldsymbol{\theta}_0) \right]}_{\text{Parameter Error}} + \underbrace{\left[ \mathcal{F}(\boldsymbol{\theta}_0) - \mathcal{F}_0 \right]}_{\text{Sampling Error}}$$
Under fixed design, the sampling error term vanishes: because $\mathcal{F}(\boldsymbol{\theta}) = \boldsymbol{\beta}^T \mathbf{C}(\boldsymbol{\theta}) \boldsymbol{\beta}$ depends only on the parameter vector and the deterministic design matrix (with no random-response dependence), $\mathcal{F}(\boldsymbol{\theta}_0) = \mathcal{F}_0$ under fixed design. Hence:
$$\hat{\mathcal{F}} - \mathcal{F}_0 = \underbrace{\left[ \mathcal{F}(\hat{\boldsymbol{\theta}}) - \mathcal{F}(\boldsymbol{\theta}_0) \right]}_{\text{Parameter Error}}$$

For the parameter error, applying the Multivariate Mean Value Theorem (MVT) yields:
$$ \mathcal{F}(\hat{\boldsymbol{\theta}}) - \mathcal{F}(\boldsymbol{\theta}_0) = \nabla_{\boldsymbol{\theta}} \mathcal{F}(\tilde{\boldsymbol{\theta}})^T (\hat{\boldsymbol{\theta}} - \boldsymbol{\theta}_0) $$
where $\tilde{\boldsymbol{\theta}}$ lies on the line segment between $\hat{\boldsymbol{\theta}}$ and $\boldsymbol{\theta}_0$. 

Recall that the joint estimator is $\hat{\boldsymbol{\theta}} = (\hat{\boldsymbol{\beta}}^T, \hat{\boldsymbol{\gamma}}^T)^T$. The $\sqrt{M}$-consistency of the nuisance parameter, $\hat{\boldsymbol{\gamma}} \xrightarrow{p} \boldsymbol{\gamma}_0$, is guaranteed by Assumption (R5). Concurrently, the consistency of the main estimator, $\hat{\boldsymbol{\beta}} \xrightarrow{p} \boldsymbol{\beta}_0$, was established previously via the topological properties of the shrinking ball $B_M$. Combining these results yields the joint convergence in probability: $\hat{\boldsymbol{\theta}} \xrightarrow{p} \boldsymbol{\theta}_0$. 

Because $\tilde{\boldsymbol{\theta}}$ is bounded between $\hat{\boldsymbol{\theta}}$ and $\boldsymbol{\theta}_0$, it follows that $\tilde{\boldsymbol{\theta}} \xrightarrow{p} \boldsymbol{\theta}_0$. By the Continuous Mapping Theorem and the smoothness conditions established in Assumption (R3), the empirical gradient continuously maps the convergent parameter to its population limit: $\nabla_{\boldsymbol{\theta}} \mathcal{F}(\tilde{\boldsymbol{\theta}}) = \nabla_{\boldsymbol{\theta}} \mathcal{F}(\boldsymbol{\theta}_0) + o_p(1)$.

$$\mathcal{F}(\hat{\boldsymbol{\theta}}) - \mathcal{F}(\boldsymbol{\theta}_0) = \left( \nabla_{\boldsymbol{\theta}} \mathcal{F}(\boldsymbol{\theta}_0) + o_p(1) \right)^T (\hat{\boldsymbol{\theta}} - \boldsymbol{\theta}_0)$$

Because the working weight matrix $\mathbf{W}^\star$ depends on the linear predictor via the marginal mean, the covariance metric $\mathbf{C}(\boldsymbol{\theta})$ is a function of both the main parameter $\boldsymbol{\beta}$ and the nuisance parameter $\boldsymbol{\gamma}$. Applying the full product rule to the variance $\mathcal{F}(\boldsymbol{\theta}) = \boldsymbol{\beta}^T \mathbf{C}(\boldsymbol{\theta}) \boldsymbol{\beta}$ resolves the parameter error into its linear components:
$$= \left[ 2\boldsymbol{\beta}_0^T \mathbf{C}_0 + \mathbf{D}_{\boldsymbol{\beta}_0}^T \right] (\hat{\boldsymbol{\beta}} - \boldsymbol{\beta}_0) + \mathbf{D}_{\boldsymbol{\gamma}_0}^T (\hat{\boldsymbol{\gamma}} - \boldsymbol{\gamma}_0) + o_p(M^{-1/2})$$
where $\mathbf{D}_{\boldsymbol{\beta}_0}$ and $\mathbf{D}_{\boldsymbol{\gamma}_0}$ represent the metric gradients evaluated at $\boldsymbol{\theta}_0$, with their $k$-th elements defined as $\boldsymbol{\beta}_0^T \left( \frac{\partial \mathbf{C}_0}{\partial \beta_k} \right) \boldsymbol{\beta}_0$ and $\boldsymbol{\beta}_0^T \left( \frac{\partial \mathbf{C}_0}{\partial \gamma_k} \right) \boldsymbol{\beta}_0$, respectively.

By Assumption (R3), the working weights $w_i^\star(\boldsymbol{\theta})$ are twice continuously differentiable in $\boldsymbol{\theta}$ with uniformly bounded derivatives on $\Theta$ . Consequently, the metric $\mathbf{C}(\boldsymbol{\theta})$ is continuously differentiable, and the partial gradients $\mathbf{D}_{\boldsymbol{\beta}_0}$ and $\mathbf{D}_{\boldsymbol{\gamma}_0}$ are well-defined and bounded.

Substituting the established asymptotic linear representations for the estimation errors gives:
$$= \left[ 2\boldsymbol{\beta}_0^T \mathbf{C}_0 + \mathbf{D}_{\boldsymbol{\beta}_0}^T \right] \left( \frac{1}{M} \sum_{j=1}^M \bar{\mathbf{I}}_0^{-1} \mathbf{s}_j^\star \right) + \mathbf{D}_{\boldsymbol{\gamma}_0}^T \left( \frac{1}{M} \sum_{j=1}^M \boldsymbol{a}_{j} \right) + o_p(M^{-1/2})$$

Combining this parameter error sequence with the sampling error yields the final asymptotic linear form:
$$\hat{\mathcal{F}} - \mathcal{F}_0 = \frac{1}{M}\sum_{j=1}^M \psi_j^{(\mathcal{F})} + o_p(M^{-1/2})$$
where $\psi_j^{(\mathcal{F})} = \left[ 2\boldsymbol{\beta}_0^T \mathbf{C}_0 + \mathbf{D}_{\boldsymbol{\beta}_0}^T \right] \bar{\mathbf{I}}_0^{-1} \mathbf{s}_j^\star + \mathbf{D}_{\boldsymbol{\gamma}_0}^T 
\boldsymbol{a}_j$.

Linearization of $\hat{\mathcal{N}} - \mathcal{N}_0$. To evaluate the variance of the residuals, we rewrite the empirical estimator as $\hat{\mathcal{N}} = \frac{1}{n} \sum_{i=1}^n A_i(\hat{\boldsymbol{\theta}}) \left( y_i - m_i(\hat{\boldsymbol{\theta}}) \right)^2$, 
where the deterministic component is defined as $A_i(\boldsymbol{\theta}) := w_i^\star(\boldsymbol{\theta}) / (m_i^\prime(\boldsymbol{\theta}))^2$. 

Using an identical MVT decomposition strategy for $\mathcal{N}(\hat{\boldsymbol{\theta}})$:

$$\hat{\mathcal{N}} - \mathcal{N}_0 = \nabla_{\boldsymbol{\theta}} \mathcal{N}(\tilde{\boldsymbol{\theta}})^T (\hat{\boldsymbol{\theta}} - \boldsymbol{\theta}_0) + \left[ \mathcal{N}(\boldsymbol{\theta}_0) - \mathcal{N}_0 \right]$$

Under the smoothness established in Assumption (R3), the empirical gradient is locally Lipschitz continuous. Coupled with the Law of Large Numbers and the consistency $\tilde{\boldsymbol{\theta}} \xrightarrow{p} \boldsymbol{\theta}_0$, evaluating the empirical gradient at the intermediate value $\tilde{\boldsymbol{\theta}}$ converges in probability to its population expectation at the true parameter $\boldsymbol{\theta}_0$. Applying the product rule and chain rule to the summation yields the sample gradient:

$$\nabla_{\boldsymbol{\theta}} \mathcal{N}(\boldsymbol{\theta}) = \frac{1}{n} \sum_{i=1}^n \left[ \left( y_i - m_i(\boldsymbol{\theta}) \right)^2 \nabla_{\boldsymbol{\theta}} A_i(\boldsymbol{\theta}) - 2 A_i(\boldsymbol{\theta}) \left( y_i - m_i(\boldsymbol{\theta}) \right) \nabla_{\boldsymbol{\theta}} m_i(\boldsymbol{\theta}) \right]$$ 

Evaluated at any fixed $\boldsymbol{\theta}$, $A_i(\boldsymbol{\theta})$ and $m_i(\boldsymbol{\theta})$ are deterministic under fixed design. The marginal residual has a mean of zero ($\E[y_i - m_i(\boldsymbol{\theta}_0)] = 0$). Consequently, the expectation of the cross-term vanishes:$$\E \left[ 2 A_i(\boldsymbol{\theta}_0) \left( y_i - m_i(\boldsymbol{\theta}_0) \right) \nabla_{\boldsymbol{\theta}} m_i(\boldsymbol{\theta}_0) \right] \propto \E \left[ y_i - m_i(\boldsymbol{\theta}_0) \right] = 0.$$

The expectation of the remaining term simplifies by definition, since $\E[(y_i - m_i(\boldsymbol{\theta}_0))^2] = \text{Var}(y_i)$. This isolates the limit gradient to the derivative of the weights multiplied by the true variance:
$$\mathbf{g}_{\mathcal{N}, \boldsymbol{\theta}} := \text{plim}_{n \to \infty} \nabla_{\boldsymbol{\theta}} \mathcal{N}(\tilde{\boldsymbol{\theta}}) = \lim_{n \to \infty} \frac{1}{n} \sum_{i=1}^n \nabla_{\boldsymbol{\theta}} A_i(\boldsymbol{\theta}_0) \cdot \text{Var}(y_i).$$

Let $\mathbf{g}_{\mathcal{N}, \boldsymbol{\beta}}$ and $\mathbf{g}_{\mathcal{N}, \boldsymbol{\gamma}}$ denote the column vectors corresponding to the partial gradients with respect to $\boldsymbol{\beta}$ and $\boldsymbol{\gamma}$, respectively. 

Returning to the parameter error error term $\mathbf{g}_{\mathcal{N}}^T (\hat{\boldsymbol{\theta}} - \boldsymbol{\theta}_0)$ from the MVT decomposition above: since the joint parameter vector is concatenated as $\boldsymbol{\theta} = (\boldsymbol{\beta}^T, \boldsymbol{\gamma}^T)^T$, we conformably partition the deterministic limit gradient into $\mathbf{g}_{\mathcal{N}} = (\mathbf{g}_{\mathcal{N}, \boldsymbol{\beta}}^T, \mathbf{g}_{\mathcal{N}, \boldsymbol{\gamma}}^T)^T$. Expanding this inner product separates the error into its main and nuisance parameter components:
$$ \text{Parameter Error} = \mathbf{g}_{\mathcal{N}, \boldsymbol{\beta}}^T (\hat{\boldsymbol{\beta}} - \boldsymbol{\beta}_0) + \mathbf{g}_{\mathcal{N}, \boldsymbol{\gamma}}^T (\hat{\boldsymbol{\gamma}} - \boldsymbol{\gamma}_0) + o_p(M^{-1/2}). $$

We now substitute the asymptotic linear representations (influence functions) derived previously for both $\hat{\boldsymbol{\beta}}$ and $\hat{\boldsymbol{\gamma}}$. Because the limit gradients $\mathbf{g}_{\mathcal{N}, \boldsymbol{\beta}}$ and $\mathbf{g}_{\mathcal{N}, \boldsymbol{\gamma}}$ are deterministic constants, the total parameter error can be expressed as a scaled sum of independent cluster influences:
$$ \text{Parameter Error} = \mathbf{g}_{\mathcal{N}, \boldsymbol{\beta}}^T \left( \frac{1}{M} \sum_{j=1}^M \bar{\mathbf{I}}_0^{-1} \mathbf{s}_j^\star \right) + \mathbf{g}_{\mathcal{N}, \boldsymbol{\gamma}}^T \left( \frac{1}{M} \sum_{j=1}^M \boldsymbol{a}_j \right) + o_p(M^{-1/2}). $$

Concurrently, the sampling error evaluated at the true parameters represents the zero-mean random fluctuation of the squared residuals around the true variance:
$$\mathcal{N}(\boldsymbol{\theta}_0) - \mathcal{N}_0 = \frac{1}{M} \sum_{j=1}^M \frac{M}{n} \sum_{i \in j} A_i(\boldsymbol{\theta}_0) \left[ (y_i - m_i(\boldsymbol{\theta}_0))^2 - \text{Var}(y_i) \right]$$

Combining the parameter error and the aggregated sampling error yields the asymptotic linear form for the dispersion component:
$$\hat{\mathcal{N}} - \mathcal{N}_0 = \frac{1}{M}\sum_{j=1}^M \psi_j^{(\mathcal{N})} + o_p(M^{-1/2})$$

where the independent cluster-$j$ influence on $\hat{\mathcal{N}}$ is mathematically assembled as:
$$\psi_j^{(\mathcal{N})} = \mathbf{g}_{\mathcal{N}, \boldsymbol{\beta}}^T \bar{\mathbf{I}}_0^{-1} \mathbf{s}_j^\star + \mathbf{g}_{\mathcal{N}, \boldsymbol{\gamma}}^T \boldsymbol{a}_{j} + \frac{M}{n} \sum_{i \in j} A_i(\boldsymbol{\theta}_0) \left[ (y_i - m_i(\boldsymbol{\theta}_0))^2 - \text{Var}(y_i) \right]$$
 
\textbf{Delta Method for $\hat R^2 - \rho^2$.}
Let $g(\mathcal{F}, \mathcal{N}) = \mathcal{F}/(\mathcal{F}+\mathcal{N})$. Applying the MVT one final time to $g$ at $(\hat{\mathcal{F}}, \hat{\mathcal{N}})$ around $(\mathcal{F}_0, \mathcal{N}_0)$:
$$\hat R^2 - \rho^2 = \nabla g(\tilde{\mathcal{F}}, \tilde{\mathcal{N}})^T \begin{pmatrix} \hat{\mathcal{F}} - \mathcal{F}_0 \\ \hat{\mathcal{N}} - \mathcal{N}_0 \end{pmatrix}$$

Since $\hat{\mathcal{F}} \xrightarrow{p} \mathcal{F}_0$ and $\hat{\mathcal{N}} \xrightarrow{p} \mathcal{N}_0$, the gradient evaluates asymptotically to $\left(\frac{\mathcal{N}_0}{D_0^2}, -\frac{\mathcal{F}_0}{D_0^2}\right)^T$ where $D_0 := \mathcal{F}_0 + \mathcal{N}_0$. Substituting our derived $O_p(M^{-1/2})$ representations yields:
$$\hat R^2 - \rho^2 = \frac{\mathcal{N}_0}{D_0^2}(\hat{\mathcal{F}} - \mathcal{F}_0) - \frac{\mathcal{F}_0}{D_0^2}(\hat{\mathcal{N}} - \mathcal{N}_0) + o_p(M^{-1/2})$$

\textbf{Stage (iii): Combined Cluster Influence and Asymptotic Normality.}

Substituting the individual linearizations into the Delta method equation and multiplying by $\sqrt{M}$, we assemble the final augmented influence function for $\hat{R}^2$:
$$ \sqrt{M}(\hat R^2 - \rho^2) = \frac{1}{\sqrt{M}}\sum_{j=1}^M \psi_j^{(R^2)} + o_p(1) $$
where the combined cluster-level $R^2$ influence is defined as:
$$ \psi_j^{(R^2)} := \frac{\mathcal{N}_0}{D_0^2}\psi_j^{(\mathcal{F})} - \frac{\mathcal{F}_0}{D_0^2}\psi_j^{(\mathcal{N})} $$

To establish the limiting distribution, we verify the conditions for the Central Limit Theorem. Note that $\psi_j^{(R^2)}$ is a linear combination of the nuisance influence function $\boldsymbol{a}_j$, the quasi-scores $\mathbf{s}_j^\star$, and the cluster-aggregated quadratic residuals. By Assumption (R3), these constituent random vectors possess uniformly bounded absolute moments of order $4+\delta$ for some $\delta > 0$. Furthermore, under the bounded domain constraints (Assumption R1) and the bounded cluster sizes (Assumption R2), the scaling coefficients and weights within $\psi_j^{(R^2)}$ are uniformly bounded constants. 

Consequently, the independent, zero-mean sequence $\{ \psi_j^{(R^2)} \}_{j=1}^M$ satisfies the Lyapunov condition for some $0 < \delta^\prime < \delta$ (i.e., $\sup_M \frac{1}{M} \sum_{j=1}^M \E[|\psi_j^{(R^2)}|^{2+\delta^\prime}] < \infty$). An application of the Lyapunov Central Limit Theorem directly establishes the desired joint asymptotic normality:
$$ \sqrt{M} \left( \hat{R}^{2} - \rho^2 \right) \xrightarrow{d} \mathcal{N}(0, \sigma^2_\rho) $$
where the asymptotic sandwich variance is given by $\sigma^2_\rho = \lim_{M \to \infty} \frac{1}{M} \sum_{j=1}^M \text{Var}(\psi_j^{(R^2)})$. $\blacksquare$

\newpage

\section*{Funding}
HQ, AM, KM, and SN were supported by the Canada Excellence Research Chairs (CERC) Program (grant CERC-2022-00074) and the Natural Sciences and Engineering Research Council of Canada (NSERC) Discovery Grants Program (grant RGPIN-2025-04813).

\section*{Competing interest}
The authors declare that no competing interests exist.

\section*{Data Availability Statement}
The data used in the two case studies are openly available in the
Dryad Digital Repository at
\url{https://doi.org/10.5061/dryad.tmpg4f4v6}
and
\url{https://datadryad.org/dataset/doi%3A10.5061/dryad.pc866t24z} respectively.

\section*{Use of AI tools}
During the preparation of this manuscript, ChatGPT (GPT-5.5, OpenAI) and Claude (Opus 5, Anthropic) were used to check mathematical proofs, improve English grammar, and assist coding. All resulting text and code were reviewed and edited as needed. The authors take full responsibility for the content of the article.

\section*{Author Contributions}
\textbf{Hao Qin}: Conceptualization, Methodology, Formal analysis, Validation, Writing – original draft, Writing – review \& editing, Software. \textbf{Ayumi Mizuno}: Conceptualization, Methodology, Formal analysis, Validation, Writing – original draft, Writing – review \& editing, Software. \textbf{Kyle Morrison}: Formal analysis, Validation, Writing – review \& editing, Software. \textbf{Allison K. White}: Formal analysis, Validation, Writing – review \& editing, Software. \textbf{Shinichi Nakagawa}: Conceptualization, Methodology, Validation, Supervision, Project administration, Funding acquisition, Writing – review \& editing.

\newpage

\printbibliography

\end{document}